\PassOptionsToPackage{table}{xcolor}
\documentclass[11pt,a4paper]{article}
\pdfoutput=1
\usepackage{jheppub}
\usepackage[T1]{fontenc}
\usepackage[utf8]{inputenc}
\usepackage{bm,mathtools,booktabs,array,graphicx,etoolbox}
\usepackage{mathrsfs,dsfont,amsthm}
\usepackage{newtxtext,newtxmath}
\usepackage{microtype}
\usepackage{enumitem}
\usepackage{multirow}
\usepackage{tikz}
\usetikzlibrary{arrows.meta,positioning,calc,decorations.markings}
\usepackage{acro}
\acsetup{single=true}
\DeclareAcronym{SM}{short=SM,long=Standard Model}
\DeclareAcronym{SUSY}{short=SUSY,long=supersymmetry}
\DeclareAcronym{SUGRA}{short=SUGRA,long=supergravity}
\DeclareAcronym{FN}{short=FN,long=Froggatt--Nielsen}
\DeclareAcronym{CKM}{short=CKM,long=Cabibbo--Kobayashi--Maskawa}
\DeclareAcronym{GUT}{short=GUT,long=grand unified theory}
\DeclareAcronym{MSSM}{short=MSSM,long=minimal supersymmetric Standard Model}
\DeclareAcronym{PDG}{short=PDG,long=Particle Data Group}
\DeclareAcronym{CG}{short=CG,long=Clebsch--Gordan}

\newcommand{\papertitle}{Jacobi flavor symmetry from the Siegel boundary}
\hypersetup{pdfauthor={Xiang-Gan Liu},pdftitle={\papertitle},bookmarksdepth=3}

\ifdefined\beforetochook
\renewcommand\beforetochook{\pagestyle{myplain}\pagenumbering{roman}\renewcommand\baselinestretch{1.0}\footnotesize}
\fi
\ifdefined\afterTocRuleSpace
\renewcommand\afterTocRuleSpace{\clearpage\renewcommand\baselinestretch{1.1}\normalsize}
\fi
\newcommand{\Z}{\mathds{Z}}

\newcommand{\C}{\mathds{C}}
\newcommand{\HH}{\mathds{H}}
\newcommand{\Sp}{\mathrm{Sp}}
\newcommand{\SL}{\mathrm{SL}}
\newcommand{\GL}{\mathrm{GL}}
\newcommand{\PSp}{\mathrm{PSp}}
\newcommand{\PSU}{\mathrm{PSU}}

\newcommand{\PGL}{\mathrm{PGL}}
\newcommand{\im}{\operatorname{Im}}
\newcommand{\re}{\operatorname{Re}}
\newcommand{\tr}{\operatorname{tr}}
\newcommand{\diag}{\operatorname{diag}}
\newcommand{\eps}{\varepsilon}
\newcommand{\si}{\mathsf{S}_1}
\newcommand{\sii}{\mathsf{S}_2}
\newcommand{\EE}{\mathsf{E}}
\newcommand{\FF}{\mathsf{F}}
\newcommand{\RR}{\mathsf{R}}
\newcommand{\PP}{\mathsf{P}}
\newcommand{\DD}{\mathsf{D}}
\newcommand{\CC}{\mathsf{C}}

\definecolor{inkblue}{RGB}{35,68,105}
\definecolor{softblue}{RGB}{237,243,249}
\definecolor{softred}{RGB}{250,236,234}
\definecolor{inkred}{RGB}{150,45,40}
\definecolor{inkgreen}{RGB}{40,110,60}
\definecolor{surfline}{RGB}{38,70,83}
\definecolor{surftop}{RGB}{241,246,248}
\definecolor{surfbot}{RGB}{198,216,224}
\definecolor{acyc}{RGB}{214,86,64}
\definecolor{bcyc}{RGB}{42,157,143}
\definecolor{modcol}{RGB}{190,132,24}
\definecolor{graphcol}{RGB}{78,96,150}
\newcommand{\tablesetup}[1][1.3]{\renewcommand{\arraystretch}{#1}}
\newcommand{\twoline}[2]{{\renewcommand{\arraystretch}{1.15}\begin{array}{@{}c@{}}#1\\\relax #2\end{array}}}
\newenvironment{tpmatrix}{\renewcommand{\arraystretch}{1}\begin{pmatrix}}{\end{pmatrix}}
\tikzset{box/.style={draw=inkblue,rounded corners=2pt,fill=softblue,align=center,inner sep=5pt},
 cbox/.style={draw=inkred,rounded corners=2pt,fill=softred,align=center,inner sep=5pt},
 arr/.style={-{Latex[length=2mm]},thick,draw=inkblue},
 note/.style={font=\footnotesize,align=center,text=black!75}}

\title{\papertitle}
\author{Xiang-Gan Liu}
\affiliation{Instituto de F\'isica, Universidad Nacional Aut\'onoma de M\'exico, Cd.~de M\'exico, C.P.~04510, M\'exico}
\emailAdd{xgliu@fisica.unam.mx}

\abstract{We extend the analysis of residual genus-two Siegel modular symmetries from interior fixed loci to the boundary of the Siegel moduli space, where one or more combinations of the Siegel moduli become large. For each boundary region we determine the surviving modular group, its finite image at level $N=3$ and its action on fields and couplings. At the Klingen cusp, the surviving modular subsymmetry realizes the Jacobi flavor symmetry proposed by Feruglio and Marrone. The shift of the third modulus supplies the Jacobi form automorphy factor, and the Fourier order in this modulus fixes the index. The five Siegel modular forms of weight 1 and level 3 reproduce the four weight-1 Jacobi multiplets of their construction, at indices $0,3,9$ and $12$. At weight 2 their products fix the relative coefficient of two Yukawa structures. At the other boundaries the same Siegel modular forms yield distinct three-family textures. In the two-ray region, where both diagonal imaginary parts are large and the off-diagonal modulus is finite, a model with real Yukawa coefficients fits the quark mass ratios and mixing parameters, with the CKM phase coming entirely from the moduli.}

\begin{document}
\maketitle
\raggedbottom
\section{Introduction}\label{sec:intro}
The \ac{SM} describes the masses and mixing of quarks and leptons through Yukawa couplings, but it does not explain their values. The charged-fermion masses span several orders of magnitude, and the mixing patterns of quarks and leptons are very different, which suggests that some underlying principle organizes the flavor structure. A traditional approach postulates a flavor symmetry acting on the three families, broken by the vacuum expectation values of additional scalar fields, the flavons \cite{Froggatt:1978nt,Altarelli:2010gt,Feruglio:2019ybq}. The predictions then depend on the vacuum alignment of the flavons, which usually requires an elaborate scalar sector. Modular flavor symmetry offers an alternative \cite{Feruglio:2017spp}: the Yukawa couplings are modular forms of a complex modulus $\tau$, and the vacuum value of the modulus takes over the role of the flavons. This idea is motivated by string compactifications, where target-space modular groups act on the moduli of the compactification space \cite{Baur:2019kwi,Nilles:2020nnc}, and it has developed into a broad program of model building \cite{Kobayashi:2023zzc,Ding:2023htn}.

The vacuum value of the modulus breaks the modular flavor symmetry, but not necessarily completely. A residual symmetry can fix the leading powers of the Yukawa couplings near a special value of a modulus. For instance, at a fixed point such as $\tau=i$ or $\tau=\omega=e^{2\pi i/3}$ a finite stabilizer remains, and fermion mass hierarchies can be read from powers of the distance to the fixed point \cite{Novichkov:2021evw,Feruglio:2021dte}. On the other hand, near the cusp $\im\tau\to\infty$ the translations remain, and the expansion is instead in powers of an exponentially small parameter~\cite{Novichkov:2021evw,Petcov:2023vws,deMedeirosVarzielas:2023crv}.

The situation becomes richer when several moduli are present, since the symmetry can survive on a curve or a surface. The simplest example is the genus-two Siegel modular group $\Sp(4,\Z)$, which acts on three complex moduli and arises, for instance, in heterotic compactifications with a Wilson line \cite{LopesCardoso:1996nc,Baur:2020yjl}. On an interior locus fixed by a finite subgroup, the transformations that preserve the locus act as a modular symmetry on its remaining moduli. We call this group a \emph{modular subsymmetry}, or simply the modular symmetry of the locus.\footnote{We use the same term at a cusp for the integral subgroup preserving the limit.} The interior fixed loci were classified by Gottschling \cite{Gottschling:1961fix}. Their application to fermion masses and mixing was developed in refs.~\cite{Ding:2020zxw,Ding:2021iqp,Ding:2024xhz,Carducci:2026shs}. In this work we extend this treatment to the boundary of the moduli space, where one or more moduli become large and, as at the cusp of genus one, the surviving translations constrain the Fourier expansion of the Yukawa couplings.

Unlike genus one, where $\SL(2,\Z)$ has a single cusp, the genus-two boundary has two types of cusps \cite{Satake:1956,Baily:1966,Igusa:1967,AMRT:1975,Namikawa:1980tor,Hulek:1993mod}. At a Klingen cusp only one diagonal modulus becomes large, and couplings near it still depend on $(\tau,z)$. At a Siegel cusp both diagonal moduli become large, and the couplings depend on the direction of approach, distinguishing a curve (the two-ray region) from the corners at its ends. Each boundary region preserves a different modular subsymmetry of $\Sp(4,\Z)$ and acts differently on matter fields, so the same Siegel modular forms generate distinct mass textures.

The Klingen boundary directly realizes the Jacobi flavor symmetry recently proposed by Feruglio and Marrone \cite{Feruglio:2026xxo}, where the elliptic variable $z$ plays the role of a flavon. While the embedding of the Jacobi group at this cusp was already noted in ref.~\cite{Feruglio:2026xxo} (section~4.2), here we show that the rest of the construction also follows from the Siegel theory: because translations of $z$ simultaneously shift the third modulus $\sigma$, each Fourier--Jacobi coefficient transforms as a Jacobi form with its automorphy factor and index fixed by the Fourier order, while canonical normalization supplies a suppression of higher orders similar to the \ac{FN} mechanism \cite{Froggatt:1978nt}.

We carry out this analysis explicitly at level $N=3$, where the finite Siegel modular group is the double cover of Burkhardt's simple group \cite{Burkhardt:1891,Todd:1947,ATLAS:1985} and five theta functions of the hexagonal lattice form a weight-1 quintet \cite{Freitag:2004burk}. For every positive-dimensional interior locus and every boundary region we determine the modular subsymmetry, its level-3 finite modular group and the branching of this quintet. At the Klingen boundary, all four weight-1 Jacobi form multiplets of Feruglio and Marrone \cite{Feruglio:2026xxo,Gritsenko:2026tb} emerge from a single Fourier--Jacobi expansion of the quintet, whereas full Siegel modular covariance first constrains the theory at weight 2 by fixing the relative coefficient of two independent Jacobi Yukawa structures. Near the Siegel cusp, the same five theta functions yield distinct three-family textures, including a two-ray quark model with CP-conserving couplings and a factorized corner model.

The rest of this paper is organized as follows. Section~\ref{sec:background} introduces $\Sp(4,\Z)$ through the geometry of genus-two surfaces and reviews the Siegel modular theory and its interior invariant subspaces. Section~\ref{sec:cusps} classifies the boundary regions together with their modular subsymmetries and their generalized CP extensions, and describes the resulting pattern of mass hierarchies. Section~\ref{sec:klingen} derives the Jacobi flavor theory at the Klingen boundary. Section~\ref{sec:models} constructs three-family models in the other boundary regions and follows a single globally covariant coupling, the Hessian of Burkhardt's quartic, from the interior to the corner. Section~\ref{sec:topdown} discusses the heterotic interpretation and other string-theoretic aspects, and section~\ref{sec:conclusions} gives our conclusions. The group presentations, the representations used in the models, the explicit Jacobi forms, the higher-level theta series and the rule for the leading Fourier powers at the corner are collected in the appendices.

\section{Siegel modular symmetry at level 3}\label{sec:background}
We first recall how $\Sp(4,\Z)$ arises from the geometry of genus-two surfaces and then collect the ingredients of a Siegel modular flavor model \cite{Ding:2020zxw,Ding:2021iqp}, describe the finite modular group at level 3 and construct its modular forms explicitly. We then turn to the elliptic invariant subspaces and the corresponding modular subsymmetries. General background on Siegel modular forms can be found in refs.~\cite{Freitag:1983,Klingen:1990,vanderGeer:2008sm}.

\subsection{Genus-two surfaces and the Siegel modular group}\label{sec:moduli}
\begin{figure}[t]
\centering
\begin{tikzpicture}[scale=1.3,
  surf/.style={thick,draw=surfline,top color=surftop,bottom color=surfbot},
  cyc/.style={very thick,postaction={decorate,decoration={markings,mark=at position #1 with {\arrow{Latex[length=1.7mm,width=1.5mm]}}}}}]
\shadedraw[surf]
  (-2.55,0) .. controls (-2.55,1.12) and (-1.08,1.22) .. (0,0.66)
            .. controls (1.08,1.22) and (2.55,1.12) .. (2.55,0)
            .. controls (2.55,-1.12) and (1.08,-1.22) .. (0,-0.66)
            .. controls (-1.08,-1.22) and (-2.55,-1.12) .. (-2.55,0) -- cycle;
\foreach \x in {-1.23,1.23}{
 \draw[cyc=0.25,bcyc] (\x,0.0) ellipse (0.80 and 0.46);
 \fill[white] ({\x-0.42},-0.02) .. controls ({\x-0.20},0.16) and ({\x+0.20},0.16) .. ({\x+0.42},-0.02)
              .. controls ({\x+0.20},-0.19) and ({\x-0.20},-0.19) .. ({\x-0.42},-0.02) -- cycle;
 \draw[thick,surfline] ({\x-0.55},0.10) .. controls ({\x-0.25},-0.22) and ({\x+0.25},-0.22) .. ({\x+0.55},0.10);
 \draw[thick,surfline] ({\x-0.42},-0.02) .. controls ({\x-0.20},0.16) and ({\x+0.20},0.16) .. ({\x+0.42},-0.02);
 \draw[thick,acyc,dashed] (\x,-0.14) .. controls ({\x-0.17},-0.45) and ({\x-0.17},-0.76) .. (\x,-0.99);
 \draw[cyc=0.55,acyc] (\x,-0.14) .. controls ({\x+0.17},-0.45) and ({\x+0.17},-0.76) .. (\x,-0.99);
}
\node[acyc,below] at (-1.23,-1.0) {$a_1\ (e_1)$};
\node[acyc,below] at (1.23,-1.0) {$a_2\ (e_2)$};
\node[bcyc] at (-1.23,0.68) {$b_1\ (f_1)$};
\node[bcyc] at (1.23,0.68) {$b_2\ (f_2)$};
\node[modcol] at (-3.05,0.62) {$\tau$};
\draw[modcol,thin,-{Latex[length=1.4mm]}] (-2.93,0.55) -- (-2.08,0.22);
\node[modcol] at (3.05,0.62) {$\sigma$};
\draw[modcol,thin,-{Latex[length=1.4mm]}] (2.93,0.55) -- (2.08,0.22);
\node[modcol] at (0,0.02) {$z$};
\end{tikzpicture}
\caption{A genus-two surface with a symplectic basis of $a$-cycles (red) and $b$-cycles (green), labeled by the unit vectors $e_i$ and $f_i$ of $\Z^4$. The periods over the $b$-cycles give the period matrix $\Omega$ of eq.~\eqref{eq:Omega}: $\tau$ and $\sigma$ belong to the two handles, and $z$ couples them.}\label{fig:genus2}
\end{figure}
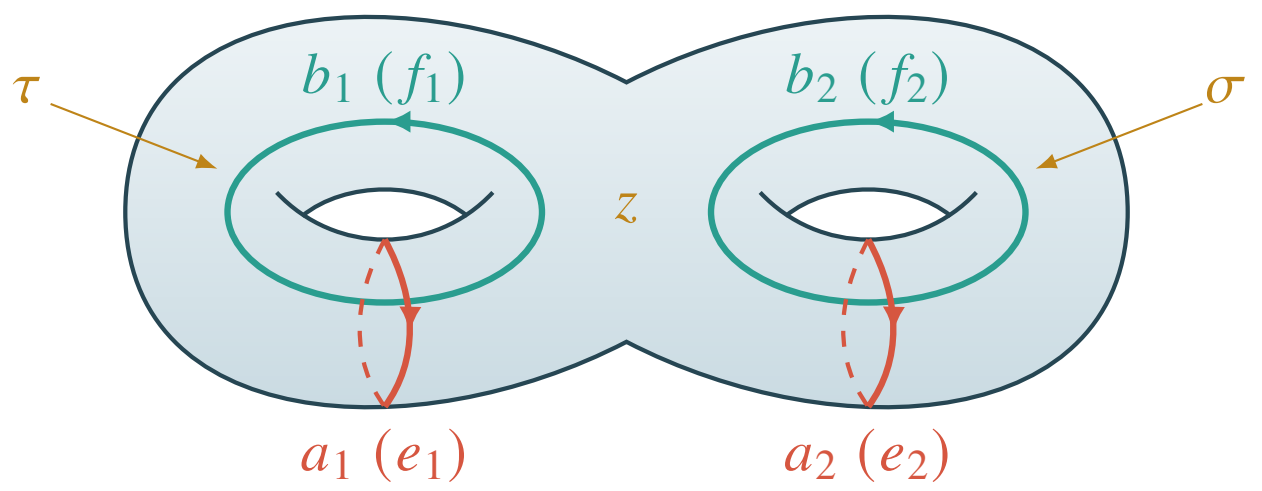

The three moduli on which the Siegel modular group acts are best viewed as the period matrix of a genus-two Riemann surface $\Sigma_2$. This picture makes the action of $\Sp(4,\Z)$, and later its boundary limits, intuitive. On $\Sigma_2$ we choose two $a$-cycles and two $b$-cycles with intersection numbers
\begin{equation}\label{eq:intersection}
 a_i\cdot b_j=\delta_{ij}\,,\qquad a_i\cdot a_j=b_i\cdot b_j=0\,,
\end{equation}
as in figure~\ref{fig:genus2}; such a basis is called symplectic. We normalize the two holomorphic differentials by $\oint_{a_i}\omega_j=\delta_{ij}$, and their periods over the $b$-cycles define the period matrix
\begin{equation}\label{eq:Omega}
 \Omega=\biggl(\oint_{b_i}\omega_j\biggr)=\begin{pmatrix}\tau&z\\z&\sigma\end{pmatrix}\,,\qquad \im\tau>0,\quad \det\im\Omega>0\,.
\end{equation}
By Riemann's bilinear relations $\Omega$ is symmetric with positive-definite imaginary part, i.e.\ it is a point of the Siegel upper half-space $\HH_2$ \cite{Siegel:1939}. The diagonal entries $\tau$ and $\sigma$ are the moduli of the two handles and $z$ couples them. For $z=0$ the surface splits into two elliptic curves $E_\tau$ and $E_\sigma$ joined at a point \cite{vanderGeer:2008sm}.

The period matrix depends on the choice of cycles. Any other symplectic basis is related to the original one by an integral matrix $\gamma$,
\begin{equation}\label{eq:basis-change}
 \begin{pmatrix}b'\\a'\end{pmatrix}=\gamma\begin{pmatrix}b\\a\end{pmatrix}\,,\qquad
 \gamma=\begin{pmatrix}A&B\\C&D\end{pmatrix}\,,\qquad
 b=\begin{pmatrix}b_1\\b_2\end{pmatrix},\quad a=\begin{pmatrix}a_1\\a_2\end{pmatrix}\,,
\end{equation}
and the intersection numbers eq.~\eqref{eq:intersection} are preserved precisely when $\gamma$ is symplectic. The Siegel modular group, also called the symplectic modular group, therefore consists of the integral symplectic matrices,
\begin{equation}\label{eq:Sp4Z}
 \Gamma_2=\Sp(4,\Z)=\{\gamma\in\operatorname{Mat}_{4\times4}(\Z):\gamma^\top J\gamma=J\}\,,\qquad
 J=\begin{pmatrix}0&I_2\\-I_2&0\end{pmatrix}\,.
\end{equation}
If we label the cycles $a_i$ and $b_i$ by the unit vectors $e_i$ and $f_i$ of $\Z^4$, $J$ is simply the matrix of intersection numbers, $e_i^\top Jf_j=a_i\cdot b_j=\delta_{ij}$. In the new basis the old differentials have $a$-periods $C\Omega+D$ and $b$-periods $A\Omega+B$, and renormalizing them by $(C\Omega+D)^{-1}$ gives the new period matrix
\begin{equation}\label{eq:siegel-action}
 \Omega\longmapsto \gamma\Omega=(A\Omega+B)(C\Omega+D)^{-1}\,.
\end{equation}
Notice that $\pm I_4$ act identically on $\Omega$. Moreover, the matrix $C\Omega+D$, which relates the old and new normalized differentials, reappears in the automorphy factor of section~\ref{sec:susy}. We use the generators and sign convention of ref.~\cite{Ding:2020zxw},
\begin{equation}\label{eq:generators}
 \mathsf{S}=\begin{pmatrix}0&I_2\\-I_2&0\end{pmatrix}\,,\qquad
 \mathsf{T}_i=\begin{pmatrix}I_2&B_i\\0&I_2\end{pmatrix}\,,\qquad i=1,2,3\,,
\end{equation}
where
\begin{equation}\label{eq:B-basis}
 B_1=\begin{pmatrix}1&0\\0&0\end{pmatrix}\,,\quad
 B_2=\begin{pmatrix}0&0\\0&1\end{pmatrix}\,,\quad
 B_3=\begin{pmatrix}0&1\\1&0\end{pmatrix}\,.
\end{equation}
Thus $\mathsf{T}_1,\mathsf{T}_2,\mathsf{T}_3$ shift $\tau,\sigma,z$ by one and $\mathsf{S}$ sends $\Omega$ to $-\Omega^{-1}$. Geometrically, $\mathsf{T}_i$ keeps the $a$-cycles and twists the $b$-cycles along them, $b\to b+B_ia$, whereas $\mathsf{S}$ exchanges the two sets of cycles, $(b';a')=(a;-b)$, see eq.~\eqref{eq:basis-change}. Two families of elements built from these generators will be used repeatedly below: the symmetric translations $\mathsf{T}_B$ and the $\GL(2,\Z)$ transformations $\gamma_A$,
\begin{equation}\label{eq:U-and-MA}
 \mathsf{T}_B=\mathsf{T}_1^{b_{11}}\mathsf{T}_2^{b_{22}}\mathsf{T}_3^{b_{12}}\,,\qquad
 \gamma_A=\operatorname{diag}(A,A^{-\top})\,,\qquad A\in\GL(2,\Z)\,,
\end{equation}
which act as $\Omega\mapsto\Omega+B$ and $\Omega\mapsto A\Omega A^\top$.\footnote{Throughout, $A^{-\top}\equiv(A^{-1})^\top=(A^\top)^{-1}$ denotes the inverse transpose.} The latter change the bases of $b$- and $a$-cycles by $A$ and $A^{-\top}$, respectively, and thus mix the two handles. Every $\GL(2,\Z)$ transformation is a word in the generators. We use in particular the two partial inversions and two elementary shears,
\begin{gather}
 \si=\mathsf{T}_1\mathsf{S}\mathsf{T}_1\mathsf{S}^{-1}\mathsf{T}_1\,,\qquad
 \sii=\mathsf{T}_2\mathsf{S}\mathsf{T}_2\mathsf{S}^{-1}\mathsf{T}_2\,,\notag\\
 \EE=\sii\mathsf{T}_3\sii^{-1}=\gamma_{\left(\begin{smallmatrix}1&1\\0&1\end{smallmatrix}\right)}\,,\qquad
 \FF=\si\mathsf{T}_3\si^{-1}=\gamma_{\left(\begin{smallmatrix}1&0\\1&1\end{smallmatrix}\right)}\,.\label{eq:partial-words}
\end{gather}
The element $\si$ acts as $\tau\to-1/\tau$, $z\to-z/\tau$, $\sigma\to\sigma-z^2/\tau$, exchanging $a_1$ and $b_1$ up to orientation, and $\sii$ is its counterpart for $\sigma$; together $\si\sii=\mathsf{S}$. The remaining generators used below are listed in appendix~\ref{app:words}.

Vacua related by eq.~\eqref{eq:siegel-action} correspond to the same point in moduli space. A fundamental domain $\mathcal{F}_2$ for the full group $\Gamma_2$ is defined by choosing a representative with $|\re\Omega_{ij}|\leq 1/2$, $0\leq 2\im z\leq\im\tau\leq\im\sigma$, and $|\det(C\Omega+D)|\geq1$ for every $\gamma\in\Gamma_2$. The first two conditions use translations and $\GL(2,\Z)$ transformations, while the last selects a representative with maximal $\det\im\Omega$ in its orbit. The finite list of inequalities at genus two and the boundary identifications are described in ref.~\cite{Ding:2020zxw}, section~6.1. The fundamental domain is not compact. Its ends describe degenerations of $\Sigma_2$ in which $a$-cycles are pinched: $\im\sigma\to\infty$ with $\tau$ and $z$ finite pinches $a_2$ (the Klingen boundary), while sending both eigenvalues of $\im\Omega$ to infinity pinches $a_1$ and $a_2$ (the Siegel cusp). We study these limits in section~\ref{sec:cusps}, where we use the equivalent choice $z\mapsto-z$ to display the cusp directions.

\subsection{Supersymmetric theories with Siegel modular symmetry}\label{sec:susy}
In a flavor model the period matrix is promoted to a set of dynamical fields. The three entries of $\Omega$ are the scalar components of chiral moduli multiplets. The quarks, leptons and Higgs fields belong to chiral matter multiplets $\Phi_I$. For each simple gauge factor $G_a$ there is a vector multiplet with chiral field strength $\mathcal{W}^a_\alpha$. At two derivatives their action with rigid $\mathcal{N}=1$ \ac{SUSY} has the superspace form
\begin{equation}\label{eq:rigid-action}
 S_{\rm rigid}=\int d^4x\,d^4\theta\,\mathscr{K}
 +\left[\int d^4x\,d^2\theta\left(\mathscr{W}+
 \frac{1}{4}\sum_a f_a\,\mathcal{W}^{a\alpha}\mathcal{W}^a_\alpha\right)+\mathrm{h.c.}\right]\,.
\end{equation}
Adjoint indices are contracted within each gauge factor. The real K\"ahler potential $\mathscr{K}$ determines the kinetic terms, while the superpotential $\mathscr{W}$ and the Wilsonian gauge kinetic functions $f_a$ are holomorphic. Expanding $\mathscr{W}$ in powers of the chiral matter fields $\Phi_I$, and anticipating the modular labels introduced below, we can write a general $n$-field interaction as
\begin{equation}\label{eq:W-expansion}
 \mathscr{W}(\Omega,\Phi)=\sum_{n}\sum_{I_1,\dots,I_n}\bigl[Y^{(k)}_{\mathbf{r}}(\Omega)\,\Phi_{I_1}\cdots\Phi_{I_n}\bigr]_{\mathbf{1}}\,,
\end{equation}
where the holomorphic coupling $Y^{(k)}_{\mathbf{r}}(\Omega)$ may depend on the moduli $\Omega$. In particular, cubic terms ($n=3$) give Yukawa interactions, quadratic terms give mass terms, and quartic terms include the Weinberg operator $(lH_u)(lH_u)/\Lambda$.

Now impose Siegel modular symmetry. With the automorphy factor
\begin{equation}\label{eq:automorphy}
 j(\gamma,\Omega)=\det(C\Omega+D)\,,
\end{equation}
we assign each $\Phi_I$ modular weight $-k_I$ and a finite modular representation $\mathbf{r}_I$. Covariance of a superpotential term requires its holomorphic coefficients to transform in multiplets,
\begin{align}\label{eq:siegel-form-field}
 \Phi_I&\longmapsto j(\gamma,\Omega)^{-k_I}\rho_{\mathbf{r}_I}(\gamma)\Phi_I\,,\notag\\
 Y^{(k)}_{\mathbf{r}}(\gamma\Omega)&=j(\gamma,\Omega)^k
 \rho_{\mathbf{r}}(\gamma)Y^{(k)}_{\mathbf{r}}(\Omega)\,.
\end{align}
At level $N$ the matrices $\rho$ factor through $\Gamma_{2,N}=\Gamma_2/\Gamma_2(N)$, where $\Gamma_2(N)$ is the principal congruence subgroup of level $N$,
\begin{equation}\label{eq:principal-congruence}
 \Gamma_2(N)=\bigl\{\gamma\in\Sp(4,\Z):\ \gamma\equiv I_4\pmod{N}\bigr\}\,.
\end{equation} Holomorphic coefficients regular in the Siegel upper half-space and at its cusps are vector-valued Siegel modular forms. We denote a component by $Y^{(k)}_{\mathbf{r},i}$. On an invariant locus or near a cusp, the full Siegel group $\Gamma_2$ is replaced by the modular subsymmetry $\Gamma_{\mathrm{sub}}\subset\Gamma_2$ preserving that locus or limit. The transformation laws in eq.~\eqref{eq:siegel-form-field} keep the same form with $\gamma\in\Gamma_{\mathrm{sub}}$, and $\rho$ factors through the finite quotient $\Gamma_{\mathrm{sub}}/(\Gamma_{\mathrm{sub}}\cap\Gamma_2(N))$. A term $[Y^{(k)}_{\mathbf{r}}\Phi_{I_1}\cdots\Phi_{I_n}]_{\mathbf{1}}$ is allowed in a modular-invariant rigid superpotential when its weights cancel and the tensor product admits a trivial singlet,
\begin{equation}\label{eq:rigid-weights}
 k=\sum_{s=1}^n k_{I_s}\,,\qquad
 \mathbf{1}\ \subset\ \mathbf{r}\otimes\mathbf{r}_{I_1}\otimes\cdots\otimes\mathbf{r}_{I_n}\,.
\end{equation}
The brackets denote contraction with an invariant tensor. Gauge indices must also form a singlet, and identical fields must be symmetrized. For instance, once the matter-field and Higgs representations are chosen, each independent trivial singlet contraction gives an allowed Yukawa structure.

In a unitary basis for the matter fields, a minimal K\"ahler potential, shown to quadratic order in the matter fields, is
\begin{equation}\label{eq:KS}
 \mathscr{K}=-h\log\det\im\Omega+\sum_I(\det\im\Omega)^{-k_I}
 \Phi_I^\dagger e^{2V_I}\Phi_I+\cdots\,,\qquad h>0\,.
\end{equation}
Here $V_I$ is the gauge vector superfield acting in the gauge representation of $\Phi_I$. We suppress vector backgrounds in the mass formulas below. In a non-unitary basis the bilinear contains the corresponding invariant Hermitian metric. The matter-field terms are invariant under modular transformations, while the moduli term gives a K\"ahler transformation,
\begin{equation}\label{eq:kahler-shift}
 \det\im(\gamma\Omega)=\frac{\det\im\Omega}{|j(\gamma,\Omega)|^2}\,,
 \qquad \mathscr{K}\longmapsto \mathscr{K}+F_\gamma+\overline{F}_\gamma\,,\qquad F_\gamma=h\log j(\gamma,\Omega)\,.
\end{equation}
In rigid \ac{SUSY} the integral of this shift over full superspace vanishes.

In \ac{SUGRA} a K\"ahler transformation acts on the superpotential as well. In reduced Planck units the chiral compensator $S_0$ makes this visible in the action \cite{Ding:2020zxw,Kugo:1982mr,Kugo:2016lum},
\begin{align}\label{eq:sugra-action}
 S_{\rm SUGRA}={}&-3\int d^4x\,d^4\theta\,E\,\overline{S}_0S_0 e^{-\mathscr{K}/3}+\left[\int d^4x\,d^2\theta\,\mathcal{E}
 \left(S_0^3 \mathscr{W}+\frac{1}{4}\sum_a f_a\,\mathcal{W}^{a\alpha}\mathcal{W}^a_\alpha\right)
 +\mathrm{h.c.}\right]\,.
\end{align}
Here $E$ and $\mathcal{E}$ are the full and chiral superspace densities. Invariance of the first term requires $S_0\mapsto e^{F_\gamma/3}S_0$. We take $\mathscr{W}$ to transform as a trivial singlet under the finite modular group, with no additional multiplier phase. Invariance of $S_0^3\mathscr{W}$ then gives
\begin{equation}\label{eq:sugra-weight}
 \mathscr{W}\longmapsto j(\gamma,\Omega)^{-h}\mathscr{W}\,,\qquad k=\sum_{s=1}^n k_{I_s}-h\,.
\end{equation}
The singlet condition in eq.~\eqref{eq:rigid-weights} still applies. The models below use the rigid weight assignment. With minimal metrics, canonical normalization supplies the factor $(\det\im\Omega)^{k/2}$ in each physical Yukawa matrix. In \ac{SUGRA} the additional factor $e^{\mathscr{K}_{\rm mod}/2}$, with $\mathscr{K}_{\rm mod}=\mathscr{K}|_{\Phi=0}$, gives the same expression with the weight in eq.~\eqref{eq:sugra-weight}.

\paragraph{Modular anomalies and gauge kinetic functions.}
The target-space Siegel modular symmetry is a discrete gauge symmetry arising from the duality of a string compactification, so its mixed gauge--modular anomaly must vanish. Under $\gamma\in\Gamma_2$, the chiral fermion path-integral measure shifts the effective action in a background of a simple gauge factor $G_a$ by an anomalous phase. Generalizing the genus-one analysis of ref.~\cite{Liu:2026ati} with $c\tau+d$ replaced by $j(\gamma,\Omega)=\det(C\Omega+D)$, cancellation of a nonzero net anomaly requires a compensating shift of the Wilsonian gauge kinetic function. In our conventions
\begin{equation}\label{eq:gauge-anomaly}
 f_a(\gamma\Omega)=f_a(\Omega)+\frac{\mathcal{A}_a}{8\pi^2}\log j(\gamma,\Omega)-\frac{1}{8\pi^2}\log\chi_a(\gamma)\pmod{\frac{i}{4\pi}\Z}\,,
\end{equation}
or equivalently, in the conventions $g_a^{-2}=\re f_a$ and $\theta_a=-8\pi^2\im f_a$,
\begin{equation}\label{eq:gauge-anomaly-exp}
 e^{-8\pi^2 f_a(\gamma\Omega)}=j(\gamma,\Omega)^{-\mathcal{A}_a}\chi_a(\gamma)\,e^{-8\pi^2 f_a(\Omega)}\,.
\end{equation}
The weight and the finite phase follow from the light spectrum. Writing $d_I=\dim\mathbf{r}_I$ and $T_a(I)$ for the Dynkin index of $\Phi_I$ under $G_a$ (summed over all other gauge quantum numbers of $\Phi_I$), one finds
\begin{equation}\label{eq:anomaly-data}
 \mathcal{A}_a=h\,C_2(G_a)+\sum_I T_a(I)d_I(2k_I-h)\,,\qquad
 \chi_a(\gamma)=\prod_I\bigl[\det\rho_{\mathbf{r}_I}(\gamma)\bigr]^{2T_a(I)}\,.
\end{equation}
Here the $h$ terms include the \ac{SUGRA} rotations of the matter fermions and gauginos. The determinant character is trivial for representations of the full level-3 finite Siegel modular group defined in section~\ref{sec:finite-group}, which has only a single (trivial) one-dimensional representation. On the other hand, a theory defined on a modular subsymmetry can have a nontrivial singlet.

Eq.~\eqref{eq:gauge-anomaly} fixes the modular variation of $f_a$. For example, away from the zeros of $\chi_{10}$, a holomorphic threshold can include $b_a\log\chi_{10}(\Omega)/(8\pi^2)$. Since $\chi_{10}$ has weight 10 and trivial character, this term shifts by $10b_a\log j/(8\pi^2)$. In a heterotic theory a Green--Schwarz shift of the dilaton can supply the remainder: if $f_a$ contains $\kappa_a S$ with $S\mapsto S-\delta_{\rm GS}\log j/(8\pi^2)$, the weight condition becomes $\mathcal{A}_a=10b_a-\kappa_a\delta_{\rm GS}$. A nontrivial $\chi_a$ requires a matching phase as well. The spectrum of charged states fixes the allowed singularities of the threshold, and modular-invariant terms may be added to $f_a$. Examples with Wilson lines and other Siegel threshold forms are discussed in ref.~\cite{Hernandez-Segura:2026oyy}.

\subsection{The finite modular group at level 3}\label{sec:finite-group}
We use $\Z_3=\Z/3\Z$ for the field of three elements and for its additive group. The finite Siegel modular groups $\Gamma_{2,N}=\Gamma_2/\Gamma_2(N)$ have orders \cite{Freitag:1983}
\begin{equation}\label{eq:finite-order-general}
 |\Gamma_{2,N}|=N^{10}\prod_{p\,|\,N}\Bigl(1-\frac{1}{p^2}\Bigr)\Bigl(1-\frac{1}{p^4}\Bigr)\,,
\end{equation}
so that $\Gamma_{2,2}\cong S_6$ has $720$ elements~\cite{Ding:2020zxw} and
\begin{equation}\label{eq:finite-orders}
 \Gamma_{2,3}\cong\Sp(4,\Z_3)\,,\qquad |\Gamma_{2,3}|=51840\,.
\end{equation}
A convenient presentation of $\Gamma_{2,3}$ in terms of the generators eq.~\eqref{eq:generators} is
\begin{gather}
 \mathsf{S}^4=\mathsf{T}_1^3=\mathsf{T}_2^3=\mathsf{T}_3^3=(\mathsf{S}\mathsf{T}_1\mathsf{T}_2)^3=1\,,\qquad
 \mathsf{T}_i\mathsf{T}_j=\mathsf{T}_j\mathsf{T}_i\,,\qquad
 \mathsf{S}=\si\sii\,,\qquad \si\mathsf{T}_2=\mathsf{T}_2\si\,,\notag\\
 \EE\mathsf{T}_2\EE^{-1}=\mathsf{T}_1\mathsf{T}_2\mathsf{T}_3,\quad
 \EE\mathsf{T}_3\EE^{-1}=\mathsf{T}_1^2\mathsf{T}_3,\quad
 \FF\mathsf{T}_1\FF^{-1}=\mathsf{T}_1\mathsf{T}_2\mathsf{T}_3,\quad
 \FF\mathsf{T}_3\FF^{-1}=\mathsf{T}_2^2\mathsf{T}_3\,,\label{eq:G23-presentation}
\end{gather}
with the elements eq.~\eqref{eq:partial-words}. All relations except $\mathsf{T}_i^3=1$ hold in $\Sp(4,\Z)$: the first line contains the commuting translations and the factorization of $\mathsf{S}$ into partial inversions, and the second line records how the shears $\EE,\FF$ act on translations, $\Omega+B\mapsto A(\Omega+B)A^\top$. A presentation of $\Sp(4,\Z)$ itself is recalled in appendix~\ref{app:words}.

The quotient $\Gamma_{2,3}/\{\pm I_4\}$ is one of the most thoroughly studied finite groups. It is the simple group of order $25920$ investigated by Burkhardt in his work on hyperelliptic modular functions \cite{Burkhardt:1891}, and it has several other incarnations \cite{ATLAS:1985,Todd:1947,Hunt:1996}:
\begin{equation}\label{eq:Burkhardt-group}
 \PSp(4,\Z_3)\cong\PSU(4,2)\cong W(E_6)^+\,.
\end{equation}
Here the prefix $\mathrm{P}$ denotes ``projective'', i.e.\ the quotient by the scalar matrices, and $W(E_6)^+$ is the index-two rotation subgroup of the Weyl group of $E_6$, which acts on the $27$ lines of a smooth cubic surface \cite{Jordan:1870}. In five dimensions the group preserves a unique quartic, Burkhardt's quartic, which has $45$ nodes \cite{Burkhardt:1891,Todd:1936,Baker:1946,Finkelnberg:1989,deJong:1990}. We call $ \PSp(4,\Z_3)\cong \Gamma_{2,3}/\{\pm I_4\}$ the Burkhardt group.

The group $\Gamma_{2,3}$, the double cover of the Burkhardt group, has $34$ irreducible representations (listed in appendix~\ref{app:group-reps}).
Since $-I_4$ acts trivially on $\Omega$ and has $j(-I_4,\Omega)=1$ at genus two, integer-weight vector-valued Siegel modular forms transform only in the $20$ irreducible representations of the Burkhardt group \cite{Ding:2020zxw}. In the notation $[\mathbf{d},n]$, meaning $n$ inequivalent irreducible representations of dimension $d$, they are
\begin{align}\label{eq:Burkhardt-irreps}
 &[\mathbf{1},1],\ [\mathbf{5},2],\ [\mathbf{6},1],\ [\mathbf{10},2],\ [\mathbf{15},2],\ [\mathbf{20},1],\ [\mathbf{24},1],\ [\mathbf{30},3],\notag\\ &[\mathbf{40},2],\ [\mathbf{45},2],\ [\mathbf{60},1],\ [\mathbf{64},1],\ [\mathbf{81},1]\,.
\end{align}
Neither group has a three-dimensional irreducible representation. Three families of leptons or quarks therefore cannot form an irreducible triplet of the full Siegel modular group at level 3. Triplets appear only after the symmetry has been reduced, on invariant subspaces and at cusps, which is one more reason to study them.

\subsection{Modular forms of level 3}\label{sec:theta}
\paragraph{Theta functions of weight 1.}
The Siegel modular forms of weight $k=1$ and level $N=3$ can be constructed from theta functions of the $A_2$ lattice \cite{Freitag:2004burk}, the two-dimensional hexagonal lattice spanned by two simple roots $\alpha_1,\alpha_2$ of length squared two at an angle of $2\pi/3$ (so that $\alpha_1\cdot\alpha_2=-1$). Let $L$ be this lattice and $\mu_0=(2\alpha_1+\alpha_2)/3$, so that $\mu_0^2=2/3$ and $L$, $L+\mu_0$, $L+2\mu_0$ are the three classes of the dual lattice. Summing over two lattice vectors $\ell\in L+a\mu_0$ and $\ell'\in L+b\mu_0$, we define
\begin{equation}\label{eq:Theta}
 \Theta_{ab}(\Omega)=\sum_{\ell\in L+a\mu_0}\ \sum_{\ell'\in L+b\mu_0}
 \exp\bigl[\pi i\bigl(\tau\,\ell^2+2z\,\ell\cdot\ell'+\sigma\,\ell'^2\bigr)\bigr]\,,\qquad a,b=0,1,2\,.
\end{equation}
In the basis $(\alpha_1,\alpha_1+\alpha_2)$ the same sum can be written
\begin{equation}\label{eq:Theta-Gram}
 \Theta_{ab}(\Omega)=\sum_{G\in\operatorname{Mat}_{2\times2}(\Z)}
  \exp\!\left\{\pi i\,\tr\!\left[(G+P_{ab}/3)^\top Q(G+P_{ab}/3)\,\Omega\right]\right\}\,.
\end{equation}
Here the Gram matrix is $Q=\left(\begin{smallmatrix}2&1\\1&2\end{smallmatrix}\right)$ and the characteristic is specified by $P_{ab}=\left(\begin{smallmatrix}a&b\\a&b\end{smallmatrix}\right)$.

Since $\Theta_{ab}=\Theta_{-a,-b}$, there are five independent functions, which we order as
\begin{equation}\label{eq:five-basis}
 Y^{(1)}_{\mathbf{5}}=\bigl(\Theta_{00},\Theta_{10},\Theta_{01},\Theta_{21},\Theta_{11}\bigr)^\top\equiv\bigl(Y_1,Y_2,Y_3,Y_4,Y_5\bigr)^\top\,,
\end{equation}
with $Y_i\equiv Y^{(1)}_{\mathbf{5},i}$. These five functions form an irreducible quintet $\mathbf{5}$ of the Burkhardt group, and its transformation matrices follow from three short calculations on the lattice sums. A shift of $\Omega$ multiplies each term of eq.~\eqref{eq:Theta} by a phase. Since $\ell^2=2a^2/3$ and $\ell\cdot\ell'=2ab/3$ modulo $2\Z$ and $\Z$, respectively,
\begin{equation}\label{eq:rho-T}
 \rho_{\mathbf{5}}(\mathsf{T}_1)=\diag(1,\omega,1,\omega,\omega),\quad
 \rho_{\mathbf{5}}(\mathsf{T}_2)=\diag(1,1,\omega,\omega,\omega),\quad
 \rho_{\mathbf{5}}(\mathsf{T}_3)=\diag(1,1,1,\omega,\omega^2)\,,
\end{equation}
with $\omega=e^{2\pi i/3}$. A $\GL(2,\Z)$ transformation $\Omega\to A\Omega A^\top$ can be undone by relabeling the pair $(\ell,\ell')$, which permutes the five functions. Since $j(\gamma_A,\Omega)=\det A$,
\begin{equation}\label{eq:rho-A}
 \rho_{\mathbf{5}}(\gamma_A)=(\det A)\,\Pi_A\,,\qquad \Theta_{(a,b)}(A\Omega A^\top)=\Theta_{(a,b)A}(\Omega)\,,
\end{equation}
where $\Pi_A$ is the permutation matrix that sends the label $(a,b)$ to $(a,b)A$. The exchange $\tau\leftrightarrow\sigma$ swaps $Y_2\leftrightarrow Y_3$, the reflection $z\to-z$ swaps $Y_4\leftrightarrow Y_5$, and $A=\left(\begin{smallmatrix}0&-1\\1&1\end{smallmatrix}\right)$ permutes $Y_2\to Y_3\to Y_5\to Y_2$ while fixing $Y_1$ and $Y_4$. Poisson summation over both lattice vectors gives the inversion,
\begin{equation}\label{eq:rho-S}
 \rho_{\mathbf{5}}(\mathsf{S})=\frac{1}{3}\begin{pmatrix}
 -1&-2&-2&-2&-2\\
 -1&1&-2&1&1\\
 -1&-2&1&1&1\\
 -1&1&1&1&-2\\
 -1&1&1&-2&1
 \end{pmatrix}\,.
\end{equation}
Eqs.~\eqref{eq:rho-T} and \eqref{eq:rho-S} specify the quintet representation $\mathbf{5}$ completely. All the matrices have determinant one and preserve the Hermitian form $G_5=\diag(1,2,2,2,2)$. We use it as the minimal matter metric in this basis. In the normalized basis $(Y_1,\sqrt{2}Y_2,\sqrt{2}Y_3,\sqrt{2}Y_4,\sqrt{2}Y_5)$ the matrices are unitary and $\rho_{\mathbf{5}}(\mathsf{S})$ is real and symmetric. We keep the basis eq.~\eqref{eq:five-basis} because the Fourier expansions are simplest there.

With the level-3 variables
\begin{equation}\label{eq:q-variables}
 q_\tau=e^{2\pi i\tau/3}\,,\qquad q_\sigma=e^{2\pi i\sigma/3}\,,\qquad \zeta=e^{2\pi iz/3}\,,
\end{equation}
a lattice pair $(\ell,\ell')$ contributes $q_\tau^{3\ell^2/2}\zeta^{3\ell\cdot\ell'}q_\sigma^{3\ell'^2/2}$, with integer exponents. Counting lattice vectors gives
\begin{align}\label{eq:theta-q}
 Y_1(\Omega)&=1+6\bigl(q_\tau^3+q_\sigma^3\bigr)+6\,q_\tau^3q_\sigma^3\bigl(\zeta^{-6}+2\zeta^{-3}+2\zeta^{3}+\zeta^{6}\bigr)+\cdots\,,\notag\\
 Y_2(\Omega)&=3q_\tau+3q_\tau^4+6\,q_\tau q_\sigma^3\bigl(\zeta^{-3}+1+\zeta^{3}\bigr)+6\,q_\tau^4q_\sigma^3\bigl(\zeta^{-6}+1+\zeta^{6}\bigr)+\cdots\,,\notag\\
 Y_3(\Omega)&=3q_\sigma+3q_\sigma^4+6\,q_\tau^3q_\sigma\bigl(\zeta^{-3}+1+\zeta^{3}\bigr)+6\,q_\tau^3q_\sigma^4\bigl(\zeta^{-6}+1+\zeta^{6}\bigr)+\cdots\,,\notag\\
 Y_4(\Omega)&=q_\tau q_\sigma\bigl(3\zeta^{-2}+6\zeta\bigr)+\bigl(q_\tau^4q_\sigma+q_\tau q_\sigma^4\bigr)\bigl(6\zeta^{-2}+3\zeta^{4}\bigr)+q_\tau^4q_\sigma^4\bigl(3\zeta^{-8}+6\zeta^{4}\bigr)+\cdots\,,\notag\\
 Y_5(\Omega)&=q_\tau q_\sigma\bigl(3\zeta^{2}+6\zeta^{-1}\bigr)+\bigl(q_\tau^4q_\sigma+q_\tau q_\sigma^4\bigr)\bigl(6\zeta^{2}+3\zeta^{-4}\bigr)+q_\tau^4q_\sigma^4\bigl(3\zeta^{8}+6\zeta^{-4}\bigr)+\cdots\,,
\end{align}
where all omitted terms contain $q_\tau^{n_1}$ or $q_\sigma^{n_2}$ with $n_1\ge7$ or $n_2\ge7$. The expansions display the translation characters eq.~\eqref{eq:rho-T} term by term, and they already anticipate the boundary structures of section~\ref{sec:cusps}: as $q_\sigma\to0$ only $Y_1$ and $Y_2$ survive, as $q_\tau,q_\sigma\to0$ only $Y_1$, and the powers of $\zeta$ are bounded by positivity of the Fourier exponents.

At $q_\sigma=0$ the pair $(Y_1,Y_2)$ reduces to the genus-one theta series $\theta_L(\tau)=1+6q_\tau^3+\cdots$ and $\theta_{L+\mu_0}(\tau)=3q_\tau+3q_\tau^4+\cdots$ of $A_2$ and its shifted class. These are the cubic theta functions of the Borweins \cite{Borwein:1991cubic,Borwein:1994cubic}, and they form the weight-1 doublet of the finite modular group $T'$ at level 3 \cite{Liu:2019khw},
\begin{align}\label{eq:D-doublet}
 Y^{(1,0)}_{\mathbf{2}^{(0)}}(\tau)&=\begin{pmatrix}-\sqrt{2}\theta_{L+\mu_0}(\tau)\\ \theta_L(\tau)\end{pmatrix}=\begin{pmatrix}-3\sqrt{2}\dfrac{\eta(3\tau)^3}{\eta(\tau)}\\[6pt] 3\dfrac{\eta(3\tau)^3}{\eta(\tau)}+\dfrac{\eta(\tau/3)^3}{\eta(\tau)}\end{pmatrix}\,,
\end{align}
with
\begin{equation}\label{eq:rho2}
 \rho_{\mathbf{2}^{(0)}}(\mathsf{S})=\frac{i}{\sqrt{3}}\begin{pmatrix}1&\sqrt{2}\\\sqrt{2}&-1\end{pmatrix}\,,\qquad
 \rho_{\mathbf{2}^{(0)}}(\mathsf{T})=\begin{pmatrix}\omega&0\\0&1\end{pmatrix}\,.
\end{equation}
Here $\mathsf{S}=\left(\begin{smallmatrix}0&-1\\1&0\end{smallmatrix}\right)$ and $\mathsf{T}=\left(\begin{smallmatrix}1&1\\0&1\end{smallmatrix}\right)$, as in ref.~\cite{Feruglio:2026xxo}. When the genus-one group is embedded in the Siegel group acting on $\tau$, $\mathsf{T}$ becomes $\mathsf{T}_1$ and $\mathsf{S}$ becomes $\mathsf{s}_J=\si^{-1}$ (the element $\si$ of appendix~\ref{app:words} realizes $\mathsf{S}^{-1}$); the same convention applies in eq.~\eqref{eq:D-fixed} and section~\ref{sec:klingen}. The same doublet will reappear as the index-0 Jacobi form of Feruglio and Marrone.

\paragraph{Higher weights.}
The dimensions of the spaces $\mathcal{M}_k(\Gamma_2(3))$ of level-3 Siegel modular forms are encoded in the Hilbert series \cite{Freitag:2004burk}
\begin{align}\label{eq:Hilbert}
 \sum_{k\ge0}\dim \mathcal{M}_k(\Gamma_2(3))\,t^k&=\frac{1+t+t^2+6t^3+6t^4+t^5+t^6+t^7}{(1-t)^4}\notag\\
 &=1+5t+15t^2+40t^3+95t^4+196t^5+\cdots\,.
\end{align}
Freitag and Salvati Manni showed that the full ring $\bigoplus_{k\ge 0}\mathcal{M}_k(\Gamma_2(3))$ is generated by the five weight-1 theta functions $Y_1,\dots,Y_5$ together with five additional forms of weight 3, subject to twenty polynomial relations in weights 5 and 6 \cite{Freitag:2004burk,Freitag:2004burk2}; these five forms alone obey a single degree-$18$ relation defining the projective dual of Burkhardt's quartic \cite{Freitag:2004burk}. The decomposition of the low-weight modular form spaces into representations of the Burkhardt group is
\begin{align}\label{eq:weight-decomposition}
 \mathcal{M}_{k=1}&=\mathbf{5}\,,\qquad \mathcal{M}_{k=2}=\mathrm{Sym}^2\mathbf{5}=\mathbf{15}\,,\qquad \mathcal{M}_{k=3}=\mathrm{Sym}^3\mathbf{5}\oplus\mathbf{5}=\bar{\mathbf{5}}\oplus\mathbf{30}\oplus\mathbf{5}\,,\notag\\
 \mathcal{M}_{k=4}&=\mathrm{Sym}^4\mathbf{5}\oplus\mathbf{5}\otimes\mathbf{5}=\bigl(\mathbf{1}\oplus\mathbf{15}\oplus\mathbf{24}\oplus\mathbf{30}^{\prime}\bigr)\oplus\bigl(\mathbf{10}\oplus\mathbf{15}\bigr)\,,
\end{align}
where $\bar{\mathbf{5}}$ is the complex conjugate of $\mathbf{5}$, the unprimed $\mathbf{30}$ is real and $\mathbf{30}^{\prime}$ is complex, and $\mathrm{Sym}^n$ and $\wedge^2$ denote symmetric and antisymmetric tensor powers. We now describe the forms used in this paper.

At weight $k=2$ the fifteen independent modular forms are the quadratic monomials in $Y_1,\dots,Y_5$, which furnish the irreducible symmetric-square representation $\mathrm{Sym}^2\mathbf{5}=\mathbf{15}$ of the Burkhardt group. In the theta basis a convenient set of components is
\begin{equation}\label{eq:weight2-basis}
 Y^{(2)}_{\mathbf{15}}=\Bigl(\bigl(Y^{(1)}_{\mathbf{5},i}\bigr)^2\ (1\le i\le5),\ \ \sqrt{2}\,Y^{(1)}_{\mathbf{5},i}Y^{(1)}_{\mathbf{5},j}\ (1\le i<j\le5)\Bigr)^\top\,.
\end{equation}
Modular form multiplets of weight 2 are therefore quadratic polynomials in $Y_1,\dots,Y_5$. For products of two quintets, $\mathbf{5}\otimes\mathbf{5}=\mathbf{10}_{A}\oplus\mathbf{15}_{S}$ and $\mathbf{5}\otimes\bar{\mathbf{5}}=\mathbf{1}\oplus\mathbf{24}$.

The quintet has a single invariant of degree four and none of lower degree,
\begin{equation}\label{eq:I4}
 I_4=Y_1^4+8Y_1\,(Y_2^3+Y_3^3+Y_4^3+Y_5^3)+48\,Y_2Y_3Y_4Y_5\,.
\end{equation}
Up to a rescaling of variables this is Burkhardt's quartic \cite{Burkhardt:1891,Hunt:1996}.\footnote{\label{fn:lattice}Evaluated on the theta functions, $I_4$ is the genus-two theta series of the $E_8$ lattice, which is obtained by gluing four copies of $A_2$ along the tetracode \cite{Conway:1999sp}. Since $E_8$ is alone in its genus, Siegel's main theorem identifies this series with the Siegel Eisenstein series of weight 4 \cite{Siegel:1935}, as noted in ref.~\cite{Freitag:2004burk}. Similarly, the components of the gradient of $I_4$ are genus-two theta series of $E_6$ with characteristics, and the index-three sublattice $E_6\oplus A_2\subset E_8$ relates the two statements through Euler's identity $I_4=\frac{1}{4}\,Y^{(1)}_{\mathbf{5}}\cdot\nabla I_4$.} The gradient of $I_4$ gives the $\bar{\mathbf{5}}$ in $\mathrm{Sym}^3\mathbf{5}$, while the five additional forms of weight 3 constructed in ref.~\cite{Freitag:2004burk} transform like the weight-1 quintet but are not polynomials in the theta functions.

\paragraph{The Hessian and $\chi_{10}$.}
Differentiating the quartic twice gives a symmetric matrix of weight-2 forms,
\begin{equation}\label{eq:Hessian-matrix}
 \mathcal{H}=\left[\frac{\partial^2 I_4}{\partial Y^{(1)}_{\mathbf{5},i}\partial Y^{(1)}_{\mathbf{5},j}}\right]
 =\begin{pmatrix}
 12Y_1^2&24Y_2^2&24Y_3^2&24Y_4^2&24Y_5^2\\
 24Y_2^2&48Y_1Y_2&48Y_4Y_5&48Y_3Y_5&48Y_3Y_4\\
 24Y_3^2&48Y_4Y_5&48Y_1Y_3&48Y_2Y_5&48Y_2Y_4\\
 24Y_4^2&48Y_3Y_5&48Y_2Y_5&48Y_1Y_4&48Y_2Y_3\\
 24Y_5^2&48Y_3Y_4&48Y_2Y_4&48Y_2Y_3&48Y_1Y_5
 \end{pmatrix}\,,
\end{equation}
which transforms as $\rho_{\mathbf{5}}(\gamma)^\top\mathcal{H}(\gamma\Omega)\rho_{\mathbf{5}}(\gamma)=j(\gamma,\Omega)^2\mathcal{H}(\Omega)$. It is the only coupling of weight 2 between two quintets, since $\mathbf{5}\otimes\mathbf{5}=\mathbf{10}\oplus\mathbf{15}$ contains the self-dual $\mathcal{M}_{k=2}=\mathbf{15}$ only once, and in section~\ref{sec:global} we use it as a coupling of two matter quintets.
Its determinant is a scalar modular form of weight 10 that vanishes when $\im\sigma\to\infty$. There is only one such form, the Igusa cusp form $\chi_{10}$ \cite{Igusa:1962}, and comparing the leading Fourier coefficients gives
\begin{equation}\label{eq:Hessian-chi10}
 \det\mathcal{H}=2^{19}3^9\,\chi_{10}(\Omega)\,,\qquad
 \chi_{10}=e^{2\pi i(\tau+\sigma)}\bigl(e^{2\pi iz}-2+e^{-2\pi iz}\bigr)+\cdots\,.
\end{equation}
The form $\chi_{10}$ vanishes exactly on the surface $z=0$ and its images under $\Sp(4,\Z)$, i.e.\ on the Humbert surface $H_1$ \cite{vanderGeer:2008sm}, where $\Omega$ describes two elliptic curves joined at a point rather than a smooth genus-two surface (section~\ref{sec:moduli}). This surface is therefore the massless locus of that coupling.

\subsection{Elliptic invariant subspaces}\label{sec:elliptic}
Let $H$ be a finite subgroup of $\Sp(4,\Z)$ and $\mathcal{F}_H$ the set of points that it leaves fixed,
\begin{equation}\label{eq:FH}
 \mathcal{F}_H=\{\Omega:\ h\Omega=\Omega\ \text{for all}\ h\in H\}\,.
\end{equation}
We take $H$ to be the full pointwise stabilizer of $\mathcal{F}_H$,\footnote{The stabilizer of a point is the subgroup of transformations that leave it fixed. Here $H$ is the subgroup that fixes every point of $\mathcal{F}_H$; it always contains $\pm I_4$.} including $\pm I_4$. Two groups are relevant for a model on $\mathcal{F}_H$. The group $H$ leaves every point of $\mathcal{F}_H$ fixed; it acts on the moduli transverse to $\mathcal{F}_H$ and on the fields. The normalizer $N(H)$, i.e.\ the set of transformations $\gamma$ with $\gamma H\gamma^{-1}=H$, maps $\mathcal{F}_H$ into itself and acts on the moduli that remain.\footnote{If $\Omega\in\mathcal{F}_H$ and $\gamma$ normalizes $H$, then $h(\gamma\Omega)=\gamma(\gamma^{-1}h\gamma)\Omega=\gamma\Omega$, so $\gamma\Omega\in\mathcal{F}_H$.} When the moduli are restricted to $\mathcal{F}_H$, the Siegel modular theory becomes a theory of the remaining moduli with modular symmetry $N(H)$, and its finite modular group at level 3 is the image of $N(H)$ modulo three,
\begin{equation}\label{eq:normalizer-quotient}
 N_3(H)=N(H)\big/\bigl(N(H)\cap\Gamma_2(3)\bigr)\,.
\end{equation}
If $u$ denotes the independent moduli on the locus and $\Omega=\Omega(u)$, the induced transformation $u\mapsto u'$ and its action on matter fields are fixed by the same matrices:
\begin{equation}\label{eq:restricted-matter}
 \Omega(u')=(A\Omega(u)+B)(C\Omega(u)+D)^{-1}\,,\qquad
 \Phi_I\longmapsto j(\gamma,\Omega(u))^{-k_I}\rho_{\mathbf{r}_I}(\gamma)\Phi_I,\quad \gamma\in N(H)\,.
\end{equation}
For a theory defined on the locus $\mathcal{F}_H$, $\rho_{\mathbf{r}_I}$ may be chosen directly as a representation of $N_3(H)$. A field of the ambient theory carries its restricted representation. The group acting faithfully on the surviving moduli is $N(H)/H$.
Near $\mathcal{F}_H$ couplings are expanded in powers of the transverse displacement, and $H$ fixes which powers can appear. The leading allowed powers can control mass ratios and mixing angles near an interior fixed locus \cite{Ding:2024xhz,Carducci:2026shs}.

\begin{table}[t!]
\centering\small\tablesetup[1.5]\setlength{\tabcolsep}{6pt}
\begin{tabular}{ccc>{$}c<{$}>{$}c<{$}}
\toprule
$\Omega\in\mathcal{F}_H$ & $\dim_{\C}$ & Stabilizer $H$ & \text{Modular symmetry}~N(H) & \text{Finite modular group}~N_3(H)\\ \midrule
$\begin{tpmatrix}u&0\\0&v\end{tpmatrix}$ & 2 & $\Z_2^2$ & (\SL(2,\Z)^2)\rtimes\Z_2 & (T'\times T')\rtimes\Z_2\cong[1152,155468]\\
$\begin{tpmatrix}u&v\\v&u\end{tpmatrix}$ & 2 & $\Z_2^2$ & (\SL(2,\Z)^2)_{(2)}\rtimes\Z_2 & (T'\times T')\rtimes\Z_2\cong[1152,155468]\\
$\begin{tpmatrix}i&0\\0&u\end{tpmatrix}$ & 1 & $\Z_4\times\Z_2$ & \Z_4\times\SL(2,\Z) & \Z_4\times T'\cong[96,69]\\
$\begin{tpmatrix}\omega&0\\0&u\end{tpmatrix}$ & 1 & $\Z_6\times\Z_2$ & \Z_6\times\SL(2,\Z) & \Z_6\times T'\cong[144,156]\\
$\begin{tpmatrix}u&0\\0&u\end{tpmatrix}$ & 1 & $D_8$ & (\SL(2,\Z)\times D_8)/\Z_2 & (T'\times D_8)/\Z_2\cong[96,202]\\
$\begin{tpmatrix}u&1/2\\1/2&u\end{tpmatrix}$ & 1 & $D_8$ & D_8\cdot\Gamma_0(2)^+ & (T'\rtimes\Z_4)\rtimes\Z_2\cong[192,989]\\
$\begin{tpmatrix}u&u/2\\u/2&u\end{tpmatrix}$ & 1 & $D_{12}$ & D_{12}\cdot\Gamma_0(3)^+ & \Z_3\times\left((\Z_6\times S_3)\rtimes\Z_2\right)\cong[216,121]\\
\bottomrule
\end{tabular}
\caption{Interior invariant surfaces and curves, their complex dimensions, their pointwise stabilizers $H$, and their modular symmetries $N(H)$, along with their finite modular groups $N_3(H)$ at level 3 (the finite modular groups at level 2 can be found in ref.~\cite{Ding:2020zxw}). The last column retains the central $-I_4$. We write $[n,m]$ for the GAP SmallGroups identifier \cite{GAP4}, $D_n$ for the dihedral group of order $n$ and $T'=\SL(2,\Z_3)$. The subscript $(2)$ requires agreement modulo two. The actions of the two Fricke groups on the curve coordinates are given in eq.~\eqref{eq:Fricke-actions}.}\label{tab:interior}
\end{table}

Up to equivalence there are two invariant surfaces, five invariant curves and six isolated fixed points \cite{Gottschling:1961fix}. The two surfaces, $z=0$ and $\tau=\sigma$, are representatives of the Humbert surfaces $H_1$ and $H_4$ \cite{vanderGeer:2008sm}. Table~\ref{tab:interior} lists the surfaces and curves together with their level-3 groups, which we obtained by reducing integral generators of $N(H)$ modulo three. Top-down realizations of these elliptic invariant subspaces and their modular subsymmetries from heterotic orbifolds are discussed in refs.~\cite{Baur:2020yjl,Nilles:2021glx}. On $z=0$, two copies of $\SL(2,\Z)$ act separately on $\tau$ and $\sigma$, and the mirror exchange $\tau\leftrightarrow\sigma$ completes the modular symmetry. On $\tau=\sigma$, the useful coordinates are $u+v$ and $u-v$. Integrality requires the two modular matrices to agree modulo two. However, this condition imposes no restriction modulo three: the Chinese remainder theorem allows arbitrary reductions in both $T^\prime$ factors. On either surface the exchange interchanges the factors, giving $(T^\prime\times T^\prime)\rtimes\Z_2$. The central $-I_4$ maps to $(-I_2,-I_2)$.

The five curves retain one modulus. On $\diag(i,u)$ and $\diag(\omega,u)$ the first modulus is frozen at a fixed point of $\SL(2,\Z)$, and its stabilizer survives as a factor $\Z_4$, generated by $\si$, or $\Z_6$, generated by $\si^{-1}\mathsf{T}_1$. These factors commute with the modular group of $u$ and act as traditional flavor symmetries, so the finite groups $\Z_4\times T'$ and $\Z_6\times T'$ contain, besides the representations of $T'$, the charges of $\Z_4$ and $\Z_6$. On $uI_2$ the stabilizer consists of the eight signed permutation matrices, which form $D_8$ and commute with the diagonally embedded $\SL(2,\Z)$. For the last two curves the faithful action on the surviving coordinate is generated by a translation and a Fricke map,
\begin{equation}\label{eq:Fricke-actions}
 t\mapsto t+1\,,\qquad t\mapsto-\frac{1}{Nt}\,,\qquad
 \begin{cases}N=2,&t=u-1/2,\\N=3,&t=u/2.\end{cases}
\end{equation}
These generate $\Gamma_0(N)^+$.\footnote{$\Gamma_0(N)$ consists of matrices in $\SL(2,\Z)$ with lower-left entry divisible by $N$. The Fricke map extends this group. Its lift in the present theory is an integral $4\times4$ symplectic matrix.} In the last two rows of table~\ref{tab:interior} the dot denotes an extension with pointwise kernel $H$.

Restricted to $\mathcal{F}_H$, the theta functions satisfy $Y^{(1)}_{\mathbf{5}}=j(h,\Omega)\,\rho_{\mathbf{5}}(h)\,Y^{(1)}_{\mathbf{5}}$ for every $h\in H$, which removes some components and organizes the rest into multiplets of $N_3(H)$ (appendix~\ref{app:interior}). On $z=0$ the lattice sum factorizes into genus-one sums, and in terms of the doublet eq.~\eqref{eq:D-doublet},
\begin{equation}\label{eq:H1-factorization}
 \bigl(2Y_4,\,-\sqrt{2}\,Y_2,\,-\sqrt{2}\,Y_3,\,Y_1\bigr)\big|_{z=0}=Y^{(1,0)}_{\mathbf{2}^{(0)}}(\tau)\otimes Y^{(1,0)}_{\mathbf{2}^{(0)}}(\sigma)\,,\qquad Y_4=Y_5\,.
\end{equation}
The quintet thus becomes the quartet $(\mathbf{2},\mathbf{2})$ of $T'\times T'$, exchanged by $\tau\leftrightarrow\sigma$, while the direction $Y_4-Y_5$ vanishes. On the curves through the fixed points one factor is frozen. Since
\begin{equation}\label{eq:D-fixed}
 \rho_{\mathbf{2}^{(0)}}(\mathsf{S})\,Y^{(1,0)}_{\mathbf{2}^{(0)}}(i)=-i\,Y^{(1,0)}_{\mathbf{2}^{(0)}}(i)\,,\qquad \rho_{\mathbf{2}^{(0)}}(\mathsf{S}\mathsf{T})\,Y^{(1,0)}_{\mathbf{2}^{(0)}}(\omega)=-\omega Y^{(1,0)}_{\mathbf{2}^{(0)}}(\omega)\,,
\end{equation}
the frozen factor is an eigenvector of the corresponding stabilizer, with eigenvalue $-i$ or $(1+\omega)^{-1}=-\omega$. The surviving multiplet on $\diag(i,u)$ is therefore the doublet $Y^{(1,0)}_{\mathbf{2}^{(0)}}(u)$ of $T'$ with a definite $\Z_4$ charge, and on $\diag(\omega,u)$ it is $Y^{(1,0)}_{\mathbf{2}^{(0)}}(u)$ with a definite $\Z_6$ charge. On the equal-diagonal surface, exchanging the lattice vectors gives $Y_2=Y_3$.

\subsection{A flavor model on the surface \texorpdfstring{$\tau=\sigma$}{tau = sigma}}\label{sec:interior-model}
On this interior surface $\Omega=\left(\begin{smallmatrix}u&v\\v&u\end{smallmatrix}\right)$ the modular symmetry $N(H)\cong (\SL(2,\Z)^2)_{(2)}\rtimes\Z_2$ is generated by~\cite{Ding:2020zxw}
\begin{equation}\label{eq:equal-diagonal-generators}
 \mathsf{G}_{1}=\mathsf{T}_1\mathsf{T}_2\,,\qquad \mathsf{G}_{2}=\mathsf{T}_3\,,\qquad \mathsf{G}_{3}=\mathsf{S}\,,\qquad \mathsf{G}_{4}=\DD=\sii^2\,.
\end{equation}
Here $\DD=\gamma_{\diag(1,-1)}$ is the reflection $z\to-z$, which acts on the surface as $v\to-v$. The full level-3 finite modular group on this surface, $N_3(H)=(T'\times T')\rtimes\Z_2\cong[1152,155468]$ (table~\ref{tab:interior}), has no irreducible triplet, so a three-family assignment can use an unfaithful $\mathbf{1}_0\oplus\mathbf{2}_1$ representation whose finite image group is $\Z_3\times S_3\cong[18,3]$ (appendix~\ref{app:reps}):
\begin{equation}\label{eq:interior-assignment}
 q,l,u^c,d^c,e^c,N^c\sim(\mathbf{1}_0\oplus\mathbf{2}_1,2)\,,\qquad
 H_u,H_d\sim(\mathbf{1}_0,0)\,.
\end{equation}
The representation matrices are
\begin{align}\label{eq:interior-matter-rep}
 \rho(\mathsf{G}_{1})&=\diag(1,\omega,\omega),&
 \rho(\mathsf{G}_{2})&=\diag(1,\omega,\omega^2)\,,\notag\\
 \rho(\mathsf{G}_{3})&=I_3,&
 \rho(\mathsf{G}_{4})&=\begin{pmatrix}1&0&0\\0&0&1\\0&1&0\end{pmatrix}\,.
\end{align}
With $Y_2=Y_3$, choose the closed weight-4 multiplets
\begin{align}\label{eq:interior-quartics}
 Y^{(4)}_{\mathbf{1}_0}&=Y_1^4+16Y_1Y_2^3+8Y_1(Y_4^3+Y_5^3)+48Y_2^2Y_4Y_5\,,\notag\\
 Y^{(4)}_{\mathbf{1}_1}&=Y_1^3Y_2+3Y_1^2Y_4Y_5+4Y_2^4-4Y_2(Y_4^3+Y_5^3)\,,\notag\\
 Y^{(4)}_{\mathbf{2}_2,1}&=Y_1^3Y_4+6Y_1Y_2Y_5^2-8Y_2^3Y_4-Y_4^4+2Y_4Y_5^3\,,\notag\\
 Y^{(4)}_{\mathbf{2}_2,2}&=Y_1^3Y_5+6Y_1Y_2Y_4^2-8Y_2^3Y_5+2Y_4^3Y_5-Y_5^4\,.
\end{align}
The first is the restriction of $I_4$. Their transformations follow by substituting the quintet matrices; for example $\mathsf{G}_{1}$ multiplies $Y^{(4)}_{\mathbf{1}_1}$ by $\omega$ and the doublet by $\omega^2$, while $\mathsf{G}_{4}$ exchanges its two components. The superpotential is
\begin{align}\label{eq:interior-W}
 \mathscr{W}={}&H_u(u^c)^\top\mathcal{Y}_u q+H_d(d^c)^\top\mathcal{Y}_d q
 +H_d(e^c)^\top\mathcal{Y}_e l\notag\\
 &+H_u(N^c)^\top\mathcal{Y}_\nu l+\frac{\Lambda}{2}(N^c)^\top\mathcal{Y}_NN^c\,,
\end{align}
where
\begin{equation}\label{eq:interior-texture}
 \mathcal{Y}_f=\begin{pmatrix}
 \alpha_fY^{(4)}_{\mathbf{1}_0}&\beta_fY^{(4)}_{\mathbf{2}_2,2}&\beta_fY^{(4)}_{\mathbf{2}_2,1}\\
 \beta'_fY^{(4)}_{\mathbf{2}_2,2}&0&\delta_fY^{(4)}_{\mathbf{1}_1}\\
 \beta'_fY^{(4)}_{\mathbf{2}_2,1}&\delta_fY^{(4)}_{\mathbf{1}_1}&0
 \end{pmatrix}\,,\qquad f=u,d,e,\nu\,.
\end{equation}
The second and third diagonal entries vanish for the selected weight-4 forms. The two off-diagonal singlet--doublet couplings are independent for Dirac fermions and equal for Majorana fermions. With identity matter metrics this gives
\begin{align}\label{eq:interior-masses}
 M_f&=v_f(\det\im\Omega)^2\,\mathcal{Y}_f\,,\qquad f=u,d,e,\nu\,,\notag\\
 M_N&=\Lambda(\det\im\Omega)^2
 \begin{pmatrix}
 \alpha_NY^{(4)}_{\mathbf{1}_0}&\beta_NY^{(4)}_{\mathbf{2}_2,2}&\beta_NY^{(4)}_{\mathbf{2}_2,1}\\
 \beta_NY^{(4)}_{\mathbf{2}_2,2}&0&\delta_NY^{(4)}_{\mathbf{1}_1}\\
 \beta_NY^{(4)}_{\mathbf{2}_2,1}&\delta_NY^{(4)}_{\mathbf{1}_1}&0
 \end{pmatrix}\,,\qquad
 m_\nu=-M_\nu^\top M_N^{-1}M_\nu\,,
\end{align}
where $v_f=v_u$ for $f=u,\nu$ and $v_f=v_d$ for $f=d,e$, and the inverse is taken at a vacuum with nonzero $\det\mathcal{Y}_N$. The vanishing $(2,2)$ and $(3,3)$ entries enforce a two-zero texture in the doublet block: along the $v=0$ locus, where $Y^{(4)}_{\mathbf{2}_2,1}=Y^{(4)}_{\mathbf{2}_2,2}$, the exchange-odd doublet combination decouples, with mass proportional to $|\delta_f Y^{(4)}_{\mathbf{1}_1}|$, while turning on $v\neq 0$ splits the doublet and generates mixing. An equivalent representative of the equal-diagonal surface is $z=-\tau/2$, which reaches the Klingen boundary along the two-torsion section\footnote{Here two-torsion refers to half-periods $2z\in\Z\tau+\Z$ of the elliptic fiber $E_\tau=\C/(\Z\tau+\Z)$. In the level-3 coordinate $w=z/3$ of section~\ref{sec:klingen}, the representative $z=-\tau/2$ is $w=-\tau/6$.} $w=-\tau/6$ (section~\ref{sec:meet}), near which the Feruglio--Marrone fits lie. This is a first example of an interior locus reaching the boundary of the moduli space, which we study systematically in the next section. Quantitative quark fits are deferred to the boundary models of section~\ref{sec:models}.
\section{Cusp invariant subspaces and modular subsymmetries}\label{sec:cusps}
In a multi-modulus space such as $\HH_2$, residual modular symmetries arise not only on the interior elliptic invariant subspaces of section~\ref{sec:elliptic}, but also on a rich set of boundary (cusp) invariant subspaces where one or more moduli become large. In what follows we use ``boundary'' and ``cusp'' interchangeably to refer to these asymptotic regions. These boundary regions have received little attention in modular flavor model building. Whereas the interior loci preserve a finite stabilizer $H$, the boundary regions preserve infinite groups containing discrete translations, namely parabolic subgroups or, for the toroidal strata, subgroups of them. We first describe the boundary regions, then determine their modular subsymmetries and the branching of the weight-1 Siegel modular forms, trace how the interior loci reach the cusps, and finally describe the resulting pattern of mass hierarchies.

\subsection{Where the moduli become large}\label{sec:regions}
In genus one there is a single way to approach the boundary of moduli space, $\im\tau\to\infty$. In genus two the imaginary part $\im\Omega$ is a positive $2\times2$ matrix, and it can grow in different directions. A convenient description uses the three matrices
\begin{equation}\label{eq:Igusa-rays}
 Q_1=\begin{pmatrix}1&0\\0&0\end{pmatrix}\,,\qquad
 Q_2=\begin{pmatrix}0&0\\0&1\end{pmatrix}\,,\qquad
 Q_3=\begin{pmatrix}1&-1\\-1&1\end{pmatrix}\,.
\end{equation}
After a suitable $\GL(2,\Z)$ transformation $\Omega\to A\Omega A^\top$, the imaginary part can be written as
\begin{equation}\label{eq:Y-decomposition}
 \im\Omega=y_1Q_1+y_2Q_2+y_3Q_3 = \begin{pmatrix}y_1+y_3&-y_3\\-y_3&y_2+y_3\end{pmatrix}\,,\qquad y_i\ge0\,,
\end{equation}
that is, $\im\tau=y_1+y_3$, $\im\sigma=y_2+y_3$ and $\im z=-y_3$.\footnote{A positive binary form can be brought by $\GL(2,\Z)$ to the reduced range $0\le-2\im z\le\im\tau\le\im\sigma$, where all three coefficients $y_i$ are non-negative. In genus two the cone generated by $Q_1,Q_2,Q_3$ and its $\GL(2,\Z)$ images give the standard genus-two perfect-cone decomposition, and the regions below are the boundary strata of the corresponding toroidal compactification, first constructed by Igusa \cite{Igusa:1967,AMRT:1975,Namikawa:1980tor,Hulek:1993mod}.} Geometrically, $Q_i=v_iv_i^\top$, where $v_1=(1,0)^\top$, $v_2=(0,1)^\top$ and $v_3=(1,-1)^\top$ label the cycles $a_1$, $a_2$ and $a_1-a_2$. A large $y_i$ then describes a long, thin neck around the corresponding cycle, which is pinched as $y_i\to\infty$. A cusp limit is reached when one or more of these cone coordinates $y_i$ become large. The corresponding modulus matrices can be written explicitly:
\begin{align}\label{eq:boundary-Omega-explicit}
 \text{Klingen:}\qquad&\Omega=\begin{pmatrix}\tau&z\\z&\re\sigma+i(y_2+y_3)\end{pmatrix}\,,
 &&y_2\to\infty,\quad \tau,z\text{ finite}\,,\notag\\
 \text{two-ray:}\qquad&\Omega=\re\Omega+i\begin{pmatrix}y_1+y_3&-y_3\\-y_3&y_2+y_3\end{pmatrix}\,,
 &&y_1,y_2\to\infty,\quad y_3\text{ bounded}\,,\notag\\
 \text{corner:}\qquad&\Omega=\re\Omega+i\begin{pmatrix}y_1+y_3&-y_3\\-y_3&y_2+y_3\end{pmatrix}\,,
 &&y_1,y_2,y_3\to\infty\,.
\end{align}
Real parts are kept bounded. In the toroidal compactification these regions are nested in closure ($\text{corner}\subset\overline{\text{two-ray}}\subset\overline{\text{Klingen}}$), whereas in the Satake compactification both the two-ray and corner limits project to the zero-dimensional Siegel cusp, and their modular subsymmetries are subgroups of the Siegel cusp group rather than of the Klingen group. In the Klingen representative $y_3=-\im z$ stays fixed. In a flavor model we work near these limits, with the small exponentials nonzero. In the Klingen region only $q_\sigma=e^{2\pi i\sigma/3}$ of eq.~\eqref{eq:q-variables} is small; $\tau$ and $z$ remain moduli. In the two-ray region $q_\tau=e^{2\pi i\tau/3}$ is also small, while $\zeta=e^{2\pi iz/3}$ remains finite. At the corner the natural local coordinates are
\begin{equation}\label{eq:corner-coordinates}
 q_1=e^{2\pi i(\tau+z)/3}\,,\qquad q_2=e^{2\pi i(\sigma+z)/3}\,,\qquad
 q_3=e^{-2\pi iz/3}\,,\qquad |q_i|=e^{-2\pi y_i/3}\,.
\end{equation}
All exponentials have the level-3 normalization. If both eigenvalues of $\im\Omega$ diverge without choosing a direction, one approaches the Siegel cusp. The Klingen and Siegel cusps are the two types of rational boundary components of the Satake compactification, associated with isotropic lines and isotropic planes in $\mathds{Q}^4$ \cite{Satake:1956,Baily:1966}; the Klingen components are copies of the modular curve, the Siegel cusps are points \cite{Klingen:1967,Hulek:2002geo}. Keeping the elliptic coordinate $z$ gives the two-dimensional toroidal boundary over the Klingen curve. Over the Siegel point the two-ray limit retains $\zeta$ and gives a curve, while the corner is a point. Table~\ref{tab:cusp-groups} writes these complex dimensions as an ordered pair (Sat., tor.). At Klingen the first entry counts $\tau$ and the second counts $(\tau,z)$. Figure~\ref{fig:cone} illustrates the corresponding limits, and figure~\ref{fig:degenerations} below shows how the genus-two surface degenerates in each of them.

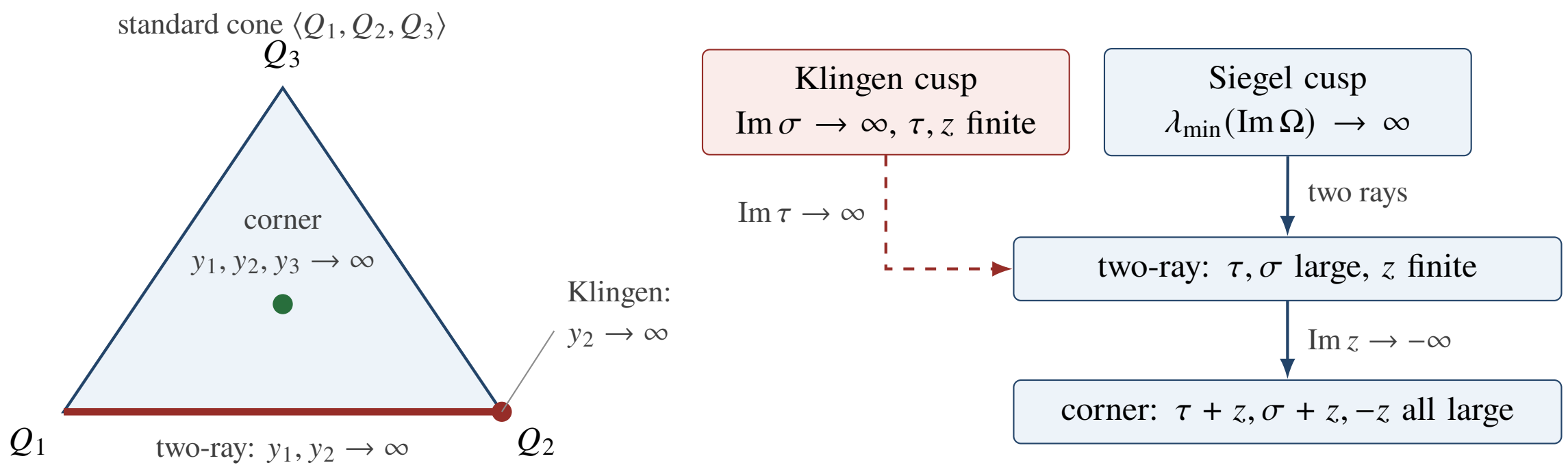
\begin{figure}[t!]
\centering
\begin{tikzpicture}[font=\small,scale=.92,transform shape]
\coordinate (A) at (0,0);\coordinate (B) at (4.6,0);\coordinate (C) at (2.3,3.4);
\fill[softblue] (A)--(B)--(C)--cycle;
\draw[inkblue,thick] (A)--(B)--(C)--cycle;
\draw[inkred,line width=2pt] (A)--(B);
\fill[inkred] (B) circle(3pt);
\fill[inkgreen] (2.3,1.13) circle(3pt);
\node[below left=1pt] at (A) {$Q_1$};\node[below right=1pt] at (B) {$Q_2$};\node[above=2pt] at (C) {$Q_3$};
\node[note] at (2.3,-0.42) {two-ray: $y_1,y_2\to\infty$};
\node[note,align=center] at (2.3,1.75) {corner\\$y_1,y_2,y_3\to\infty$};
\draw[black!45] (B)--(5.15,0.85);
\node[note,anchor=west,align=left] at (5.15,1.02) {Klingen:\\$y_2\to\infty$};
\node[note] at (2.3,4.05) {standard cone $\langle Q_1,Q_2,Q_3\rangle$};
\node[cbox,text width=35mm,anchor=west] (k) at (6.7,3.25) {Klingen cusp\\$\im\sigma\to\infty$, $\tau,z$ finite};
\node[box,text width=35mm,anchor=west] (s) at (10.92,3.25) {Siegel cusp\\$\lambda_{\min}(\im\Omega)\to\infty$};
\node[box,text width=54mm,anchor=center] (r) at (12.85,1.5) {two-ray: $\tau,\sigma$ large, $z$ finite};
\node[box,text width=54mm,anchor=center] (c) at (12.85,0) {corner: $\tau+z,\sigma+z,-z$ all large};
\draw[arr] (s.south)--node[right,note,xshift=2pt]{two rays}(r.north);
\draw[arr,dashed,inkred] (k.south) |- node[pos=0.25,left,note,xshift=-2pt]{$\im\tau\to\infty$} (r.west);
\draw[arr] (r.south)--node[right,note,xshift=2pt]{$\im z\to-\infty$}(c.north);
\end{tikzpicture}
\caption{Cusp limits and the data they retain. On the left the triangle represents the standard cone generated by $Q_1,Q_2,Q_3$: a vertex selects a rank-one degeneration, an edge selects two rays, and the interior selects all three. On the right the two rational Satake boundary types, namely the one-dimensional Klingen cusp and the zero-dimensional Siegel cusp, appear in parallel at the top. The upper solid arrow selects the one-dimensional two-ray toroidal stratum over the Siegel cusp, the lower solid arrow passes to the corner point in its closure (both carrying subgroups of $\Z^3\rtimes\GL(2,\Z)$ in table~\ref{tab:cusp-groups}), and the dashed arrow indicates the toroidal closure relation from Klingen to two-ray upon taking $\im\tau\to\infty$.}\label{fig:cone}
\end{figure}

\subsection{Modular subsymmetries and their actions}\label{sec:stabilisers}
In each boundary region, the surviving modular transformations are those elements of $\Sp(4,\Z)$ that map the region into itself. Unlike the interior loci of section~\ref{sec:elliptic}, which are fixed pointwise by a finite subgroup $H$ with normalizer $N(H)$, a boundary region is defined by a degeneration limit rather than by a finite fixed-point group; its modular subsymmetry is simply the stabilizer of that boundary component (or toroidal cone) in $\Sp(4,\Z)$, generalizing the translation $\tau\to\tau+1$ of the genus-one cusp.

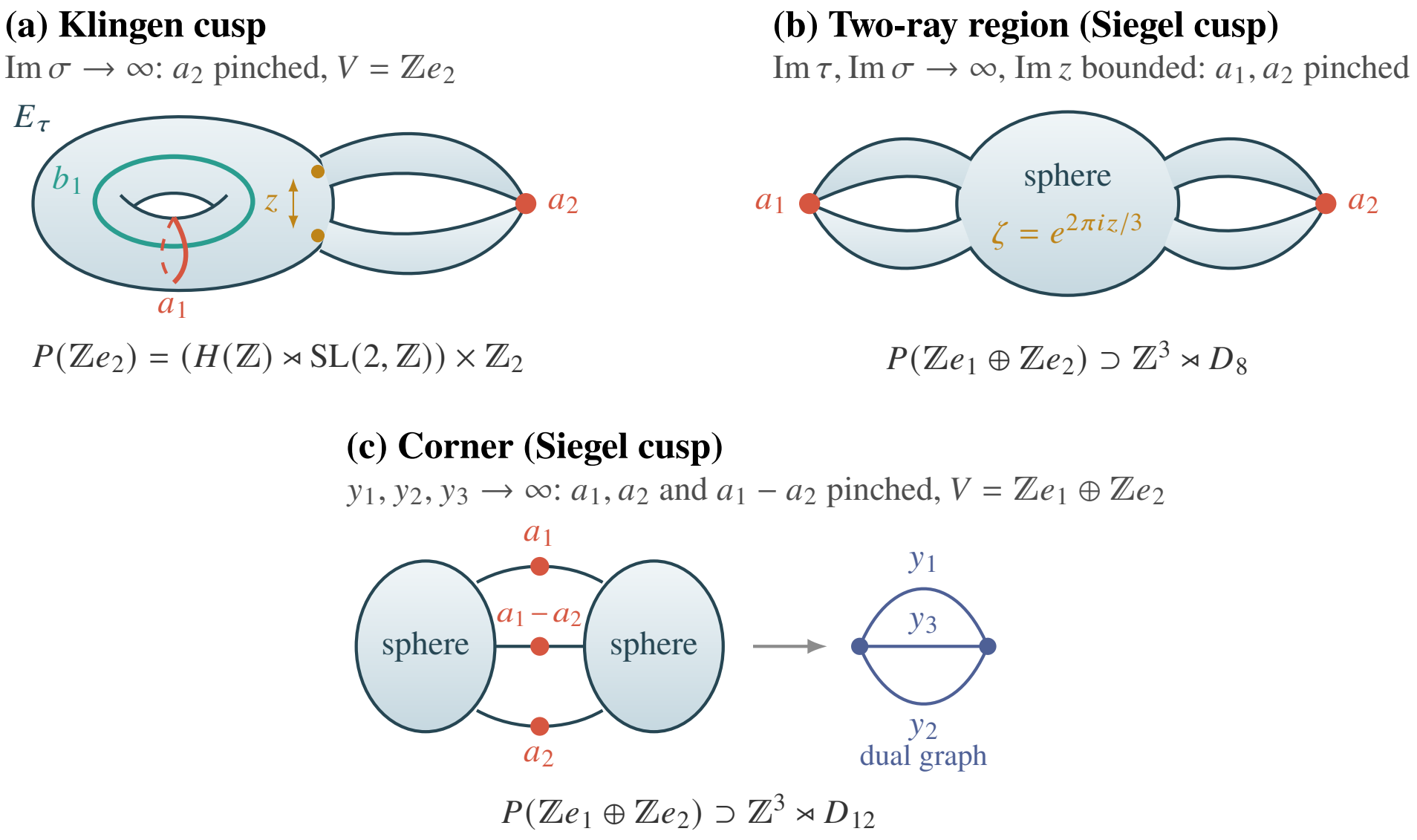
\begin{figure}[t!]
\centering
\begin{tikzpicture}[font=\footnotesize, scale=0.95, transform shape,
  surf/.style={thick,draw=surfline,top color=surftop,bottom color=surfbot},
  ttl/.style={anchor=west,font=\small\bfseries},
  sub/.style={anchor=west,text=black!70},
  stab/.style={text=black!80}]
\begin{scope}[shift={(0,0)}]
\node[ttl] at (-2.9,1.65) {(a) Klingen cusp};
\node[sub] at (-2.9,1.25) {$\im\sigma\to\infty$:\ $a_2$ pinched,\ $V=\Z e_2$};
\shadedraw[surf]
  (-2.50,0) .. controls (-2.50,1.02) and (-0.25,1.02) .. (0.24,0.34)
            .. controls (0.34,0.12) and (0.34,-0.12) .. (0.24,-0.34)
            .. controls (-0.25,-1.02) and (-2.50,-1.02) .. (-2.50,0) -- cycle;
\draw[very thick,bcyc] (-1.18,0.02) ellipse (0.74 and 0.42);
\fill[white] (-1.56,-0.03) .. controls (-1.36,0.14) and (-1.00,0.14) .. (-0.80,-0.03)
             .. controls (-1.00,-0.185) and (-1.36,-0.185) .. (-1.56,-0.03) -- cycle;
\draw[thick,surfline] (-1.68,0.08) .. controls (-1.41,-0.21) and (-0.95,-0.21) .. (-0.68,0.08);
\draw[thick,surfline] (-1.56,-0.03) .. controls (-1.36,0.14) and (-1.00,0.14) .. (-0.80,-0.03);
\draw[thick,acyc,dashed] (-1.18,-0.13) .. controls (-1.33,-0.38) and (-1.33,-0.60) .. (-1.18,-0.74);
\draw[very thick,acyc] (-1.18,-0.13) .. controls (-1.03,-0.38) and (-1.03,-0.60) .. (-1.18,-0.74);
\node[bcyc] at (-2.17,0.22) {$b_1$};
\node[acyc] at (-1.18,-0.98) {$a_1$};
\node[text=surfline] at (-2.5,0.8) {$E_\tau$};
\shadedraw[surf]
  (0.19,0.44) .. controls (1.12,0.92) and (1.92,0.52) .. (2.12,0)
              .. controls (1.72,0.18) and (1.02,0.44) .. (0.28,0.16);
\shadedraw[surf]
  (0.19,-0.44) .. controls (1.12,-0.92) and (1.92,-0.52) .. (2.12,0)
               .. controls (1.72,-0.18) and (1.02,-0.44) .. (0.28,-0.16);
\fill[acyc] (2.12,0) circle (2.8pt);
\node[acyc,right=2pt] at (2.12,0) {$a_2$};
\fill[modcol] (0.17,0.30) circle (1.8pt);
\fill[modcol] (0.17,-0.30) circle (1.8pt);
\draw[{Latex[length=1.2mm]}-{Latex[length=1.2mm]},thin,modcol] (-0.06,0.24) -- (-0.06,-0.24) node[midway,left] {$z$};
\node[stab] at (-0.2,-1.45) {$P(\Z e_2)=\left(H(\Z)\rtimes\SL(2,\Z)\right)\times\Z_2$};
\end{scope}
\begin{scope}[shift={(7.2,0)}]
\node[ttl] at (-2.9,1.65) {(b) Two-ray region (Siegel cusp)};
\node[sub] at (-2.9,1.25) {$\im\tau,\im\sigma\to\infty$,\ $\im z$ bounded:\ $a_1,a_2$ pinched};
\shadedraw[surf] (0,0) ellipse (1.04 and 0.86);
\foreach \s in {-1,1}{
\shadedraw[surf]
  ({\s*0.90},0.41) .. controls ({\s*1.62},0.86) and ({\s*2.22},0.46) .. ({\s*2.42},0)
               .. controls ({\s*2.02},0.18) and ({\s*1.48},0.39) .. ({\s*1.00},0.14);
\shadedraw[surf]
  ({\s*0.90},-0.41) .. controls ({\s*1.62},-0.86) and ({\s*2.22},-0.46) .. ({\s*2.42},0)
                .. controls ({\s*2.02},-0.18) and ({\s*1.48},-0.39) .. ({\s*1.00},-0.14);
\fill[acyc] ({\s*2.42},0) circle (2.8pt);
}
\node[acyc,left=2pt] at (-2.42,0) {$a_1$};
\node[acyc,right=2pt] at (2.42,0) {$a_2$};
\node[text=surfline] at (0,0.24) {sphere};
\node[text=modcol] at (0,-0.26) {$\zeta=e^{2\pi iz/3}$};
\node[stab] at (0,-1.45) {$P(\Z e_1\oplus\Z e_2)\supset \Z^3\rtimes D_8$};
\end{scope}
\begin{scope}[shift={(2.8,-4.15)}]
\node[ttl] at (-2.5,1.85) {(c) Corner (Siegel cusp)};
\node[sub] at (-2.5,1.45) {$y_1,y_2,y_3\to\infty$:\ $a_1,a_2$ and $a_1-a_2$ pinched,\ $V=\Z e_1\oplus\Z e_2$};
\shadedraw[surf] (-1.62,0) ellipse (0.65 and 0.80);
\shadedraw[surf] (0.52,0) ellipse (0.65 and 0.80);
\node[text=surfline] at (-1.62,0) {sphere};
\node[text=surfline] at (0.52,0) {sphere};
\draw[thick,surfline] (-1.14,0.60) .. controls (-0.78,0.80) and (-0.30,0.80) .. (0.04,0.60);
\draw[thick,surfline] (-0.97,0.0) -- (-0.13,0.0);
\draw[thick,surfline] (-1.14,-0.60) .. controls (-0.78,-0.80) and (-0.30,-0.80) .. (0.04,-0.60);
\fill[acyc] (-0.55,0.75) circle (2.5pt);
\fill[acyc] (-0.55,0.0) circle (2.5pt);
\fill[acyc] (-0.55,-0.75) circle (2.5pt);
\node[acyc,above=1pt] at (-0.55,0.77) {$a_1$};
\node[acyc,above=1pt,font=\scriptsize] at (-0.55,0.02) {$a_1\!-\!a_2$};
\node[acyc,below=1pt] at (-0.55,-0.77) {$a_2$};
\draw[-{Latex[length=1.8mm]},thick,black!45] (1.45,0) -- (2.15,0);
\fill[graphcol] (2.45,0) circle (2.3pt);
\fill[graphcol] (3.65,0) circle (2.3pt);
\draw[thick,graphcol] (2.45,0) .. controls (2.75,0.72) and (3.35,0.72) .. (3.65,0) node[midway,above] {$y_1$};
\draw[thick,graphcol] (2.45,0) -- (3.65,0) node[midway,above=-1pt] {$y_3$};
\draw[thick,graphcol] (2.45,0) .. controls (2.75,-0.72) and (3.35,-0.72) .. (3.65,0) node[midway,below] {$y_2$};
\node[graphcol,font=\scriptsize] at (3.05,-1.05) {dual graph};
\node[stab] at (0.85,-1.55) {$P(\Z e_1\oplus\Z e_2)\supset \Z^3\rtimes D_{12}$};
\end{scope}
\end{tikzpicture}
\caption{Boundary degenerations of the genus-two surface of figure~\ref{fig:genus2}, pinching one, two, or three cycles. Pinched cycles are shown as red dots; their vectors span the isotropic sublattice $V$ of eq.~\eqref{eq:parabolic}, and the stabilizer of each region is given below the panel. (a)~Klingen cusp: pinching $a_2$ leaves the elliptic curve $E_\tau$ with two points glued into a node, separated by $z$ (cf.\ figure~1 of ref.~\cite{Feruglio:2026xxo}). (b)~Two-ray region: pinching $a_1$ and $a_2$ leaves a sphere with two nodes, parametrized by $\zeta$. (c)~Corner: a third cycle, in the class $a_1-a_2$, is pinched and the surface splits into two spheres joined at three nodes, whose dual graph is the theta graph with edge parameters $y_i$; its symmetries permute the three edges and give the finite factor $D_{12}\cong S_3\times\Z_2$ of the corner stabilizer (appendix~\ref{sec:corner-rule}). Panels~(b) and~(c) are the two toroidal strata over the Siegel cusp, whose full parabolic stabilizer is $P(\Z e_1\oplus\Z e_2)=\Z^3\rtimes\GL(2,\Z)$ (figure~\ref{fig:cone}).}\label{fig:degenerations}
\end{figure}

Notice that this definition follows a general recipe, which applies equally to $\Sp(2g,\Z)$ at higher genus. In the language of section~\ref{sec:moduli}, a boundary limit pinches a set of $a$-cycles. Their vectors span a primitive isotropic sublattice $V\subset\Z^{2g}$ of rank $r\in\{1,\dots,g\}$, and the modular subsymmetry is the parabolic stabilizer\footnote{With the cycles ordered as $(b;a)$, a change of basis $(b';a')=\gamma\,(b;a)$ maps the span of the pinched $a$-cycles to itself precisely when $\gamma V=V$, as follows from $\gamma^\top J\gamma=J$.}
\begin{equation}\label{eq:parabolic}
 P(V)=\bigl\{\gamma\in\Sp(2g,\Z):\ \gamma V=V\bigr\}\,.
\end{equation}
Because every parabolic subgroup is self-normalizing, $N_{\Sp(2g,\Z)}(P(V))=P(V)$, no second step analogous to passing from $H$ to $N(H)$ is needed: $P(V)$ already contains all transformations preserving the boundary component. For a toroidal region one keeps in addition only the elements of the $\GL(r,\Z)$ factor of $P(V)$ that preserve the chosen cone, and these form a finite group. At genus two, $r=1$ gives the line $\Z e_2$, for which $a_2$ is pinched (the Klingen cusp, $\im\sigma\to\infty$), and $r=2$ gives the plane $\Z e_1\oplus\Z e_2$, for which $a_1$ and $a_2$ are pinched (the Siegel cusp, where both eigenvalues of $\im\Omega$ are large). The two-ray region and the corner refine the latter, and at the corner a cycle in the class $a_1-a_2$ is pinched as well (figure~\ref{fig:degenerations}). Matter fields always transform as in eq.~\eqref{eq:siegel-form-field}, so we only need the action on the moduli and the automorphy factor $j$. Below we determine the modular subsymmetry for each boundary region explicitly; the resulting groups are summarized in table~\ref{tab:cusp-groups}.

\paragraph{Klingen region.}
The surviving transformations are generated by the modular transformations of $\tau$, the translations of $z$, the shift of $\sigma$ and $-I_4$:
\begin{align}\label{eq:klingen-action}
 G_{\gamma_1}:&\quad \tau\to\frac{a\tau+b}{c\tau+d},\quad z\to\frac{z}{c\tau+d},\quad \sigma\to\sigma-\frac{cz^2}{c\tau+d},&& j=c\tau+d\,,\notag\\
 H(\lambda,\mu,\kappa):&\quad \tau\to\tau,\quad z\to z+\lambda\tau+\mu,\quad \sigma\to\sigma+\lambda^2\tau+2\lambda z+\kappa,&& j=1\,,
\end{align}
with $\gamma_1=\left(\begin{smallmatrix}a&b\\c&d\end{smallmatrix}\right)\in\SL(2,\Z)$ and $\lambda,\mu,\kappa$ integers. They are realized inside $\Sp(4,\Z)$ by the symplectic matrices
\begin{equation}\label{eq:embedded}
 G_{\gamma_1}=\begin{pmatrix}a&0&b&0\\0&1&0&0\\c&0&d&0\\0&0&0&1\end{pmatrix}\,,\qquad
 H(\lambda,\mu,\kappa)=\begin{pmatrix}1&0&0&\mu\\\lambda&1&\mu&\kappa-\lambda\mu\\0&0&1&-\lambda\\0&0&0&1\end{pmatrix}\,.
\end{equation}
The pair $(\tau,z)$ transforms exactly as under the Jacobi group of Feruglio and Marrone, and the elements $H(\lambda,\mu,\kappa)$ form the integral Heisenberg group in the convention eq.~\eqref{eq:H-law}. In the Siegel theory these transformations also shift $\sigma$, and $H(0,0,1)$ is simply $\sigma\to\sigma+1$. For the Klingen isotropic line $\Z e_2$, an invertible integral matrix preserving $\Z e_2$ must send the primitive generator $e_2$ to $\pm e_2$ (so its second column is $\pm e_2$): $G_{\gamma_1}$ and $H(\lambda,\mu,\kappa)$ fix $e_2$ itself, while $-I_4$ sends $e_2\mapsto -e_2$. Geometrically, these transformations preserve the pinched cycle $a_2$ up to orientation. The modular symmetry is therefore $P(\Z e_2)=\Gamma^{\mathrm{J}}\times\Z_2$, where $\Gamma^{\mathrm{J}}=H(\Z)\rtimes\SL(2,\Z)$ is the integral Jacobi group (the stabilizer of $e_2$) and the $\Z_2$ factor is $\{\pm I_4\}$. It is generated by
\begin{equation}\label{eq:Klingen-generator-words}
 P(\Z e_2)=\left(H(\Z)\rtimes\SL(2,\Z)\right)\times\Z_2 = \left\langle~\si,\quad \mathsf{T}_1,\quad \mathsf{X}=\si \mathsf{T}_3\si^{-1},\quad \mathsf{Y}=\mathsf{T}_3,\quad \mathsf{Z}=\mathsf{T}_2,\quad \mathsf{S}^2~\right\rangle\,.
\end{equation}
Since $\sigma\to\sigma+3$ belongs to $\Gamma_2(3)$, a Siegel modular form of level 3 admits a Fourier--Jacobi expansion $Y^{(k)}_{\mathbf{r}}(\Omega)=\sum_{n_2\ge 0}q_\sigma^{n_2}\,\phi_{n_2}(\tau,z)$ (eq.~\eqref{eq:FJ}), whose coefficients transform as vector-valued Jacobi forms of index $m=3n_2$. Section~\ref{sec:klingen} develops this case in detail.

\paragraph{Siegel cusp.}
When both cycles degenerate without a specified direction, preserving the maximal isotropic plane $\Z e_1\oplus\Z e_2$ forces the lower-left block $C=0$, so the parabolic stabilizer $P(\Z e_1\oplus\Z e_2)$ consists of the symmetric translations and the $\GL(2,\Z)$ transformations,
\begin{equation}\label{eq:siegel-parabolic}
 \Omega\to A\Omega A^\top+B\,,\qquad A\in\GL(2,\Z),\quad B=B^\top\ \text{integer}\,,\qquad j=\det A=\pm1\,.
\end{equation}
Together they form the group $\Z^3\rtimes\GL(2,\Z)$, generated by
\begin{equation}\label{eq:Siegel-generator-words}
 P(\Z e_1\oplus\Z e_2)=\Z^3\rtimes\GL(2,\Z) = \left\langle~\mathsf{T}_1,\mathsf{T}_2,\mathsf{T}_3,\quad \RR=\EE\FF^{-1}\EE,\quad \EE=\sii \mathsf{T}_3\sii^{-1},\quad \DD=\sii^2 ~\right\rangle\,,
\end{equation}
where $\FF=\si \mathsf{T}_3\si^{-1}$. 
Both translations and $\GL(2,\Z)$ transformations preserve $\det\im\Omega$, so the kinetic terms are invariant. A level-3 Siegel modular form has the Fourier expansion
\begin{equation}\label{eq:full-Fourier}
 Y^{(k)}_{\mathbf{r}}(\Omega)=\sum_{n_1,\ell,n_2}c(n_1,\ell,n_2)\,e^{2\pi i(n_1\tau+\ell z+n_2\sigma)/3}\,,\qquad n_1,n_2\ge0,\quad \ell^2\le4n_1n_2\,,
\end{equation}
where $n_1,\ell,n_2$ are integers. The exponents can be collected in a matrix $T=\left(\begin{smallmatrix}n_1&\ell/2\\\ell/2&n_2\end{smallmatrix}\right)$, with $n_1\tau+\ell z+n_2\sigma=\tr(T\Omega)$, and the condition $\ell^2\le4n_1n_2$ says that $T$ is positive semi-definite. It is required by holomorphy at the boundary.\footnote{This is the Koecher principle \cite{Koecher:1954}: a holomorphic Siegel modular form of genus two is automatically bounded near the boundary, and only exponents with $T$ positive semi-definite occur.} Comparing both sides of eq.~\eqref{eq:siegel-form-field} term by term gives
\begin{equation}\label{eq:Fourier-actions}
 \rho_Y(\mathsf{T}_B)\,c(T)=e^{2\pi i\tr(TB)/3}\,c(T)\,,\qquad
 c(A^\top TA)=(\det A)^k\,\rho_Y(\gamma_A)^{-1}\,c(T)\,.
\end{equation}
The translations assign each Fourier coefficient a phase, i.e.\ a charge under $\Z_3^3$ determined by $(n_1,\ell,n_2)$ modulo three, while the $\GL(2,\Z)$ transformations relate the coefficients of different exponents. For instance, the exchange $\tau\leftrightarrow\sigma$ maps $(n_1,\ell,n_2)\to(n_2,\ell,n_1)$ and $z\to-z$ maps $\ell\to-\ell$.

\paragraph{Two-ray region.}
Here the surviving $\GL(2,\Z)$ transformations are the eight signed permutation matrices, which form $D_8$ and are generated by the exchange $\PP$ ($\tau\leftrightarrow\sigma$) and the reflection $\DD$ ($z\to-z$). On the local coordinates,
\begin{equation}\label{eq:curve-action}
 \mathsf{T}_B:\ (q_\tau,q_\sigma,\zeta)\to(\omega^{b_{11}}q_\tau,\ \omega^{b_{22}}q_\sigma,\ \omega^{b_{12}}\zeta)\,,\qquad
 \PP:\ q_\tau\leftrightarrow q_\sigma\,,\qquad \DD:\ \zeta\to\zeta^{-1}\,,
\end{equation}
with $j=-1$ for $\PP$ and $\DD$. The modular symmetry is thus the subgroup $\Z^3\rtimes D_8\subset\Z^3\rtimes\GL(2,\Z)$, generated by
\begin{equation}\label{eq:curve-generator-words}
 \Z^3\rtimes D_8 = \left\langle~ \mathsf{T}_1,\mathsf{T}_2,\mathsf{T}_3,\quad \PP=\DD\RR,\quad \DD=\sii^2~\right\rangle\,.
\end{equation} A coupling is a double power series in $q_\tau$ and $q_\sigma$,
\begin{equation}\label{eq:curve-expansion}
 Y^{(k)}_{\mathbf{r}}=\sum_{n_1,n_2\ge0}q_\tau^{n_1}q_\sigma^{n_2}\,\psi_{n_1n_2}(\zeta)\,,\qquad \psi_{n_1n_2}(\zeta)=\sum_{\ell^2\le4n_1n_2}c(n_1,\ell,n_2)\,\zeta^\ell\,,
\end{equation}
whose coefficients are polynomials in $\zeta$ and $\zeta^{-1}$. The remaining modulus $\zeta$ is acted on only by the rotations $\zeta\to\omega\zeta$ and the inversion $\zeta\to\zeta^{-1}$.

\paragraph{Corner.}
Here the surviving $\GL(2,\Z)$ transformations permute the three matrices $Q_i$. They form the dihedral group $D_{12}$, generated by $\PP$ and $\CC=\gamma_{A_{\CC}}$ with $A_{\CC}=\left(\begin{smallmatrix}0&-1\\1&1\end{smallmatrix}\right)$. Their action and that of the translations are
\begin{gather}
 \mathsf{T}_B:\ q_i\to\omega^{t_i}q_i\,,\quad \text{with} \quad (t_1,t_2,t_3)=(b_{11}+b_{12},\,b_{22}+b_{12},\,-b_{12})\,,\notag\\
 \PP:\ q_1\leftrightarrow q_2\,,\qquad \CC:\ (q_1,q_2,q_3)\to(q_3,q_1,q_2)\,,\label{eq:corner-action}
\end{gather}
with $j=-1$ for $\PP$ and $j=1$ for $\CC$. The modular symmetry is thus the subgroup $\Z^3\rtimes D_{12}\subset\Z^3\rtimes\GL(2,\Z)$, generated by
\begin{equation}\label{eq:corner-generator-words}
 \Z^3\rtimes D_{12} = \left\langle~ \mathsf{T}_1,\mathsf{T}_2,\mathsf{T}_3,\quad \CC=\RR^{-1}\EE,\quad \PP=\DD\RR ~\right\rangle\,.
\end{equation} Each Fourier term becomes a monomial,
\begin{equation}\label{eq:corner-monomial}
 e^{2\pi i(n_1\tau+\ell z+n_2\sigma)/3}=q_1^{n_1}q_2^{n_2}q_3^{n_3}\,,\qquad n_3=n_1+n_2-\ell\,,
\end{equation}
and positivity of $T$ makes all exponents non-negative. Appendix~\ref{sec:corner-rule} turns this condition into a simple rule for the leading powers.

\subsection{Finite modular groups at level 3}\label{sec:groups3}
Reducing the modular subsymmetries modulo three gives the finite modular groups of the boundary regions, listed in table~\ref{tab:cusp-groups}. The indices of their congruence images count the inequivalent boundary components: at level 3 there are $40$ inequivalent Klingen regions, $40$ Siegel cusps, $240$ two-ray regions and $160$ corners. The first two count the respective sets of isotropic lines and isotropic planes in $\Z_3^4$.

The groups in table~\ref{tab:cusp-groups} are images modulo $\{\pm I_4\}$. As noted in section~\ref{sec:finite-group}, $-I_4$ fixes every $\Omega$ with $j(-I_4,\Omega)=1$, so integer-weight Siegel modular forms have $\rho_Y(-I_4)=I$ and see only this quotient. Matter fields may still carry double-cover representations with $\rho_\psi(-I_4)=-I$, on which the central $\Z_2$ acts as a parity; the full groups including $-I_4$ are given in appendix~\ref{app:cusp-groups}.

\begin{table}[t]
\centering\small\tablesetup[1.5]
\setlength{\tabcolsep}{4.5pt}
\begin{tabular}{lcccc}
\toprule
\multirow{2}{*}{Region} & \multirow{2}{*}{Representative $\Omega$} & $\dim_{\C}$ & \multirow{2}{*}{Modular symmetry} & Finite modular group\\ [-4pt]
& & (Sat., tor.) & & (projective, $N=3$)\\
\midrule
Klingen &
$\twoline{\begin{tpmatrix}\tau&z\\z&\re\sigma+i(y_2+y_3)\end{tpmatrix}}
 {y_2\to\infty,\ \tau,z\ \text{fixed}}$ &
$(1,\,2)$ &
$\twoline{\left(H(\Z)\rtimes\SL(2,\Z)\right)}{\times\,\Z_2}$ &
$\twoline{J_3=\Delta(27)\rtimes T'}{\cong[648,533]}$\\[8pt] \midrule
Siegel cusp &
$\twoline{\begin{tpmatrix}\tau&z\\z&\sigma\end{tpmatrix}}
 {\lambda_{\min}(\im\Omega)\to\infty}$ &
$(0,\,\text{--})$ &
$\Z^3\rtimes\GL(2,\Z)$ &
$\twoline{\Z_3^3\rtimes S_4}{\cong[648,704]}$\\[8pt]
Two-ray &
$\twoline{\re\Omega+i\begin{tpmatrix}y_1+y_3&-y_3\\-y_3&y_2+y_3\end{tpmatrix}}
 {y_1,y_2\to\infty,\ y_3\ \text{bounded}}$ &
$(0,\,1)$ &
$\Z^3\rtimes D_8$ &
$\twoline{\Z_3\times S_3\times S_3}{\cong[108,38]}$\\[8pt]
Corner &
$\twoline{\re\Omega+i\begin{tpmatrix}y_1+y_3&-y_3\\-y_3&y_2+y_3\end{tpmatrix}}
 {y_1,y_2,y_3\to\infty}$ &
$(0,\,0)$ &
$\Z^3\rtimes D_{12}$ &
$\twoline{\Z_3^3\rtimes S_3}{\cong[162,10]}$\\[3pt]
\bottomrule
\end{tabular}
\caption{Boundary regions, their representative moduli $\Omega$ with bounded real parts, their complex dimensions, their modular symmetries, and their (projective) finite modular groups at level 3 (modulo $\{\pm I_4\}$; the full groups retaining $-I_4$ are given in appendix~\ref{app:cusp-groups}). In $\dim_{\C}$ (Sat., tor.), the first entry counts the moduli left on the Satake boundary and the second counts the coordinates on the corresponding toroidal stratum; the Siegel cusp alone specifies no toroidal stratum. Here $\lambda_{\min}$ is the smaller eigenvalue of $\im\Omega$, and $\Z^3=\operatorname{Sym}_2(\Z)$ acts by translations.}\label{tab:cusp-groups}
\end{table}

For the Klingen vector stabilizer $\Gamma^{\mathrm{J}}=H(\Z)\rtimes\SL(2,\Z)$ of section~\ref{sec:stabilisers}, reduction modulo three gives
\begin{equation}\label{eq:finite-Jacobi}
 J_3=H(\Z_3)\rtimes\SL(2,\Z_3)
    \cong\Delta(27)\rtimes T'\cong[648,533]\,.
\end{equation}
Here $H(\Z_3)\cong \Delta(27)$ is the Heisenberg group of order $27$ and $T'\cong \SL(2,\Z_3)$. Because $P(\Z e_2)=\Gamma^{\mathrm{J}}\times\{\pm I_4\}$, the quotient by $\{\pm I_4\}$ leaves $J_3$ intact; in particular, the central involution of $T'$ is $\mathsf{s}_J^2=\si^2$, not $-I_4$.
\begin{itemize}
\item The Klingen and Siegel modular subsymmetries give different finite modular groups. In $J_3$ the translations form the non-abelian group $\Delta(27)$, because translations of $z$ by $\tau$ and by $1$ commute only up to a shift of $\sigma$. At the Siegel cusp the three translations commute and form $\Z_3^3$, and the $\GL(2,\Z)$ transformations contribute $\PGL(2,\Z_3)\simeq S_4$.
\item At fixed $\tau$, the Heisenberg translations move $w=z/3$ by the nine three-torsion points $(\lambda\tau+\mu)/3$ of $\C/(\Z\tau+\Z)$. Together with the modular action on $\tau$, they act on the level-3 family with effective group $J_3/Z(J_3)\cong[216,153]$.\footnote{This is the Hessian group acting on the Hesse family $x^3+y^3+z^3=3\lambda xyz$ \cite{Artebani:2009hes}. The completed Klingen boundary surface at level 3 is this elliptic family \cite{Hoffman:2001lt}. The modular action moves between fibers, rather than acting only on one fixed elliptic curve.} The center, $\sigma\to\sigma+1$, rotates the normal parameter, $q_\sigma\to\omega q_\sigma$. Thus the Heisenberg action on fields retains a central phase that is invisible on $(\tau,w)$.
\end{itemize}

\subsection{Branching of the weight-1 Siegel modular forms}\label{sec:branching}
In an open neighborhood of each boundary region, all three moduli remain finite and the five weight-1 Siegel modular forms stay linearly independent, splitting into multiplets of the corresponding modular symmetry (only in the strict asymptotic limit $q_a\to 0$ do the subleading multiplets vanish). The splitting follows directly from eqs.~\eqref{eq:rho-T} and \eqref{eq:rho-A} and is listed in table~\ref{tab:boundary-branching}. We give the representation matrices explicitly, since they are needed for model building.

\begin{table}[t]
\centering\small\tablesetup[1.4]
\begin{tabular}{ccll}\toprule
Region & Finite modular group & $Y^{(1)}_{\mathbf{5}}$ & Multiplets\\ \midrule
Klingen & $J_3=\Delta(27)\rtimes T'$ & $\mathbf{2}\oplus\mathbf{3}$ & $Y^{(1)}_{\mathbf{2}^{(0)}}=(-\sqrt{2}\,Y_2,\,Y_1)$; $Y^{(1)}_{\mathbf{3}^{(0)}}=(Y_3,Y_4,Y_5)$\\ \midrule
Siegel cusp & $\Z_3^3\rtimes S_4$ & $\mathbf{1}\oplus\mathbf{4}$ & $Y^{(1)}_{\mathbf{1}}=Y_1$; $Y^{(1)}_{\mathbf{4}}=(Y_2,Y_3,Y_4,Y_5)$\\
two-ray & $\Z_3\times S_3\times S_3$ & $\mathbf{1}\oplus\mathbf{2}\oplus\mathbf{2}'$ & $Y^{(1)}_{\mathbf{1}}=Y_1$; $Y^{(1)}_{\mathbf{2}}=(Y_2,Y_3)$; $Y^{(1)}_{\mathbf{2}'}=(Y_4,Y_5)$\\
corner & $\Z_3^3\rtimes S_3$ & $\mathbf{1}_0^-\oplus\mathbf{1}_1^-\oplus\mathbf{3}_{0,1}^-$ & $Y^{(1)}_{\mathbf{1}_0^-}=Y_1$; $Y^{(1)}_{\mathbf{1}_1^-}=Y_4$; $Y^{(1)}_{\mathbf{3}_{0,1}^-}=(Y_3,Y_2,Y_5)$\\\bottomrule
\end{tabular}
\caption{Branching of the weight-1 Siegel quintet $Y^{(1)}_{\mathbf{5}}(\Omega)$ under the (projective) finite modular groups near the boundary. Because we work in a neighborhood of each cusp rather than strictly at $q_a=0$, all five functions remain linearly independent and enter at different Fourier orders; the same five functions provide different modular form multiplets in different regions.}\label{tab:boundary-branching}
\end{table}

\paragraph{Klingen region.}
In the basis $\left(Y^{(1)}_{\mathbf{2}^{(0)}};\,Y^{(1)}_{\mathbf{3}^{(0)}}\right)=\left(-\sqrt{2}\,Y_2,\,Y_1;\ Y_3,\,Y_4,\,Y_5\right)$ the quintet transforms with the matrices of Feruglio and Marrone,
\begin{gather}
 \rho(\mathsf{s}_J)=\rho_{\mathbf{2}^{(0)}}(\mathsf{s}_J)\oplus\rho_{\mathbf{3}^{(0)}}(\mathsf{s}_J)\,,\qquad \rho(\mathsf{t}_J)=\rho_{\mathbf{2}^{(0)}}(\mathsf{t}_J)\oplus\rho_{\mathbf{3}^{(0)}}(\mathsf{t}_J)\,,\notag\\
 \rho(\mathsf{X})=I_2\oplus\rho_{\mathbf{3}^{(0)}}(\mathsf{X})\,,\qquad \rho(\mathsf{Y})=I_2\oplus\rho_{\mathbf{3}^{(0)}}(\mathsf{Y})\,,\qquad \rho(\mathsf{Z})=I_2\oplus\omega I_3\,,\label{eq:klingen-matrices}
\end{gather}
where $\mathsf{s}_J=\si^{-1}$, $\mathsf{t}_J=\mathsf{T}_1$, and $\mathsf{X}=H(1,0,0)$, $\mathsf{Y}=H(0,1,0)$, $\mathsf{Z}=H(0,0,1)$. The doublet matrices are those of eq.~\eqref{eq:rho2}, and the triplet matrices are
\begin{gather}
 \rho_{\mathbf{3}^{(0)}}(\mathsf{s}_J)=-\frac{i}{\sqrt{3}}\begin{pmatrix}1&1&1\\1&\omega&\omega^2\\1&\omega^2&\omega\end{pmatrix}\,,\qquad
 \rho_{\mathbf{3}^{(0)}}(\mathsf{t}_J)=\diag(1,\omega,\omega)\,,\notag\\
 \rho_{\mathbf{3}^{(0)}}(\mathsf{X})=\begin{pmatrix}0&0&1\\1&0&0\\0&1&0\end{pmatrix}\,,\qquad
 \rho_{\mathbf{3}^{(0)}}(\mathsf{Y})=\diag(1,\omega,\omega^2)\,,\qquad
 \rho_{\mathbf{3}^{(0)}}(\mathsf{Z})=\omega I_3\,.\label{eq:rho3}
\end{gather}

The doublet is neutral under $\Delta(27)$, whereas the triplet is faithful on $J_3$, with $\mathsf{Z}$ acting as $\omega$.

\paragraph{Siegel cusp.}
In the basis $\left(Y_1;\,Y_2,Y_3,Y_4,Y_5\right)$, the generators eq.~\eqref{eq:Siegel-generator-words} act by
\begin{gather}
 \rho(\mathsf{T}_B)=\diag\bigl(1,\,\omega^{b_{11}},\,\omega^{b_{22}},\,\omega^{b_{11}+b_{12}+b_{22}},\,\omega^{b_{11}+2b_{12}+b_{22}}\bigr)\,,\notag\\
 \rho(\RR)=1\oplus\sigma_1\oplus\sigma_1\,,\qquad
 \rho(\DD)=-\bigl(1\oplus I_2\oplus\sigma_1\bigr)\,,\qquad
 \rho(\EE)=1\oplus\begin{pmatrix}0&0&0&1\\0&1&0&0\\1&0&0&0\\0&0&1&0\end{pmatrix}\,,\label{eq:Siegel-cusp-matrices}
\end{gather}
where $\sigma_1=\left(\begin{smallmatrix}0&1\\1&0\end{smallmatrix}\right)$ and the $\GL(2,\Z)$ matrices follow from eq.~\eqref{eq:rho-A}. The four non-trivial characters are permuted transitively, so $(Y_2,Y_3,Y_4,Y_5)$ is an irreducible quadruplet. The rank-one exponents include $T=vv^\top$ for $v=(1,0)$, $(0,1)$, $(1,-1)$ and $(1,1)$. Modulo three these vectors span the four lines in $\Z_3^2$, and hence their permutations give the factor $S_4\cong\PGL(2,\Z_3)$.

\paragraph{Two-ray region.}
In the basis $(Y_1;\,Y_2,Y_3;\,Y_4,Y_5)$,
\begin{gather}
 \rho(\mathsf{T}_B)=\diag\bigl(1,\,\omega^{b_{11}},\,\omega^{b_{22}},\,\omega^{b_{11}+b_{12}+b_{22}},\,\omega^{b_{11}+2b_{12}+b_{22}}\bigr)\,,\notag\\
 \rho(\PP)=-\bigl(1\oplus\sigma_1\oplus I_2\bigr)\,,\qquad
 \rho(\DD)=-\bigl(1\oplus I_2\oplus\sigma_1\bigr)\,.\label{eq:two-ray-matrices}
\end{gather}

\paragraph{Corner.}
In the basis $(Y_1;\,Y_4;\,Y_3,Y_2,Y_5)$,
\begin{gather}
 \rho(\mathsf{T}_B)=\diag\bigl(1,\,\omega^{t_1+t_2+t_3},\,\omega^{t_2+t_3},\,\omega^{t_1+t_3},\,\omega^{t_1+t_2}\bigr)\,,\notag\\
 \rho(\PP)=-\left(1\oplus1\oplus\begin{pmatrix}0&1&0\\1&0&0\\0&0&1\end{pmatrix}\right)\,,\qquad
 \rho(\CC)=1\oplus1\oplus\begin{pmatrix}0&0&1\\1&0&0\\0&1&0\end{pmatrix}\,,\label{eq:corner-matrices}
\end{gather}
with $t_i$ as in eq.~\eqref{eq:corner-action}. The triplet $(Y_3,Y_2,Y_5)$ is irreducible and faithful on $\Z_3^3\rtimes S_3$.

The table shows that the same Siegel modular form multiplet supplies different multiplets in different regions. The triplet $(Y_3,Y_4,Y_5)$ of the Klingen region, which will contain the Feruglio--Marrone (Jacobi form) triplets, is not the triplet $(Y_3,Y_2,Y_5)$ of the corner. A model built in one region therefore does not continue with the same multiplet structure into another region, although the underlying Siegel theory does.

\subsection{Where the interior subspaces meet the cusps}\label{sec:meet}
The interior subspaces of section~\ref{sec:elliptic} are not compact: letting their remaining moduli grow, each of them runs into one of the cusp regions. Table~\ref{tab:meet} lists the limits.

The curve $\Omega=uI_2$ reaches the two-ray region along its symmetric direction, $q_\tau=q_\sigma$ and $\zeta=1$, and its pointwise stabilizer, the eight signed permutation matrices, is the finite part $D_8$ of the two-ray modular symmetry. In the same way the curve $\Omega=\frac{u}{2}\left(\begin{smallmatrix}2&1\\1&2\end{smallmatrix}\right)$ becomes, after $z\to-z$, $\im\Omega=\frac{\im u}2(Q_1+Q_2+Q_3)$: it reaches the corner along its central ray, with
\begin{equation}\label{eq:corner-central}
 q_1=q_2=q_3=e^{\pi iu/3}\,,
\end{equation}
and its stabilizer $D_{12}$, the automorphism group of the hexagonal form, acts on the three matrices $Q_i$ through $D_{12}/\{\pm I_2\}\simeq S_3$. The dihedral curves are therefore the axes of symmetry of the two deepest cusp regions. On these axes the symmetry equates components of the modular form multiplets: the doublet $(Y_2,Y_3)\simeq3(q_\tau,q_\sigma)$ and the corner triplet $(Y_3,Y_2,Y_5)\simeq3(q_2q_3,q_1q_3,q_1q_2)$ have equal components. Hierarchies near the boundary thus arise from moving away from the dihedral curves, and they are exponentials of differences of the cone coordinates, for instance $|q_\sigma/q_\tau|=e^{-2\pi(\im\sigma-\im\tau)/3}$. This is the boundary counterpart of the powers of the distance from a fixed point that control hierarchies in the interior \cite{Novichkov:2021evw,Feruglio:2021dte,Ding:2024xhz,Carducci:2026shs}.

On the other hand, the surfaces reach the Klingen region. The split surface $z=0$ meets it along the zero section of the elliptic variable, where $\chi_{10}$ vanishes. The surface $\tau=\sigma$ is equivalent to $z=-\tau/2$ (section~\ref{sec:weight2}) and meets the Klingen region along the two-torsion section $w=-\tau/6$. Its modular images include the sections discussed in the fits of ref.~\cite{Feruglio:2026xxo}. The curves through $\tau=i$ and $\tau=\omega$ end at the points $z=0$ of the Klingen fibers over these values of $\tau$, and their $\Z_4$ and $\Z_6$ stabilizers survive there.

\begin{table}[t]
\centering\footnotesize\tablesetup[1.6]\setlength{\tabcolsep}{3pt}
\begin{tabular}{>{\raggedright\arraybackslash}m{34mm}>{\raggedright\arraybackslash}m{17mm}>{\raggedright\arraybackslash}m{38mm}>{\raggedright\arraybackslash}m{50mm}}
\toprule
Interior subspace & Limit & Cusp region reached & Surviving structure\\
\midrule
$\diag(u,v)$ & $\im v\to\infty$ & Klingen, $z=0$ & $\chi_{10}=0$: exactly massless state\\
$\left(\begin{smallmatrix}u&v\\v&u\end{smallmatrix}\right)\simeq\{z=-\tau/2\}$ & $\im\sigma\to\infty$ & Klingen, $w=-\tau/6$ & Feruglio--Marrone zero mode (section~\ref{sec:weight2})\\
$\diag(i,u)$, $\diag(\omega,u)$ & $\im u\to\infty$ & Klingen, $\tau=i,\omega$, $z=0$ & stabilizers $\Z_4$, $\Z_6$\\
$uI_2$ & $\im u\to\infty$ & two-ray, $q_\tau=q_\sigma$, $\zeta=1$ & $H=D_8$ = finite part of $\Z^3\rtimes D_8$\\
$\left(\begin{smallmatrix}u&1/2\\1/2&u\end{smallmatrix}\right)$ & $\im u\to\infty$ & two-ray, $q_\tau=q_\sigma$, $\zeta=e^{\pi i/3}$ & $H=D_8$, combined with a translation\\
$\frac{u}{2}\left(\begin{smallmatrix}2&1\\1&2\end{smallmatrix}\right)$ & $\im u\to\infty$ & corner, $q_1=q_2=q_3$ & $H=D_{12}$ = finite part of $\Z^3\rtimes D_{12}$\\\bottomrule
\end{tabular}
\caption{Limits of the interior invariant subspaces. The dihedral curves are the symmetry axes of the two-ray and corner regions, and their pointwise stabilizers become the finite parts of the cusp modular symmetries.}\label{tab:meet}
\end{table}

\subsection{Mass hierarchies near the boundary}\label{sec:hierarchies}
Near an interior fixed point the residual stabilizer acts on a local coordinate by a phase, and the entries of the mass matrices are suppressed by powers of the distance from the point, with exponents fixed by the residual charges \cite{Novichkov:2021evw,Feruglio:2021dte}. In genus two this behavior was analyzed near isolated fixed points and along interior invariant loci \cite{Ding:2024xhz,Carducci:2026shs}. In genus one the cusp $\tau\to i\infty$ is treated in the same way, with the residual translations acting on $q=e^{2\pi i\tau/N}$ \cite{Novichkov:2021evw,Petcov:2023vws,deMedeirosVarzielas:2023crv}. The boundary regions of $\HH_2$ provide the corresponding expansion in genus two. Since, with minimal matter metrics, canonical normalization only multiplies a weight-$k$ coupling by $(\det\im\Omega)^{k/2}$, the exponential suppression of the Fourier terms survives in the physical couplings. In a basis in which the translations are diagonal, each entry of a Yukawa matrix therefore behaves, up to powers of the imaginary parts, as
\begin{equation}\label{eq:boundary-scaling}
 \bigl|(\mathcal{Y}_f)_{ij}\bigr|\simeq|c_{ij}|\prod_a|q_a|^{n_a^{(ij)}}\,,
\end{equation}
where $q_a$ runs over the small parameters of the region: $q_\sigma$ at Klingen, $(q_\tau,q_\sigma)$ in the two-ray region and $(q_1,q_2,q_3)$ at the corner. The translation characters of the two fields fix each exponent $n_a^{(ij)}$ modulo three, and positivity of the Fourier exponents fixes its smallest allowed value (appendix~\ref{sec:corner-rule}). The coefficient $c_{ij}$ depends on the moduli that stay finite. It can vanish on an interior locus that reaches the boundary, and then a power of the distance from that locus multiplies eq.~\eqref{eq:boundary-scaling}, as in eq.~\eqref{eq:two-ray-ratio} below. Near the boundary the couplings thus have the structure of a \ac{FN} model \cite{Froggatt:1978nt} with one, two or three expansion parameters, in which the Fourier orders play the role of the charges and the moduli replace the flavons.

The regions differ in how these orders can be distributed among the families. At the Klingen cusp the only small parameter is $q_\sigma$, which is rotated by the central element $\mathsf{Z}$ of $J_3$. Since $\mathsf{Z}$ acts as a scalar on every irreducible representation and trivially on all one- and two-dimensional ones (appendix~\ref{app:J3}), the three families of each matter field carry the same character. Hence all entries of a three-family Yukawa matrix have Fourier orders $n_2$ with the same residue modulo three, and the factor $e^{-2\pi n_2\varrho/3}$ of eq.~\eqref{eq:FN-identity} suppresses all of them equally, apart from factors $e^{-2\pi\varrho}$ between orders that differ by three. This factor can distinguish different sectors or multiplets, as in the Hessian coupling of section~\ref{sec:global}, where the doublet and the triplet differ by one order. Family hierarchies at the Klingen cusp, however, must come from the dependence on $(\tau,w)$, as in the fits of Feruglio and Marrone near the two-torsion sections (section~\ref{sec:weight2}), unless the leading order has reduced rank and a family obtains its mass only three orders higher. In the two-ray region $q_\tau$ and $q_\sigma$ are rotated independently and exchanged by $\PP$. For the singlet and doublet of section~\ref{sec:two-ray-model} the diagonal entries carry the orders $(0,0)$, $(1,0)$ and $(0,1)$, which give the leading masses $1:|q_\tau|:|q_\sigma|$. At the corner the components of a triplet carry different characters under all three translations, and positivity can raise an exponent beyond two, as for the classes with a vanishing entry in table~\ref{tab:corner-rule}. Apart from such cases, at level $N=3$ the residues fix the smallest allowed power of each parameter to $0$, $1$ or $2$, so larger hierarchies require large imaginary parts or large differences between them, such as $|q_\sigma/q_\tau|=e^{-2\pi(\im\sigma-\im\tau)/3}$, or a vanishing leading coefficient. At level $N$ the small parameters become $e^{-2\pi y/N}$ for a large imaginary part $y$, and the translations fix the exponents modulo $N$.

\subsection{Generalized CP}\label{sec:CP}
Each modular subsymmetry can be extended by CP. In the Siegel theory CP acts on the moduli as $\Omega\to-\overline{\Omega}$ \cite{Ding:2021iqp}. This is the action of
\begin{equation}\label{eq:CP-matrix}
 \mathsf{K}_*=\diag(I_2,-I_2)\,,\qquad \mathsf{K}_*^\top J\mathsf{K}_*=-J\,,
\end{equation}
since a matrix with multiplier $-1$ acts as $\Omega\to(A\overline{\Omega}+B)(C\overline{\Omega}+D)^{-1}$. Together with $\Gamma_2$ it generates the general symplectic group $\mathrm{GSp}(4,\Z)=\{\gamma:\gamma^\top J\gamma=\pm J\}\cong\Gamma_2\rtimes\Z_2$.
Geometrically, CP reverses the orientation of the genus-two surface $\Sigma_2$ of section~\ref{sec:moduli}, replacing it by the conjugate surface $\overline{\Sigma}_2$ and flipping the sign of the intersection form, $J\to -J$. Acting on homology by $(b';a')=\mathsf{K}_*(b;a)=(b;-a)$ restores a symplectic basis on $\overline{\Sigma}_2$. The holomorphic differentials normalized by $\oint_{a'_i}\omega'_j=\delta_{ij}$ are then $\omega'_j=-\overline{\omega}_j$, whose $b'$-periods give $\Omega'_{ij}=\oint_{b_i}(-\overline{\omega}_j)=-\overline{\Omega}_{ij}$. Conjugation by $\mathsf{K}_*$ maps $\gamma=\left(\begin{smallmatrix}A&B\\C&D\end{smallmatrix}\right)$ to $\bar{\gamma}=\left(\begin{smallmatrix}A&-B\\-C&D\end{smallmatrix}\right)$, so that $\mathsf{T}_i\to\mathsf{T}_i^{-1}$, $\mathsf{S}\to\mathsf{S}^{-1}$ and the $\GL(2,\Z)$ transformations $\gamma_A$ are unchanged. A matter field transforms as $\Phi_I\to X_I\overline{\Phi}_I$, and compatibility with eq.~\eqref{eq:siegel-form-field} requires
\begin{equation}\label{eq:CP-consistency}
 X_{\mathbf{r}}\,\rho_{\mathbf{r}}(\gamma)^*X_{\mathbf{r}}^{-1}=\rho_{\mathbf{r}}(\bar{\gamma})\,.
\end{equation}
In genus two, a second inequivalent CP automorphism of $\Sp(4,\Z)$ exists~\cite{Ding:2021iqp}; we use the standard one above.\footnote{The second automorphism multiplies $\bar{\gamma}$ by the sign character of $\Sp(4,\Z)$, so it sends $\mathsf{T}_i$ to $-\mathsf{T}_i^{-1}$ \cite{Ding:2021iqp}. In particular $\mathsf{T}_1^3\in\Gamma_2(3)$ maps to $-\mathsf{T}_1^{-3}\equiv-I_4\bmod 3$. Thus it induces the same action on the projective quotient, but does not descend to the full quotient at level 3.} The theta functions have real Fourier coefficients, so that $Y^{(1)}_{\mathbf{5}}(-\overline{\Omega})=Y^{(1)}_{\mathbf{5}}(\Omega)^*$, in agreement with eq.~\eqref{eq:CP-consistency} for $X_{\mathbf{5}}=I$: the matrices eq.~\eqref{eq:rho-T} are diagonal phases, those in eq.~\eqref{eq:rho-A} are real, and $\rho_{\mathbf{5}}(\mathsf{S})$ is real with $\rho_{\mathbf{5}}(\mathsf{S})^2=\rho_{\mathbf{5}}(-I_4)=I$. Polynomials with real coefficients in the theta functions inherit this property. The representation matrices of all assignments in this paper are diagonal phases, real signed permutations or, at the Klingen cusp, the symmetric unitary matrices discussed below, so $X_I=I$ is consistent for every field. With CP imposed, the couplings in this basis are real. A vacuum preserves CP if $-\overline{\Omega}$ is related to $\Omega$ by a modular transformation of the model.

\paragraph{Restriction to the boundary.}
The element $\mathsf{K}_*$ leaves $\im\Omega$ unchanged and therefore maps every cusp region of section~\ref{sec:regions} into itself. The extended modular symmetry is $G\rtimes\Z_2$, where $G$ is the integral modular symmetry in table~\ref{tab:cusp-groups}, and $\mathsf{K}_*$ acts on the local coordinates by complex conjugation,
\begin{equation}\label{eq:CP-local}
 (q_\tau,q_\sigma,\zeta)\to(\bar{q}_\tau,\bar{q}_\sigma,\bar{\zeta})\,,\qquad q_i\to\bar{q}_i\,.
\end{equation}
In the Klingen region it acts as $(\tau,z)\to(-\bar{\tau},-\bar{z})$ and maps $G_{\gamma_1}$ to $G_{\bar{\gamma}_1}$ and $H(\lambda,\mu,\kappa)$ to $H(\lambda,-\mu,-\kappa)$. Hence $\mathsf{s}_J\to\mathsf{s}_J^{-1}$, $\mathsf{t}_J\to\mathsf{t}_J^{-1}$, $\mathsf{X}\to\mathsf{X}$, $\mathsf{Y}\to\mathsf{Y}^{-1}$ and $\mathsf{Z}\to\mathsf{Z}^{-1}$. Inserting the Fourier--Jacobi expansion eq.~\eqref{eq:FJ} into $Y(-\overline{\Omega})=X_{\mathbf{r}}Y(\Omega)^*$ and comparing powers of $\bar{q}_\sigma$ gives
\begin{equation}\label{eq:CP-FJ}
 \phi_{n_2}(-\bar{\tau},-\bar{z})=X_{\mathbf{r}}\,\phi_{n_2}(\tau,z)^*\,.
\end{equation}
Each Fourier--Jacobi coefficient therefore inherits the CP law of the parent Siegel modular form, and the generalized CP of the Jacobi theory is the restriction of the Siegel CP. For the multiplets of section~\ref{sec:FJ3}, $X=I$ is consistent with eq.~\eqref{eq:CP-consistency}, because $\rho(\mathsf{s}_J)$ in eqs.~\eqref{eq:rho2} and \eqref{eq:rho3} is symmetric and unitary, $\rho(\mathsf{X})$ is real and the remaining matrices are diagonal phases. The scalar multipliers used in the lepton model also respect this CP transformation, since $\eta(-\bar\tau)=\overline{\eta(\tau)}$.

\paragraph{Restriction to the interior.}
All loci of table~\ref{tab:interior} are CP invariant up to a modular transformation. The element $\mathsf{K}_*$ itself preserves $\diag(u,v)$, $\left(\begin{smallmatrix}u&v\\v&u\end{smallmatrix}\right)$, $\diag(i,u)$, $uI_2$ and the hexagonal curve, while $\mathsf{T}_1^{-1}\mathsf{K}_*$ preserves $\diag(\omega,u)$, since $-\bar{\omega}=\omega+1$, and $\mathsf{T}_3\mathsf{K}_*$ preserves the curve with $z=1/2$. On the remaining coordinates these elements act as the genus-one CP, $u\to-\bar{u}$, or $t\to-\bar{t}$ up to a translation on the Fricke curves of eq.~\eqref{eq:Fricke-actions}, and they extend $N(H)$ by a factor $\Z_2$. The consequences for the models of sections~\ref{sec:klingen} and \ref{sec:models} are noted there.
\section{Jacobi flavor symmetry from the Klingen region}\label{sec:klingen}
We now examine the Klingen region of section~\ref{sec:cusps} in detail. The Klingen modular symmetry acts on two surviving variables $(\tau,z)$, and in this section we show that it reproduces the Jacobi flavor symmetry of Feruglio and Marrone. The embedding of the Jacobi group in $\Sp(4,\Z)$ was already pointed out in ref.~\cite{Feruglio:2026xxo}; here we follow this limit through at level $N=3$. We first recall the Jacobi theory of Feruglio and Marrone, then derive its transformation laws and its four multiplets of weight 1 from the Siegel theory at level 3. Moreover, at weight 2 the Fourier coefficients relate Yukawa structures that are independent in the Jacobi theory.

\subsection{The Feruglio--Marrone construction in brief}\label{sec:FMbrief}
We follow ref.~\cite{Feruglio:2026xxo}; for the mathematics of the Jacobi group and of Jacobi forms see refs.~\cite{Eichler:1985,Berndt:1998}. The Jacobi group acts on $\tau$ and an elliptic variable $z$. It contains the modular transformations $\tau\to(a\tau+b)/(c\tau+d)$, $z\to z/(c\tau+d)$, and the translations $z\to z+\lambda\tau+\mu$. On fields the translations are lifted to the integral Heisenberg group, with elements $h(\lambda,\mu,\kappa)$ and product
\begin{equation}\label{eq:H-law}
 h(\lambda_1,\mu_1,\kappa_1)\,h(\lambda_2,\mu_2,\kappa_2)=h(\lambda_1+\lambda_2,\ \mu_1+\mu_2,\ \kappa_1+\kappa_2+2\lambda_1\mu_2)\,.
\end{equation}
The generators $\mathsf{X}=h(1,0,0)$, $\mathsf{Y}=h(0,1,0)$ and $\mathsf{Z}=h(0,0,1)$ satisfy $\mathsf{X}\mathsf{Y}=\mathsf{Z}^2\mathsf{Y}\mathsf{X}$, and $\mathsf{Z}$ commutes with everything and leaves $(\tau,z)$ unchanged. With $w=z/3$, the Jacobi automorphy factor of weight $k$ and index $m$ is
\begin{equation}\label{eq:JFM}
 J_{k,m}(g;\tau,w)=(c\tau+d)^k\exp\left\{2\pi i m\left[\frac{c\bigl(w+(\lambda\tau+\mu)/3\bigr)^2}{c\tau+d}-\frac{\lambda^2\tau+6\lambda w+\kappa}{9}\right]\right\}\,,
\end{equation}
and couplings and matter fields transform as
\begin{equation}\label{eq:form-field}
 Y^{(k,m)}_{\mathbf{r}}(gx)=J_{k,m}(g;x)\,\rho_{\mathbf{r}}(g)\,Y^{(k,m)}_{\mathbf{r}}(x)\,,\qquad
 \Phi_I\to J_{k_I,m_I}(g;x)^{-1}\rho_I(g)\,\Phi_I\,,\qquad x=(\tau,w)\,.
\end{equation}
Since $\mathsf{Z}$ does not move $x=(\tau,w)$, a nonzero irreducible coupling satisfies $\rho_{\mathbf{r}}(\mathsf{Z})=e^{2\pi im/9}I_{\dim\mathbf{r}}$. At level 3 $\rho_{\mathbf{r}}(\mathsf{Z})^3=I_{\dim\mathbf{r}}$, so $m$ is a multiple of three. An invariant superpotential term requires that both the weights and the indices add up,
\begin{equation}\label{eq:selection}
 k_Y=\sum_ak_{I_a}\,,\qquad m_Y=\sum_am_{I_a}\,.
\end{equation}
The finite modular group is $J_3$ of eq.~\eqref{eq:finite-Jacobi}, with the doublet and triplet matrices given in eqs.~\eqref{eq:rho2} and \eqref{eq:rho3}. In $Y^{(k,m)}_{\mathbf{r}}$ the superscripts give the weight and index. Table~\ref{tab:FM} lists the low-weight spaces used in constructing the couplings. Their tensor products are collected in appendix~\ref{app:reps}.

\begin{table}[t]
\centering\small\tablesetup[1.3]
\begin{tabular}{ccccl}
\toprule
$k$ & $m$ & dimension & $J_3$ content & construction\\\midrule
1&0&2&$\mathbf{2}^{(0)}$&$Y^{(1,0)}_{\mathbf{2}^{(0)}}$\\
1&3&3&$\mathbf{3}^{(0)}$&$Y^{(1,3)}_{\mathbf{3}^{(0)}}$\\
1&6&0&--&--\\
1&9&2&$\mathbf{2}^{(0)}$&$Y^{(1,9)}_{\mathbf{2}^{(0)}}$\\
1&12&3&$\mathbf{3}^{(0)}$&$Y^{(1,12)}_{\mathbf{3}^{(0)}}$\\\midrule
2&0&3&$\mathbf{3}_{\rm mod}$&$Y^{(1,0)}_{\mathbf{2}^{(0)}}\otimes_S Y^{(1,0)}_{\mathbf{2}^{(0)}}$\\
2&3&6&$\mathbf{6}^{(0)}$&$Y^{(1,0)}_{\mathbf{2}^{(0)}}\otimes Y^{(1,3)}_{\mathbf{3}^{(0)}}$\\
2&6&6&$\bar{\mathbf{6}}^{(0)}$&$Y^{(1,3)}_{\mathbf{3}^{(0)}}\otimes_S Y^{(1,3)}_{\mathbf{3}^{(0)}}$\\
2&9&4&$\mathbf{1}^{(1)}\oplus\mathbf{3}_{\rm mod}$&$Y^{(1,0)}_{\mathbf{2}^{(0)}}\otimes Y^{(1,9)}_{\mathbf{2}^{(0)}}$\\
2&12&12&$\mathbf{6}^{(0)}\oplus\mathbf{6}^{(0)}$&$(Y^{(1,0)}_{\mathbf{2}^{(0)}}\otimes Y^{(1,12)}_{\mathbf{3}^{(0)}})\oplus(Y^{(1,9)}_{\mathbf{2}^{(0)}}\otimes Y^{(1,3)}_{\mathbf{3}^{(0)}})$\\\bottomrule
\end{tabular}
\caption{Low-weight Jacobi forms at level 3 with trivial scalar multiplier on the level-3 kernel, as in ref.~\cite{Feruglio:2026xxo}. The doublet $\mathbf{2}^{(0)}$ is neutral under $\Delta(27)$; the triplet $\mathbf{3}^{(0)}$ has $\rho(\mathsf{Z})=\omega$.}\label{tab:FM}
\end{table}

\subsection{The Jacobi group inside the Siegel group}\label{sec:embedding}
The symplectic matrices $G_{\gamma_1}$ and $H(\lambda,\mu,\kappa)$ of eq.~\eqref{eq:embedded} realize the Klingen transformations eq.~\eqref{eq:klingen-action}, with $H(\lambda,\mu,\kappa)$ obeying eq.~\eqref{eq:H-law}. As noted in section~\ref{sec:stabilisers}, both matrices fix $e_2$, the class of the pinched cycle $a_2$ (section~\ref{sec:moduli}). Conversely, an integral symplectic matrix fixing $e_2$ acts by $\SL(2,\Z)$ on the rank-two lattice $e_2^\perp/\Z e_2$, spanned by the classes of the cycles $a_1$ and $b_1$ of the surviving elliptic curve $E_\tau$. Removing this action leaves a matrix of the displayed Heisenberg form. Allowing $e_2\mapsto-e_2$ adds the central $-I_4$ and gives the full stabilizer of the cusp line.

Combining eq.~\eqref{eq:klingen-action}, the element $\gamma_g=G_{\gamma_1}H(\lambda,\mu,\kappa)$ acts as
\begin{equation}\label{eq:Delta}
 (\tau,z)\to\Bigl(\frac{a\tau+b}{c\tau+d},\ \frac{z+\lambda\tau+\mu}{c\tau+d}\Bigr)\,,\qquad
 \sigma\to\sigma+\Delta_g\,,\qquad
 \Delta_g=\lambda^2\tau+2\lambda z+\kappa-\frac{c\,(z+\lambda\tau+\mu)^2}{c\tau+d}\,,
\end{equation}
with $j=c\tau+d$. The pair $(\tau,z)$ has the Jacobi action, and the shift $\Delta_g$ of the third modulus does not depend on $\sigma$. The central element $\mathsf{Z}=H(0,0,1)$, which fixes $(\tau,z)$, is the ordinary translation $\sigma\to\sigma+1$. Reducing modulo three gives exactly $J_3$, so the level-3 structure of the Jacobi theory is inherited from the Siegel theory. The inclusion of the integral vector stabilizer induces the embedding $J_3\hookrightarrow\Sp(4,\Z_3)$.

The Klingen limit also determines the fundamental domain of the Jacobi group directly from $\mathcal{F}_2$ in section~\ref{sec:moduli}. Keeping $\tau$ and $z$ finite as $\im\sigma\to\infty$, the condition $\im\tau\le\im\sigma$ and every determinant inequality $|\det(C\Omega+D)|\ge 1$ involving $\sigma$ are satisfied automatically. The surviving conditions are $|\re\tau|\le\frac{1}{2}$ and $|\tau|\ge 1$ for $\tau$ (from $G_{\gamma_1}$), together with $|\re z|\le\frac{1}{2}$ and $0\le\im z\le\frac{1}{2}\im\tau$ (from the translations and the embedded sign $G_{-I_2}=\si^2: z\mapsto -z$). In the rescaled coordinate $w=z/3$, this reproduces the fundamental domain of $\Gamma^{\mathrm{J}}$ used in ref.~\cite{Feruglio:2026xxo} (up to the equivalent horizontal shift between the rectangular slice $|\re w|\le\frac{1}{6}$ and the lattice-aligned parallelogram, and the twofold folding by $w\mapsto -w$).

\subsection{Derivation of the Jacobi automorphy factor}\label{sec:derivation}
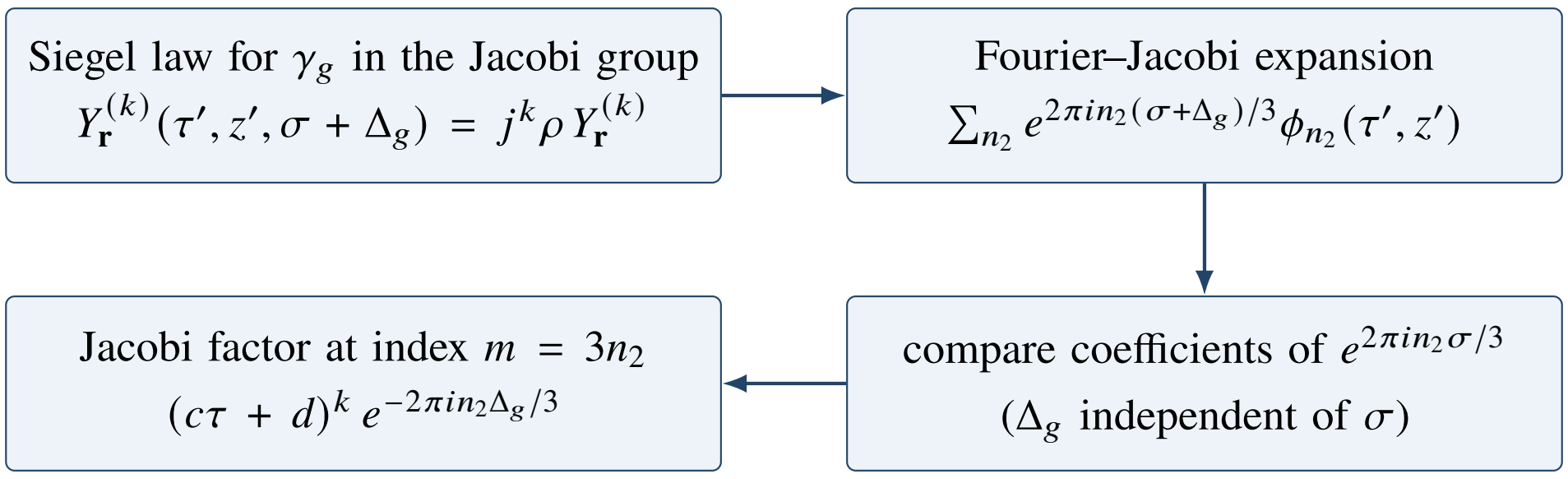
\begin{figure}[t]
\centering
\begin{tikzpicture}[font=\small]
\node[box,text width=54mm,minimum height=14mm] (s) {Siegel law for $\gamma_g$ in the Jacobi group\\[2pt] $Y^{(k)}_{\mathbf{r}}(\tau',z',\sigma+\Delta_g)=j^k\rho\,Y^{(k)}_{\mathbf{r}}$};
\node[box,text width=54mm,minimum height=14mm,right=10mm of s] (f) {Fourier--Jacobi expansion\\[2pt] $\sum_{n_2} e^{2\pi in_2(\sigma+\Delta_g)/3}\phi_{n_2}(\tau',z')$};
\node[box,text width=54mm,minimum height=14mm,below=9mm of f] (c) {compare coefficients of $e^{2\pi in_2\sigma/3}$\\[2pt] ($\Delta_g$ independent of $\sigma$)};
\node[box,text width=54mm,minimum height=14mm,below=9mm of s] (j) {Jacobi factor at index $m=3n_2$\\[2pt] $(c\tau+d)^k\,e^{-2\pi in_2\Delta_g/3}$};
\draw[arr] (s.east)--(f.west);
\draw[arr] (f.south)--(c.north);
\draw[arr] (c.west)--(j.east);
\end{tikzpicture}
\caption{How the Jacobi automorphy factor $J_{k,m}$ arises from the Siegel automorphy factor $j^k$. The shift $\sigma\mapsto\sigma+\Delta_g$ of the third modulus is absorbed by the coefficient $\phi_{n_2}$ of order $n_2$, producing the exponential factor of index $m=3n_2$ in $w=z/3$ together with the restricted representation $\rho_Y(\gamma_g)$. Explicitly, $\Delta_g=\lambda^2\tau+2\lambda z+\kappa$ for translations, $\Delta_g=-cz^2/(c\tau+d)$ for modular transformations, and $\Delta_g=1$ for the center $\mathsf{Z}$ (so $\rho(\mathsf{Z})\phi_{n_2}=\omega^{n_2}\phi_{n_2}$).}\label{fig:derivation}
\end{figure}

Since $\sigma\to\sigma+3$ is in $\Gamma_2(3)$, a Siegel modular form of level 3 can be expanded in powers of $q_\sigma=e^{2\pi i\sigma/3}$,
\begin{equation}\label{eq:FJ}
 Y^{(k)}_{\mathbf{r}}(\Omega)=\sum_{n_2\ge0}q_\sigma^{n_2}\,\phi_{n_2}(\tau,z)\,,
\end{equation}
which is called the Fourier--Jacobi expansion. Only $n_2\ge0$ occurs, since the Fourier exponents are positive semi-definite. For $\gamma_g=G_{\gamma_1}H(\lambda,\mu,\kappa)$, insert this expansion on both sides of the Siegel transformation law eq.~\eqref{eq:siegel-form-field}. One obtains
\begin{equation}\label{eq:FJ-step}
 \sum_{n_2\geq0} q_\sigma^{n_2}e^{2\pi i n_2\Delta_g/3}\,
       \phi_{n_2}\bigl(g(\tau,z)\bigr)
 =(c\tau+d)^k\rho_Y(\gamma_g)
   \sum_{n_2\geq0}q_\sigma^{n_2}\phi_{n_2}(\tau,z)\,.
\end{equation}
The shift $\Delta_g$ is independent of $\sigma$. Equating the coefficients of $q_\sigma^{n_2}$ gives
\begin{equation}\label{eq:FJ-law}
 \phi_{n_2}\bigl(g(\tau,z)\bigr)=(c\tau+d)^k\,e^{-2\pi i n_2\Delta_g/3}\,\rho_Y(\gamma_g)\,\phi_{n_2}(\tau,z)\,.
\end{equation}
For the translations $\Delta_g=\lambda^2\tau+2\lambda z+\kappa$, and for the modular transformations $\Delta_g=-cz^2/(c\tau+d)$. Writing $z=3w$, the factor $e^{-2\pi in_2\Delta_g/3}$ becomes exactly the exponential in eq.~\eqref{eq:JFM}, with
\begin{equation}\label{eq:index-dictionary} m=3n_2\,.
\end{equation}
For $\mathsf{Z}$, $\Delta_g=1$ and eq.~\eqref{eq:FJ-law} gives $\rho_Y(\mathsf{Z})\phi_{n_2}=\omega^{n_2}\phi_{n_2}$, which reproduces the central character $e^{2\pi im/9}$ of the Jacobi multiplet. These steps are summarized in figure~\ref{fig:derivation}. They give the vector-valued, level-3 form of the classical statement that the Fourier--Jacobi coefficients of a Siegel modular form are Jacobi forms \cite{Eichler:1985,Ziegler:1989}. For flavor model building, the same computation fixes in addition the finite representation and the index of each coefficient.

\paragraph{What the Jacobi index counts.}
The translations $w\to w+\tau$ and $w\to w+1$ ($\lambda,\mu\in3\mathds{Z}$) lie in the level-3 kernel, so the finite matrices drop out and each component of an index-$m$ multiplet obeys
\begin{equation}\label{eq:quasi-period}
 \phi(\tau,w+1)=\phi(\tau,w)\,,\qquad \phi(\tau,w+\tau)=e^{-2\pi i m(\tau+2w)}\phi(\tau,w)\,.
\end{equation}
It is therefore a holomorphic section of a line bundle of degree $2m$ on the curve $\mathds{C}/(\mathds{Z}\tau+\mathds{Z})$, and by the argument principle it has exactly $2m$ zeros there unless it vanishes identically. For example, the odd combination of the triplet of index $m=3$ is
\begin{equation}\label{eq:F-odd}
 Y^{(1,3)}_{\mathbf{3}^{(0)},3}-Y^{(1,3)}_{\mathbf{3}^{(0)},2}=-\frac{\vartheta(\tau,w)^2\,\vartheta(\tau,2w)}{\eta(\tau)}\,,
\end{equation}
where $\vartheta$ is the odd Jacobi theta function of appendix~\ref{app:forms}; this is minus the theta block $Q_{1,1}$ from which Feruglio and Marrone build the triplet. It has a triple zero at $w=0$ and simple zeros at the three other half-periods of $\mathds{C}/(\mathds{Z}\tau+\mathds{Z})$, in total $3+3=6=2m$. The same quasi-periodicity characterizes the lowest-Landau-level wavefunctions on a torus with $M=2m$ flux quanta \cite{Cremades:2004wa}, and the Gaussian $e^{-2\pi m(\im w)^2/\im\tau}$ that appears in the canonical normalization below, eq.~\eqref{eq:FN-identity}, is the corresponding Landau-level Gaussian. In this sense the Jacobi index is half a magnetic flux, while in the Siegel picture it is three times a Fourier order.

\subsection{Jacobi forms from Siegel modular forms}\label{sec:FJ3}
We now expand the five weight-1 Siegel modular forms in the Klingen region. The quintet splits into the doublet and triplet of table~\ref{tab:boundary-branching},
\begin{equation}\label{eq:NC}
 Y^{(1)}_{\mathbf{2}^{(0)}}=\begin{pmatrix}-\sqrt{2}\,\Theta_{10}\\\Theta_{00}\end{pmatrix}\,,\qquad
 Y^{(1)}_{\mathbf{3}^{(0)}}=\begin{pmatrix}\Theta_{01}\\\Theta_{21}\\\Theta_{11}\end{pmatrix}\,.
\end{equation}
In eq.~\eqref{eq:Theta} a vector $\ell'$ contributes at order $q_\sigma^{n_2}$ with $n_2=3\ell'^2/2$, so the coefficients are obtained by collecting the vectors $\ell'$ of equal length. For fixed $\ell'$ the sum over $\ell$ is a Jacobi theta function in $z$. The orders $n_2=0,1,3,4$ receive $1$, $3$, $6$ and $3$ vectors, namely $\ell'=0$, the shortest vectors of $L+\mu_0$, the roots of $L$ and $-2$ times the shortest vectors of $L+\mu_0$, while no vector gives $n_2=2$. Each of these shells forms a single orbit under the symmetries preserving its lattice class, and the result is
\begin{equation}\label{eq:NC-expansion}
 Y^{(1)}_{\mathbf{2}^{(0)}}=Y^{(1,0)}_{\mathbf{2}^{(0)}}+6\,q_\sigma^3Y^{(1,9)}_{\mathbf{2}^{(0)}}+\mathcal{O}(q_\sigma^9)\,,\qquad
 Y^{(1)}_{\mathbf{3}^{(0)}}=-3\,q_\sigma Y^{(1,3)}_{\mathbf{3}^{(0)}}+3\,q_\sigma^4Y^{(1,12)}_{\mathbf{3}^{(0)}}+\mathcal{O}(q_\sigma^7)\,,
\end{equation}
where $Y^{(1,0)}_{\mathbf{2}^{(0)}},Y^{(1,9)}_{\mathbf{2}^{(0)}},Y^{(1,3)}_{\mathbf{3}^{(0)}},Y^{(1,12)}_{\mathbf{3}^{(0)}}$ are the Feruglio--Marrone Jacobi form multiplets, including normalizations and signs. Explicitly, with
\begin{equation}
 f_j(\tau,z)=\sum_{\ell\in L+j\mu_0}e^{\pi i\tau\ell^2+2\pi i z\,\ell\cdot\mu_0}\,,
\end{equation}
one finds
\begin{equation}\label{eq:FG-f}
 Y^{(1,3)}_{\mathbf{3}^{(0)}}=-\begin{pmatrix}f_0(\tau,3w)\\f_2(\tau,3w)\\f_1(\tau,3w)\end{pmatrix}\,,\qquad
 Y^{(1,12)}_{\mathbf{3}^{(0)}}=\begin{pmatrix}f_0(\tau,6w)\\f_1(\tau,6w)\\f_2(\tau,6w)\end{pmatrix}\,,
\end{equation}
while $Y^{(1,0)}_{\mathbf{2}^{(0)}}$ is the doublet eq.~\eqref{eq:D-doublet} and $Y^{(1,9)}_{\mathbf{2}^{(0)}}$ comes from the six vectors of length squared two; explicit formulae are in appendix~\ref{app:forms}.

All four building blocks are therefore Fourier coefficients of ordinary Siegel modular forms of level 3. Moreover, the empty index-6 space agrees with the independent Feruglio--Marrone construction in table~\ref{tab:FM}. The Siegel multiplets contain them in pairs: $Y^{(1,0)}_{\mathbf{2}^{(0)}}$ and $Y^{(1,9)}_{\mathbf{2}^{(0)}}$ sit in the same doublet $Y^{(1)}_{\mathbf{2}^{(0)}}$, and $Y^{(1,3)}_{\mathbf{3}^{(0)}}$ and $Y^{(1,12)}_{\mathbf{3}^{(0)}}$ in the same triplet $Y^{(1)}_{\mathbf{3}^{(0)}}$.

\subsection{The Feruglio--Marrone flavor model and its Siegel realization}\label{sec:FN}
The charged-fermion model of Feruglio and Marrone assigns weight $k_I=1$ and index $m_I=6$ to each matter triplet. Its neutrino extension uses the following assignments \cite{Feruglio:2026xxo}:
\begin{align}\label{eq:matter}
 q,l&\sim(\mathbf{3}^{(2)},1,6)_{\eps}\,,\notag\\
 u^c,d^c,e^c,N^c&\sim(\mathbf{3}^{(1)},1,6)_{\eps^{-1}}\,,\qquad
 H_u,H_d\sim(\mathbf{1}^{(0)},0,0)\,.
\end{align}
Here the entries are $(\mathbf{r},k_I,m_I)$, and the subscript is a scalar multiplier in the transformation law.\footnote{$\eps$ is the Dedekind eta multiplier. Its odd powers are defined on the metaplectic double cover, which records the sign in $\sqrt{c\tau+d}$. Its square descends to a character of order twelve of $\SL(2,\Z)$, as follows from the phases of $\eta^2$ under $\mathsf{S}$ and $\mathsf{T}$. Appendix~\ref{app:cusp-groups} gives its presentation, and table~\ref{tab:matter-images} records the finite matrix image including the multiplier.} The multipliers cancel between the two fields in each Dirac term. The bilinear $N^cN^c$ carries the multiplier $\eps^{-2}$, so the Majorana term requires a Jacobi form sextet $Y^{(2,12)}_{\mathbf{6}^{(1)}}[\eps^2]$ of weight $k=2$ and index $m=12$ with multiplier $\eps^2$. Feruglio and Marrone introduced this assignment because using the same two holomorphic textures in all three lepton matrices gave too little mixing in their numerical fits. On the section $w=(1-\tau)/6$ (which is modular-equivalent to the two-torsion section $w=-\tau/6$ analyzed in section~\ref{sec:weight2}), the charged-lepton and Dirac matrices annihilate the odd-parity direction. The Majorana matrix, which carries the multiplier $\eps^2$, couples that direction to the even subspace and generically has rank two there. The seesaw must therefore be evaluated away from the exact section \cite{Feruglio:2026xxo}.
In the right--left convention, the superpotential is
\begin{align}\label{eq:FM-W}
 \mathscr{W}={}&H_u(u^c)^\top\mathcal{Y}_uq+H_d(d^c)^\top\mathcal{Y}_dq
 +H_d(e^c)^\top\mathcal{Y}_el\notag\\
 &+H_u(N^c)^\top\mathcal{Y}_\nu l+\frac{\Lambda_R}{2}(N^c)^\top\mathcal{Y}_NN^c\,.
\end{align}
In the ordered doublet--triplet basis $R=X\otimes W=(X_1W_1,X_1W_2,X_1W_3,X_2W_1,X_2W_2,X_2W_3)^\top$, the same symmetric contraction applies to every sextet,
\begin{equation}\label{eq:MXY}
 \mathcal{Y}[R]=\begin{pmatrix}
 R_4&-R_3/\sqrt{2}&-R_2/\sqrt{2}\\
 -R_3/\sqrt{2}&R_5&-R_1/\sqrt{2}\\
 -R_2/\sqrt{2}&-R_1/\sqrt{2}&R_6
 \end{pmatrix}\,.
\end{equation}
We write $\mathcal{Y}[X,W]=\mathcal{Y}[X\otimes W]$. Each Dirac sector uses the same two sextets, with its own overall strength $y_f$ and relative coefficient $r_f$. In the ordering of eq.~\eqref{eq:MXY} their linear combination is
\begin{align}
 R_f=Y^{(1,0)}_{\mathbf{2}^{(0)}}\otimes Y^{(1,12)}_{\mathbf{3}^{(0)}}+r_f\,Y^{(1,9)}_{\mathbf{2}^{(0)}}\otimes Y^{(1,3)}_{\mathbf{3}^{(0)}}\,.
\end{align}
The Yukawa matrices entering the superpotential are
\begin{align}\label{eq:FM-Yukawa}
 \mathcal{Y}_f&=y_f\,\mathcal{Y}[R_f]\,,\qquad f=u,d,e,\nu\,,\notag\\
 \mathcal{Y}_N&=y_N\,\mathcal{Y}\!\left[Y^{(2,12)}_{\mathbf{6}^{(1)}}[\eps^2]\right]\,.
\end{align}
For example the $(1,1)$ entry of $\mathcal{Y}_f/y_f$ is $Y^{(1,0)}_{\mathbf{2}^{(0)},2}Y^{(1,12)}_{\mathbf{3}^{(0)},1}+r_f Y^{(1,9)}_{\mathbf{2}^{(0)},2}Y^{(1,3)}_{\mathbf{3}^{(0)},1}$. The three diagonal entries all use the second component of each doublet, while the off-diagonal entries use the first. The components of the sextet $Y^{(2,12)}_{\mathbf{6}^{(1)}}[\eps^2]$ in the Majorana term are given in ref.~\cite{Feruglio:2026xxo}, appendix~C. The untwisted assignment $(\mathbf{3}^{(0)},1,6)$ gives the same Dirac matrices.

For the minimal kinetic terms
\begin{align}\label{eq:KFM}
 \mathscr{K}=-h\log\im\tau+4\pi\nu\frac{(\im w)^2}{\im\tau}
 +\sum_I (\im\tau)^{-k_I}e^{4\pi m_I(\im w)^2/\im\tau}\Phi_I^\dagger\Phi_I\,,
\end{align}
all matter-field metrics in eq.~\eqref{eq:matter} are equal, and the physical mass matrices follow directly from eq.~\eqref{eq:FM-Yukawa}:
\begin{align}\label{eq:FM-mass-matrices}
 M_f&=v_f y_f\,\im\tau\,e^{-24\pi(\im w)^2/\im\tau}
 \begin{pmatrix}
 R_{f,4}&-R_{f,3}/\sqrt{2}&-R_{f,2}/\sqrt{2}\\
 -R_{f,3}/\sqrt{2}&R_{f,5}&-R_{f,1}/\sqrt{2}\\
 -R_{f,2}/\sqrt{2}&-R_{f,1}/\sqrt{2}&R_{f,6}
 \end{pmatrix}\,,\notag\\
 M_N&=\Lambda_R\im\tau\,e^{-24\pi(\im w)^2/\im\tau}\mathcal{Y}_N\,,\notag\\
 m_\nu&=-\frac{v_u^2\im\tau\,e^{-24\pi(\im w)^2/\im\tau}}{\Lambda_R}
 \mathcal{Y}_\nu^\top\mathcal{Y}_N^{-1}\mathcal{Y}_\nu\,.
\end{align}
We use the normalization of rigid \ac{SUSY} throughout; a \ac{SUGRA} treatment also includes the common $e^{\mathscr{K}/2}$ factor and the corresponding superpotential weight. In their phenomenological fits, Feruglio and Marrone \cite{Feruglio:2026xxo} supplement eq.~\eqref{eq:KFM} by non-universal kinetic terms, built from the same Jacobi multiplets. In the reference charged-fermion fit only the right-handed quark metrics are modified, with condition numbers (ratios of the largest to the smallest eigenvalue of the metric) $\simeq7.1\times10^2$ for $u^c$ and $\simeq4.2$ for $d^c$.  In the fit with real couplings the left-handed quark metric acquires a condition number $\simeq3.6\times10^5$, and in the corresponding fit including neutrinos $\simeq2.3\times10^5$ \cite{Feruglio:2026xxo}. The Jacobi flavor symmetry does not forbid such terms, and we return in section~\ref{sec:topdown} to why the K\"ahler potential is the least protected part of a theory near this boundary.

The three-modulus realization follows directly from the transformation law eq.~\eqref{eq:FJ-law} of the Fourier--Jacobi coefficients. A Jacobi multiplet and matter field can be written as
\begin{equation}\label{eq:local-lift}
 \widehat Y^{(k,m)}_{\mathbf{r}}(\Omega)
 =e^{2\pi im\sigma/9}Y^{(k,m)}_{\mathbf{r}}(\tau,z/3)\,,\qquad
 \Phi_S=e^{-2\pi im_I\sigma/9}\Phi_J\,.
\end{equation}
The first transforms with $j^k\rho_{\mathbf{r}}$ and the second with $j^{-k_I}\rho_{\mathbf{r}_I}$ under the embedded Jacobi modular symmetry. Weight and index conservation then give the same superpotential in the two sets of variables:
\begin{equation}
 \widehat Y\prod_I\Phi_{S,I}=Y\prod_I\Phi_{J,I}\,.
\end{equation}
The scalar multipliers in eq.~\eqref{eq:matter} are retained by this change of field variables.

Following ref.~\cite{Feruglio:2026xxo}, we use the invariant height
\begin{equation}\label{eq:height}
 \varrho=\operatorname{Im}\sigma-\frac{(\operatorname{Im}z)^2}{\operatorname{Im}\tau}
 =\frac{\det\im\Omega}{\im\tau}\,,\qquad \det\im\Omega=\varrho\,\im\tau\,.
\end{equation}
For a weight-$k$ Siegel coupling, canonical normalization gives
\begin{align}\label{eq:FN-identity}
 (\det\im\Omega)^{k/2}Y^{(k)}_{\mathbf{r}}
 &=\varrho^{k/2}\sum_{n_2\ge0}e^{-2\pi n_2\varrho/3} e^{2\pi in_2\operatorname{Re}\sigma/3}
 \phi_{n_2}^{\rm can}(\tau,w)\,,\notag\\
 \phi_{n_2}^{\rm can}&=(\im\tau)^{k/2}e^{-2\pi m(\im w)^2/\im\tau}\phi_{n_2}(\tau,3w)\,,\qquad m=3n_2\,.
\end{align}
This is the Klingen case of eq.~\eqref{eq:boundary-scaling}: the order $n_2$ acts as an \ac{FN} charge with expansion parameter $e^{-2\pi\varrho/3}$. The field redefinition in eq.~\eqref{eq:local-lift} reproduces the Gaussian in eq.~\eqref{eq:KFM}, up to an overall factor at fixed height. The moduli part of eq.~\eqref{eq:KFM} follows in the same way. In the Klingen coordinates the moduli K\"ahler potential is $\mathscr{K}_S=-h\log y-h\log(\im\sigma-9v^2/y)$, with $y=\im\tau$ and $v=\im w$, and the shift $\re\sigma\to\re\sigma+c$ is an isometry. Fixing the moment map in the convention $-\partial_{\im\sigma}\mathscr{K}_S=h/\varrho_0$, and extremizing $\mathscr{K}_S+(h/\varrho_0)\im\sigma$ gives $\im\sigma-9v^2/y=\varrho_0$ and
\begin{equation}\label{eq:K-reduced}
 \mathscr{K}_{\rm red}=-h\log y+\frac{9h}{\varrho_0}\frac{v^2}{y}+\text{const}\,,
\end{equation}
which is eq.~\eqref{eq:KFM} with $4\pi\nu=9h/\varrho_0$. This is the K\"ahler quotient by the shift symmetry. Setting the height $\varrho$ to a constant directly in $\mathscr{K}_S$ gives a different result, and neither procedure is in general equivalent to integrating out a heavy multiplet; in \ac{SUGRA} the K\"ahler transformation must also be matched by the superpotential \cite{Ding:2020zxw,Yang:2007jacobi}. Thus the transformation laws of the matter fields, the Yukawa couplings and the kinetic terms of the Jacobi flavor theory all have a direct interpretation in a Klingen neighborhood.

The simplest Siegel coupling at level 3 with the Feruglio--Marrone structure is the doublet--triplet contraction of the two parts of the quintet. It is also the triplet block of the Hessian eq.~\eqref{eq:Hessian-matrix}, which we follow through all regions in section~\ref{sec:global}:
\begin{equation}\label{eq:Hessian-block}
 \frac{1}{48}\mathcal{H}_{\{3,4,5\}}=\mathcal{Y}[Y^{(1)}_{\mathbf{2}^{(0)}},Y^{(1)}_{\mathbf{3}^{(0)}}]
 =\begin{pmatrix}Y_1Y_3&Y_2Y_5&Y_2Y_4\\Y_2Y_5&Y_1Y_4&Y_2Y_3\\Y_2Y_4&Y_2Y_3&Y_1Y_5\end{pmatrix}\,.
\end{equation}
Its first two charged Fourier orders are
\begin{align}\label{eq:global-Y-exp}
 \mathcal{Y}[Y^{(1)}_{\mathbf{2}^{(0)}},Y^{(1)}_{\mathbf{3}^{(0)}}]&=-3q_\sigma\mathcal{Y}[Y^{(1,0)}_{\mathbf{2}^{(0)}},Y^{(1,3)}_{\mathbf{3}^{(0)}}]\notag\\
 &\quad+3q_\sigma^4\left(\mathcal{Y}[Y^{(1,0)}_{\mathbf{2}^{(0)}},Y^{(1,12)}_{\mathbf{3}^{(0)}}]-6\mathcal{Y}[Y^{(1,9)}_{\mathbf{2}^{(0)}},Y^{(1,3)}_{\mathbf{3}^{(0)}}]\right)+\mathcal{O}(q_\sigma^7)\,.
\end{align}
In this coupling the index-12 structure, which Feruglio and Marrone use for the Dirac Yukawa couplings, appears only at order $q_\sigma^4$. It is therefore suppressed by $e^{-2\pi\varrho}$ relative to the index-3 term at order $q_\sigma$. In the Jacobi theory the index-3 term is forbidden by the index condition eq.~\eqref{eq:selection}, since the Dirac matter fields have total index $m=12$. In the Siegel theory, however, the finite modular group cannot separate the two terms: $\mathsf{Z}$ acts as $\omega^{n_2}$ on the coefficient of $q_\sigma^{n_2}$, so the orders $n_2=1$ and $n_2=4$ carry the same central character. Hence an index-12 Dirac texture can be the leading term of a level-3 Siegel coupling only if the index-3 coefficient vanishes. This is simply the Fourier--Jacobi expansion eq.~\eqref{eq:FJ} at work: a single Siegel coupling contains Jacobi forms of all indices $m=3n_2$, and after canonical normalization the order $n_2$ comes with the factor $e^{-2\pi n_2\varrho/3}$ of eq.~\eqref{eq:FN-identity}. The Jacobi flavor theory of Feruglio and Marrone keeps a single index for each coupling and contains no height $\varrho$. Each of its couplings can therefore be viewed as a truncation of a Siegel coupling in the Klingen region to one Fourier--Jacobi order, in line with the interpretation of their theory in ref.~\cite{Feruglio:2026xxo} as an effective description of this asymptotic region. The truncation reproduces the leading term of a Siegel coupling when its lower orders vanish, whereas the relative suppression of different orders, and the relations between them at weight 2 discussed next, are visible only in the Siegel theory.

\subsection{Compatibility with full Siegel covariance}\label{sec:weight2}
While the weight-1 Jacobi multiplets all emerge freely from the Siegel quintet, embedding the theory in the full ring of level-3 Siegel modular forms imposes nontrivial relations at weight 2. Because the Fourier--Jacobi coefficient of a product is a sum over pairs of orders,
\begin{equation}\label{eq:convolution}
 [q_\sigma^{n}]\,(Y^{(k_1)}_{\mathbf{r}_1}Y^{(k_2)}_{\mathbf{r}_2})=\sum_{s+t=n}[q_\sigma^{s}]Y^{(k_1)}_{\mathbf{r}_1}\;[q_\sigma^{t}]Y^{(k_2)}_{\mathbf{r}_2}\,,
\end{equation}
where $[q_\sigma^n]F$ denotes the coefficient $\phi_n$ of $q_\sigma^n$ in the Fourier--Jacobi expansion eq.~\eqref{eq:FJ} of $F$. For the products of $Y^{(1)}_{\mathbf{2}^{(0)}}$ and $Y^{(1)}_{\mathbf{3}^{(0)}}$ this gives, at the orders relevant to Feruglio and Marrone,
\begin{align}\label{eq:fixed-sextet}
 [q_\sigma^3]\bigl(Y^{(1)}_{\mathbf{2}^{(0)},i}Y^{(1)}_{\mathbf{2}^{(0)},j}\bigr)
 &=6\bigl(Y^{(1,0)}_{\mathbf{2}^{(0)},i}Y^{(1,9)}_{\mathbf{2}^{(0)},j}+Y^{(1,9)}_{\mathbf{2}^{(0)},i}Y^{(1,0)}_{\mathbf{2}^{(0)},j}\bigr)\,,\notag\\
 [q_\sigma^4]\bigl(Y^{(1)}_{\mathbf{2}^{(0)},i}Y^{(1)}_{\mathbf{3}^{(0)},j}\bigr)
 &=3\bigl(Y^{(1,0)}_{\mathbf{2}^{(0)},i}Y^{(1,12)}_{\mathbf{3}^{(0)},j}-6Y^{(1,9)}_{\mathbf{2}^{(0)},i}Y^{(1,3)}_{\mathbf{3}^{(0)},j}\bigr)\,.
\end{align}
Two pairs of orders contribute to the same coefficient, $0+4$ and $3+1$, with a relative coefficient fixed by the Siegel theory. Since the space of weight 2 consists of quadratic theta polynomials, these convolutions constrain its coefficients. The antisymmetric singlet
\begin{equation}\label{eq:missing-singlet}
 Y^{(2,9)}_{\mathbf{1}^{(1)}}=Y^{(1,0)}_{\mathbf{2}^{(0)},1}Y^{(1,9)}_{\mathbf{2}^{(0)},2}-Y^{(1,0)}_{\mathbf{2}^{(0)},2}Y^{(1,9)}_{\mathbf{2}^{(0)},1}\,,
\end{equation}
of weight $k=2$ and index $m=9$, is absent from the Fourier--Jacobi image of weight-2 forms on $\Gamma_2(3)$ with trivial multiplier. For the Feruglio--Marrone Yukawa structures in eq.~\eqref{eq:FM-Yukawa}, the second line of eq.~\eqref{eq:fixed-sextet} fixes $r_f=-6$ in each Dirac sector, $f=u,d,e,\nu$, if its index-12 coupling comes from this coefficient. Feruglio and Marrone find $r_u\simeq-1.96-0.15i$, $r_d\simeq-3.07-0.71i$ and $r_e\simeq-2.08-0.25i$ in their reference fit to charged fermions and $(r_u,r_d,r_e)\simeq(-1.32,0.46,1.13)$ in the corresponding charged-fermion fit with real couplings \cite{Feruglio:2026xxo}. These values require two independently adjustable structures. The value $r_f=-6$ obtained in the Siegel theory is real, as CP requires in the basis of section~\ref{sec:CP}, so the completion at level 3 is compatible with CP but differs from the fitted values. In addition, their neutrino sextet presents a different constraint. Its eta multiplier gives the phase $i$ under $\tau\to\tau+3$, whereas an ordinary level-3 coefficient is invariant under this translation. Appendix~\ref{app:coefficient} constructs higher-level (Siegel) theta series that supply the independent Yukawa structures and the Majorana sextet with multiplier $\eps^2$.

\paragraph{The lightest generation near a two-torsion point.}
All Dirac entries of the Feruglio--Marrone model have index $m=12$, i.e.\ $n_2=4$, so the factor $e^{-2\pi n_2\varrho/3}$ of section~\ref{sec:hierarchies} is common to the three families and the light family must come from the dependence on $(\tau,w)$. The fits of Feruglio and Marrone lie near $z=(a\tau+b)/2$, with $a,b$ not both even \cite{Feruglio:2026xxo}. Consider $z=-\tau/2$, or $w=-\tau/6$. Completing the square in the lattice sums and using the reflection symmetry of $A_2$ gives the same null vector for both index-12 Dirac structures,
\begin{equation}\label{eq:torsion-kernel}
 \mathcal{Y}[Y^{(1,0)}_{\mathbf{2}^{(0)}},Y^{(1,12)}_{\mathbf{3}^{(0)}}]\,v_0=\mathcal{Y}[Y^{(1,9)}_{\mathbf{2}^{(0)}},Y^{(1,3)}_{\mathbf{3}^{(0)}}]\,v_0=0\,,\qquad
 v_0=(1,0,-1)^\top/\sqrt{2}\,.
\end{equation}
Thus one generation is massless on the section for every $r_f$. Moreover, modular transformations carry the result to the other two nonzero half-periods. A Siegel image of the interior surface $\tau=\sigma$ meets the boundary along this section.

For the leading term of the globally covariant coupling in eq.~\eqref{eq:global-Y-exp}, the same lattice identities give
\begin{align}\label{eq:DF-torsion}
 \mathcal{Y}[Y^{(1,0)}_{\mathbf{2}^{(0)}},Y^{(1,3)}_{\mathbf{3}^{(0)}}]v_0&=-\frac{q^{1/6}}{\sqrt{2}}\,
 Y^{(2,9)}_{\mathbf{1}^{(1)}}(\tau,-\tau/6)\,v_0\,,\notag\\
 \det\mathcal{Y}[Y^{(1,0)}_{\mathbf{2}^{(0)}},Y^{(1,3)}_{\mathbf{3}^{(0)}}]\big|_{w=-\tau/6}&=-2q^{1/6}+\mathcal{O}(q^{2/3})\,,\qquad q=e^{2\pi i\tau}\,.
\end{align}
The missing index-9 singlet therefore measures the lifting of the Jacobi zero mode. In other words, the leading Siegel texture generically gives all three generations a mass on this section.
\section{Flavor models near the other boundary regions}\label{sec:models}
Having identified the Klingen region with the Jacobi flavor theory, we now turn to the boundary regions over the Siegel cusp. We use the five weight-1 Siegel modular forms of eq.~\eqref{eq:five-basis} to construct quark and lepton mass matrices near this cusp. The two-ray model assigns the three families to a singlet and a doublet. At the corner they can form a triplet, and the relative assignments of the left- and right-handed fields determine the mass hierarchy through the Fourier orders of section~\ref{sec:hierarchies}.

For matter fields with $k_I=1$ the Yukawa matrices have weight $k=2$. We take their entries from the quadratic polynomials in $Y_1,\ldots,Y_5$, i.e.\ from the full Siegel theory at level 3. Their modular covariance condition is
\begin{equation}\label{eq:boundary-Y-covariance}
 \rho_R(\gamma)^\top\mathcal{Y}_f(\gamma\Omega)\rho_L(\gamma)=j(\gamma,\Omega)^2\mathcal{Y}_f(\Omega)\,.
\end{equation}
Translations determine which monomials can occupy each entry, and the remaining generators equate some of their coefficients. The relevant contractions are given in appendix~\ref{app:reps}; the restriction on Fourier powers is derived in appendix~\ref{sec:corner-rule}. Notice that in all models the Higgs fields have weight 0 and are trivial singlets, and the superpotential has the form
\begin{align}\label{eq:boundary-W}
 \mathscr{W}={}&H_u(u^c)^\top\mathcal{Y}_uq+H_d(d^c)^\top\mathcal{Y}_dq
 +H_d(e^c)^\top\mathcal{Y}_el
 +H_u(N^c)^\top\mathcal{Y}_\nu l+\frac{\Lambda}{2}(N^c)^\top\mathcal{Y}_NN^c\,.
\end{align}
No nontrivial weight-2 three-family texture arises directly from the unrestricted Siegel cusp group $\Z_3^3\rtimes S_4$ of table~\ref{tab:cusp-groups}. As listed in appendix~\ref{app:cusp-groups}, its irreducible representations of dimension at most three are $[\mathbf{1},2]$, $[\mathbf{2},1]$ and $[\mathbf{3},2]$; since $2\cdot 1^2+1\cdot 2^2+2\cdot 3^2=24=|S_4|$, every such representation is trivial on the translation subgroup $\Z_3^3$ and factors through $S_4$. Meanwhile, the space of modular forms of weight 2 decomposes as $\mathcal{M}_{k=2}=\mathrm{Sym}^2(\mathbf{1}\oplus\mathbf{4})=\mathbf{1}\oplus\mathbf{4}\oplus\bar{\mathbf{4}}\oplus\mathbf{6}$, spanned respectively by $Y_1^2$, $Y_1(Y_2,Y_3,Y_4,Y_5)$, $(Y_2^2,Y_3^2,Y_4^2,Y_5^2)$ and the six pairwise products of $Y_2,\dots,Y_5$. All fourteen components outside $Y_1^2$ carry nontrivial $\Z_3^3$ characters, so any weight-2 Yukawa coupling of three matter families in representations of dimension $\le 3$ can involve only $Y_1^2$ and produces no modular hierarchy. Generating an exponential hierarchy over the Siegel cusp therefore requires choosing a directional limit inside the cone $\langle Q_1,Q_2,Q_3\rangle$, namely the two-ray or corner region, which reduces $\GL(2,\Z)$ to $D_8$ or $D_{12}$ and allows low-dimensional matter multiplets to carry nontrivial translation charges.

\subsection{A flavor model in the two-ray region}\label{sec:two-ray-model}
In the two-ray region, the (projective) finite modular group is $\Z_3\times S_3\times S_3\cong[108,38]$. Assign each set of three matter fields to a singlet and a doublet:
\begin{equation}\label{eq:curve-assignment}
 q,l,u^c,d^c,e^c,N^c\sim(\mathbf{1}_3\oplus\mathbf{2}_5,1)\,,\qquad
 H_u,H_d\sim(\mathbf{1}_0,0)\,.
\end{equation}
We order the fields as the singlet followed by the two components of the doublet. The labels in this equation refer to the $\Z_6\times S_3$ representation image, as explained in appendix~\ref{app:reps}. Explicitly,
\begin{align}\label{eq:curve-model-rep}
 \rho(\mathsf{T}_B)&=\diag(1,\omega^{b_{11}},\omega^{b_{22}})\,,\notag\\
 \rho(\PP)&=-\begin{pmatrix}1&0&0\\0&0&1\\0&1&0\end{pmatrix}\,,\qquad
 \rho(\DD)=-I_3\,.
\end{align}
The weight-1 automorphy factor cancels the minus signs on the two $\GL(2,\Z)$ transformations. Thus $\PP$ exchanges the last two families, while $\DD$ acts trivially on these fields and inverts $\zeta$.

The superpotential is eq.~\eqref{eq:boundary-W}, and the modular covariance condition eq.~\eqref{eq:boundary-Y-covariance} gives
\begin{equation}\label{eq:curve-model-Y}
 \mathcal{Y}_f=\begin{pmatrix}
 \alpha_f Y_1^2&\beta_f Y_2^2&\beta_f Y_3^2\\
 \beta'_f Y_2^2&\delta_f Y_1Y_2&\delta'_f Y_4Y_5\\
 \beta'_f Y_3^2&\delta'_f Y_4Y_5&\delta_f Y_1Y_3
 \end{pmatrix}\,,\qquad f=u,d,e,\nu\,.
\end{equation}
Each entry has one allowed quadratic monomial in $Y_1,\dots,Y_5$, and the exchange then leaves five independent coefficients. The Majorana matrix has the same texture with $\beta'_N=\beta_N$, and hence four coefficients. Choosing the minimal metric $\diag(1,2,2)$, as in $G_5$ for $(Y_1,Y_2,Y_3)$, canonical normalization gives the physical matrices:
\begin{equation}\label{eq:curve-mass-matrices}
 M_f=v_f\det\im\Omega\begin{pmatrix}
 \alpha_fY_1^2&\beta_fY_2^2/\sqrt{2}&\beta_fY_3^2/\sqrt{2}\\
 \beta'_fY_2^2/\sqrt{2}&\delta_fY_1Y_2/2&\delta'_fY_4Y_5/2\\
 \beta'_fY_3^2/\sqrt{2}&\delta'_fY_4Y_5/2&\delta_fY_1Y_3/2
 \end{pmatrix}\,,\qquad f=u,d,e,\nu\,.
\end{equation}
The heavy-neutrino matrix is
\begin{equation}\label{eq:curve-Majorana-mass}
 M_N=\Lambda\det\im\Omega\begin{pmatrix}
 \alpha_NY_1^2&\beta_NY_2^2/\sqrt{2}&\beta_NY_3^2/\sqrt{2}\\
 \beta_NY_2^2/\sqrt{2}&\delta_NY_1Y_2/2&\delta'_NY_4Y_5/2\\
 \beta_NY_3^2/\sqrt{2}&\delta'_NY_4Y_5/2&\delta_NY_1Y_3/2
 \end{pmatrix}\,,\qquad m_\nu=-M_\nu^\top M_N^{-1}M_\nu\,.
\end{equation}
A simple hierarchy already follows before any fit, as anticipated in section~\ref{sec:hierarchies}. The leading terms of the weight-1 Siegel modular forms near this two-ray region are
\begin{align}\label{eq:Siegel-leading-theta}
 (Y_1,Y_2,Y_3,Y_4,Y_5)&\simeq(1,\,3q_\tau,\,3q_\sigma,\,q_\tau q_\sigma P_4,\,q_\tau q_\sigma P_5)\,,\notag\\
 P_4&=3\zeta^{-2}+6\zeta\,,\qquad P_5=3\zeta^2+6\zeta^{-1}\,.
\end{align}
If the off-diagonal entries are perturbations and $|Y_3|<|Y_2|\ll|Y_1|$, the third, second and first masses are approximately
\begin{gather}
 m_{f_3}\simeq v_f\det\im\Omega\,|\alpha_fY_1^2|\,,\quad
 m_{f_2}\simeq\frac{v_f\det\im\Omega}{2}|\delta_fY_1Y_2|\,,\quad
 m_{f_1}\simeq\frac{v_f\det\im\Omega}{2}|\delta_fY_1Y_3|\,.
\label{eq:curve-model-masses}
\end{gather}
The light-family block also shows how this first approximation changes:
\begin{equation}\label{eq:two-ray-light-block}
 \frac{v_f\det\im\Omega}{2}
 \begin{pmatrix}\delta_fY_1Y_2&\delta'_fY_4Y_5\\\delta'_fY_4Y_5&\delta_fY_1Y_3\end{pmatrix}\,,\qquad
 \det=\frac{(v_f\det\im\Omega)^2}{4}
 \left(\delta_f^2Y_1^2Y_2Y_3-(\delta'_f)^2Y_4^2Y_5^2\right)\,.
\end{equation}
Its smaller mass can be reduced by cancellation between the two displayed terms. For a real light block, before mixing with the third family, the rotation obeys
\begin{equation}\label{eq:two-ray-angle}
 \tan2\theta_f=\frac{2\delta'_fY_4Y_5}{\delta_fY_1(Y_2-Y_3)}\,.
\end{equation}
Thus the moduli supply a common hierarchy, while the independent up and down coefficients supply different masses and mixing. The following fit uses the full matrices, including the heavy-family entries.

\paragraph{Numerical fit.}
For the numerical fits in this section we use the quark parameters at the \ac{GUT} scale from ref.~\cite{Antusch:2025fpm}, table~6. They are obtained with the 2024 \ac{PDG} input in the \ac{MSSM} with $\tan\beta=10$, $M_{\rm SUSY}=3\,$TeV and $M_{\rm GUT}=2\times10^{16}\,$GeV, without \ac{SUSY} threshold corrections. The resulting mass ratios and mixing parameters are listed in table~\ref{tab:quark-fit}, where the errors of the mass ratios are propagated in quadrature from the quoted Yukawa uncertainties. To reduce the parameter count and test spontaneous CP violation, we impose generalized CP symmetry (section~\ref{sec:CP}), which forces all coefficients $\alpha_f,\beta_f,\beta'_f,\delta_f,\delta'_f$ in eq.~\eqref{eq:curve-mass-matrices} to be real. CP can then be broken only by $\re\Omega$: the rephasing-invariant combinations $M_{11}M_{22}/(M_{12}M_{21})\propto(Y_1/Y_2)^3$, $M_{11}M_{33}/(M_{13}M_{31})\propto(Y_1/Y_3)^3$ and $M_{22}M_{33}/(M_{23}M_{32})\propto Y_1^2Y_2Y_3/(Y_4Y_5)^2$ carry the CP phases, with the phases of the first two given, modulo $\pi$, by $-2\pi\re\tau$ and $-2\pi\re\sigma$ at leading order. Fitting the eight quark observables gives the following best-fit values of the input parameters:
\begin{gather}
 \Omega=\begin{pmatrix}0.03443+1.1909i&-0.32668-0.29766i\\-0.32668-0.29766i&0.16164+3.1629i\end{pmatrix}\,,\qquad \det\im\Omega=3.6781\,,\notag\\[4pt]
 \begin{array}{c|cccc}
 &\beta_f/\alpha_f&\beta'_f/\alpha_f&\delta_f/\alpha_f&\delta'_f/\alpha_f\\\hline
 f=u&-6.8302&0.27626&-0.0038705&-255.15\\
 f=d&-5.8846&8.8105&-0.78593&6974.1
 \end{array}\label{eq:quark-fit-CP}
\end{gather}
with real overall scales $\alpha_u,\alpha_d$ fixed by $m_t$ and $m_b$. The resulting mass ratios and \ac{CKM} parameters are listed in table~\ref{tab:quark-fit}; with the displayed digits every pull is below $0.05$ ($\chi^2\simeq0.003$). Even with CP imposed, the model has more real parameters than observables (eight real coefficient ratios plus six real moduli), so this benchmark demonstrates that the texture is viable rather than predicting the quark data. The benchmark also uses hierarchical coefficient ratios, so the fit does not by itself explain the full quark mass hierarchy through the moduli.
\begin{table}[t]
\centering\small\tablesetup[1.3]
\begin{tabular}{lc|cc}\toprule
observable & input \cite{Antusch:2025fpm} & two-ray fit & distinct-triplet corner fit (pull)\\\midrule
$m_u/m_c$&$(1.986\pm0.066)\times10^{-3}$&$1.985\times10^{-3}$&$2.005\times10^{-3}$ \ $(+0.3)$\\
$m_c/m_t$&$(2.810\pm0.075)\times10^{-3}$&$2.810\times10^{-3}$&$3.454\times10^{-3}$ \ $(+8.5)$\\
$m_d/m_s$&$(5.00\pm0.11)\times10^{-2}$&$4.997\times10^{-2}$&$4.816\times10^{-2}$ \ $(-1.6)$\\
$m_s/m_b$&$(1.782\pm0.027)\times10^{-2}$&$1.782\times10^{-2}$&$2.513\times10^{-3}$ \ $(-57.5)$\\
$\theta_{12}$&$0.22702\pm0.00082$&$0.22704$&$0.22577$ \ $(-1.5)$\\
$\theta_{13}$&$(3.41\pm0.07)\times10^{-3}$&$3.410\times10^{-3}$&$4.044\times10^{-3}$ \ $(+9.1)$\\
$\theta_{23}$&$(3.874\pm0.038)\times10^{-2}$&$3.875\times10^{-2}$&$3.792\times10^{-2}$ \ $(-2.2)$\\
$\delta_{\rm CKM}$&$1.139\pm0.023$&$1.138$&$1.050$ \ $(-3.9)$\\\midrule
$\chi^2$ & & $0.003$ & $3486$\\
\bottomrule\end{tabular}
\caption{Quark observables at the \ac{GUT} scale. Inputs are from table~6 of ref.~\cite{Antusch:2025fpm}, with the running and threshold assumptions specified at the beginning of the numerical fit in section~\ref{sec:two-ray-model}. The two-ray column uses the real coefficients of eq.~\eqref{eq:quark-fit-CP}. For the corner model the signed pulls are given in parentheses. We compute $\chi^2$ with diagonal uncertainties, forming the mass ratios before rounding. Angles are in radians.}\label{tab:quark-fit}
\end{table}

At leading order the two light masses come from the same coefficient $\delta_f$, see eq.~\eqref{eq:curve-model-masses}, so that $m_{f_1}/m_{f_2}\simeq|Y_3/Y_2|\simeq|q_\sigma/q_\tau|$ in every sector. The observed difference between $m_u/m_c$ and $m_d/m_s$ is accommodated by the term $\delta'_fY_4Y_5$ in the light block eq.~\eqref{eq:two-ray-light-block}, which is of relative order $q_\tau q_\sigma^2$ compared with $Y_1Y_2$ at fixed $\zeta$. In the fit this suppression is compensated by $|\delta'_u/\delta_u|\simeq6.6\times10^4$ and $|\delta'_d/\delta_d|\simeq8.9\times10^3$.

\subsection{A common-triplet model at the corner}\label{sec:corner-common}
The (projective) finite modular group at the corner region $\Z_3^3\rtimes S_3\cong[162,10]$ has twelve inequivalent triplets. We label them by $\mathbf{3}_{a,b}^{\pm}$ in appendix~\ref{app:reps}; the two entries specify translation characters and the sign specifies a transposition. Put all matter species in the same triplet,
\begin{equation}\label{eq:corner-common-assignment}
 q,l,u^c,d^c,e^c,N^c\sim(\mathbf{3}_{0,1}^{-},1)\,,\qquad H_u,H_d\sim(\mathbf{1}_0^+,0)\,.
\end{equation}
The triplet has the same transformation matrices as $(Y_3,Y_2,Y_5)^\top$ in eq.~\eqref{eq:corner-matrices}.
In this order its representation matrices are
\begin{align}\label{eq:corner-model-rep}
 \rho(\mathsf{T}_B)&=\diag(\omega^{t_2+t_3},\omega^{t_1+t_3},\omega^{t_1+t_2})\,,\notag\\
 \rho(\PP)&=-\begin{pmatrix}0&1&0\\1&0&0\\0&0&1\end{pmatrix}\,,\qquad
 \rho(\CC)=\begin{pmatrix}0&0&1\\1&0&0\\0&1&0\end{pmatrix}\,.
\end{align}
For this common triplet assignment the two symmetric contractions in eq.~\eqref{eq:boundary-W} contribute separately to the diagonal and off-diagonal entries,
\begin{equation}\label{eq:corner-model-Y}
 \mathcal{Y}_f=\alpha_f Y_1\begin{pmatrix}Y_3&0&0\\0&Y_2&0\\0&0&Y_5\end{pmatrix}
 +\beta_f Y_4\begin{pmatrix}0&Y_5&Y_2\\Y_5&0&Y_3\\Y_2&Y_3&0\end{pmatrix}\,,
 \quad f=u,d,e,\nu,N\,.
\end{equation}
The antisymmetric contraction has no weight-2 coupling with the required character. Since the triplet is irreducible, its invariant metric is proportional to $I_3$. Choosing $2I_3$, as in $G_5$ for $(Y_3,Y_2,Y_5)$, the physical matrices are
\begin{align}\label{eq:corner-common-mass-matrices}
 M_f&=\frac{v_f\det\im\Omega}{2}\begin{pmatrix}
 \alpha_fY_1Y_3&\beta_fY_4Y_5&\beta_fY_4Y_2\\\beta_fY_4Y_5&\alpha_fY_1Y_2&\beta_fY_4Y_3\\\beta_fY_4Y_2&\beta_fY_4Y_3&\alpha_fY_1Y_5
 \end{pmatrix}\,,\qquad f=u,d,e,\nu\,,\notag\\
 M_N&=\frac{\Lambda\det\im\Omega}{2}\begin{pmatrix}
 \alpha_NY_1Y_3&\beta_NY_4Y_5&\beta_NY_4Y_2\\\beta_NY_4Y_5&\alpha_NY_1Y_2&\beta_NY_4Y_3\\\beta_NY_4Y_2&\beta_NY_4Y_3&\alpha_NY_1Y_5
 \end{pmatrix}\,,\qquad m_\nu=-M_\nu^\top M_N^{-1}M_\nu\,.
\end{align}
The leading masses, apart from $v_f(\det\im\Omega)$, are
\begin{equation}\label{eq:corner-model-masses}
 \frac{3}{2}|\alpha_fq_2q_3|\,,\qquad\frac{3}{2}|\alpha_fq_1q_3|\,,\qquad\frac{3}{2}|\alpha_fq_1q_2|\,.
\end{equation}
Their ordering depends on the three cone coordinates $y_i$. The off-diagonal terms are smaller, and the leading mass ratios are again common to all sectors.

For coefficients of comparable size the rotations are also small. Since this common assignment neither generates large mixing nor distinguishes the up and down sectors at leading order, we next assign the right-handed fields to a different triplet.

\subsection{A distinct-triplet model at the corner}\label{sec:factorised}
Different left and right triplets allow sector-dependent hierarchies. The assignments are
\begin{align}\label{eq:factorised-assignment}
 q,l&\sim(\mathbf{3}_{1,0}^{-},1)\,,\notag\\
 u^c,d^c,e^c,N^c&\sim(\mathbf{3}_{2,1}^{-},1)\,,\qquad H_u,H_d\sim(\mathbf{1}_0^+,0)\,.
\end{align}
Both triplets use the permutation matrices of eq.~\eqref{eq:corner-model-rep}, while the translation matrices of the left-handed fields $q,l$ and of the right-handed fields are
\begin{equation}\label{eq:factorised-reps}
 \rho_L(\mathsf{T}_B)=\diag(\omega^{t_1},\omega^{t_2},\omega^{t_3})\,,\qquad
 \rho_R(\mathsf{T}_B)=\omega^{t_1+t_2+t_3}\rho_L(\mathsf{T}_B)\,.
\end{equation}
In eq.~\eqref{eq:boundary-W} the different left and right triplets force each Dirac entry to contain one modular form from its row and one from its column, so the modular dependence factorizes exactly as $\mathcal{Y}_f=\alpha_f\,\mathcal{D}\,C_f\,\mathcal{D}$ with $\mathcal{D}=\diag(Y_3,Y_2,Y_5)$ and a constant matrix $C_f$ whose diagonal entries are $1$ and off-diagonal entries are a single complex ratio $r_f$ (we refer to this below as a factorized texture). For the right-handed triplet chosen here, the translation charges allow only the three diagonal Majorana entries at weight 2. For these assignments the coupling matrices are
\begin{align}\label{eq:factorised-Y}
 \mathcal{Y}_f&=\alpha_f\begin{pmatrix}
 Y_3^2&r_fY_2Y_3&r_fY_3Y_5\\r_fY_2Y_3&Y_2^2&r_fY_2Y_5\\r_fY_3Y_5&r_fY_2Y_5&Y_5^2
 \end{pmatrix}\,,\qquad f=u,d,e,\nu\,,\notag\\
 \mathcal{Y}_N&=\alpha_N\begin{pmatrix}Y_2Y_5&0&0\\0&Y_3Y_5&0\\0&0&Y_2Y_3\end{pmatrix}\,.
\end{align}
In particular,
\begin{equation}\label{eq:factorised-determinants}
 \det\mathcal{Y}_f=\alpha_f^3Y_2^2Y_3^2Y_5^2(1-r_f)^2(1+2r_f)\,,\qquad
 \det\mathcal{Y}_N=\alpha_N^3Y_2^2Y_3^2Y_5^2\,.
\end{equation}
Thus $r_f=-1/2$ gives rank two and $r_f=1$ gives rank one at generic moduli. Choosing identity matter metrics, the physical matrices are
\begin{align}\label{eq:factorised-mass-matrices}
 M_f&=v_f\det\im\Omega\,\alpha_f\begin{pmatrix}
 Y_3^2&r_fY_2Y_3&r_fY_3Y_5\\r_fY_2Y_3&Y_2^2&r_fY_2Y_5\\r_fY_3Y_5&r_fY_2Y_5&Y_5^2
 \end{pmatrix}\,,\notag\\
 M_N&=\Lambda\det\im\Omega\,\alpha_N\begin{pmatrix}Y_2Y_5&0&0\\0&Y_3Y_5&0\\0&0&Y_2Y_3\end{pmatrix}\,,
 \qquad m_\nu=-M_\nu^\top M_N^{-1}M_\nu\,.
\end{align}
The heavy Majorana matrix is diagonal in the same basis. Inserting its three eigenvalues into the seesaw gives, for example,
\begin{align}\label{eq:factorised-neutrino-entries}
 (m_\nu)_{11}&=-\frac{v_u^2\det\im\Omega\,\alpha_\nu^2}{\Lambda\alpha_NY_2Y_3Y_5}
 Y_3^2\bigl[Y_3^3+r_\nu^2(Y_2^3+Y_5^3)\bigr]\,,\notag\\
 (m_\nu)_{12}&=-\frac{v_u^2\det\im\Omega\,\alpha_\nu^2}{\Lambda\alpha_NY_2Y_3Y_5}
 Y_2Y_3\bigl[r_\nu(Y_3^3+Y_2^3)+r_\nu^2Y_5^3\bigr]\,.
\end{align}
The remaining entries follow by simultaneously cycling the row and column labels and $(Y_3,Y_2,Y_5)$. This gives a concrete charged-lepton and neutrino model with the same moduli.

\subsubsection{Masses and mixing near the corner}
The leading terms of the weight-1 Siegel modular forms near this corner are
\begin{equation}\label{eq:corner-leading-theta-main}
 (Y_1,Y_2,Y_3,Y_4,Y_5)\simeq(1,\,3q_1q_3,\,3q_2q_3,\,6q_1q_2q_3,\,3q_1q_2)\,.
\end{equation}
In the quark sector diagonal field redefinitions remove the three phases of $Y_2,Y_3,Y_5$ from both matrices. Only their magnitudes and the complex $r_u,r_d$ affect the quark mass ratios and mixing. For $|Y_3|\ll|Y_2|\ll|Y_5|$, eliminating the heaviest direction gives the light block, apart from $v_f\det\im\Omega\,\alpha_f$,
\begin{equation}\label{eq:factorised-Schur}
 \begin{pmatrix}(1-r_f^2)Y_3^2&(r_f-r_f^2)Y_2Y_3\\(r_f-r_f^2)Y_2Y_3&(1-r_f^2)Y_2^2\end{pmatrix}\,.
\end{equation}
The resulting mass ratios are
\begin{equation}\label{eq:corner-masses}
 \frac{m_{f_1}}{m_{f_2}}\simeq\frac{\left|1+2r_f\right|}{\left|1+r_f\right|^2}\left|\frac{Y_3}{Y_2}\right|^2\,,
 \qquad
 \frac{m_{f_2}}{m_{f_3}}\simeq\left|1-r_f^2\right|\left|\frac{Y_2}{Y_5}\right|^2\,.
\end{equation}
The approximation requires the displayed denominators and light-block coefficients to be larger than the omitted hierarchy corrections. A value near $r_u=-1/2$ can suppress the up mass, in agreement with the exact determinant. The leading \ac{CKM} magnitudes are
\begin{align}\label{eq:factorised-CKM}
 |V_{us}|&\simeq\left|\frac{r_d}{1+r_d}-\frac{r_u}{1+r_u}\right|\left|\frac{Y_3}{Y_2}\right|\,,\quad |V_{cb}|\simeq|r_d-r_u|\left|\frac{Y_2}{Y_5}\right|\,,\quad
 |V_{ub}|\simeq\left|\frac{r_d-r_u}{1+r_u}\right|\left|\frac{Y_3}{Y_5}\right|\,.
\end{align}
The Cabibbo angle is linear in $|Y_3/Y_2|$, while $m_d/m_s$ is quadratic. Their leading ratio still depends on both quark sectors,
\begin{equation}\label{eq:Cabibbo-mdms}
 \frac{|V_{us}|^2}{m_d/m_s}\simeq\frac{|r_d-r_u|^2}{|1+r_u|^2\,|1+2r_d|}\,.
\end{equation}
 Since $Y_3/Y_2\simeq q_2/q_1$ and $Y_2/Y_5\simeq q_3/q_2$, these expressions connect the masses and mixing directly to the moduli: the two hierarchy parameters are $|q_2/q_1|=e^{-2\pi(y_2-y_1)/3}$ and $|q_3/q_2|=e^{-2\pi(y_3-y_2)/3}$, exponentials of the cone-coordinate differences from the central ray $y_1=y_2=y_3$ of section~\ref{sec:meet}.

\subsubsection{Quark mass ratios and the mixing phase}
Unlike the two-ray example, this texture has only six effective real parameters for the eight quark observables: the two positive ratios $|Y_3/Y_2|$, $|Y_2/Y_5|$ and the complex couplings $r_u$, $r_d$. Eliminating the hierarchy parameters from eqs.~\eqref{eq:corner-masses} and \eqref{eq:factorised-CKM} gives three relations that involve only $r_u$ and $r_d$,
\begin{equation}\label{eq:corner-sum-rules}
 \frac{|V_{ub}|}{|V_{us}||V_{cb}|}\simeq\frac{|1+r_d|}{|r_d-r_u|}\,,\qquad
 \frac{m_c}{m_t}\simeq|V_{cb}|^2\,\frac{|1-r_u^2|}{|r_d-r_u|^2}\,,\qquad
 \frac{m_s}{m_b}\simeq|V_{cb}|^2\,\frac{|1-r_d^2|}{|r_d-r_u|^2}\,.
\end{equation}
The data require $|1+r_d|\simeq0.39\,|r_d-r_u|$, $|1-r_u^2|\simeq1.9\,|r_d-r_u|^2$ and $|1-r_d^2|\simeq12\,|r_d-r_u|^2$. The last ratio is particularly restrictive. To test the texture where the leading expansion may fail, we diagonalize eq.~\eqref{eq:factorised-Y} directly. The smallest $\chi^2$ found in a search over its six parameters is shown in table~\ref{tab:quark-fit}. The corresponding point is
\begin{align}\label{eq:corner-best-fit}
 |Y_3/Y_2|&=0.0523686\,,\qquad |Y_2/Y_5|=0.0680733\,,\notag\\
 r_u&=-0.517679-0.086422\,i\,,\qquad r_d=-1.059246-0.245056\,i\,.
\end{align}
For example, at vanishing real parts the full Siegel modular forms give these ratios for $(y_1,y_2,y_3)\simeq(2,3.40826,4.69129)$. The leading approximation gives the cone-coordinate differences $y_2-y_1\simeq1.41$ and $y_3-y_2\simeq1.28$.

The light-mass ratios and the Cabibbo angle are within two standard deviations of the inputs, but the strange-to-bottom ratio comes out too small by a factor of seven, with $\chi^2\simeq3.5\times10^3$. With the stated \ac{GUT}-scale inputs this fit is incompatible with the data, and the relations eq.~\eqref{eq:corner-sum-rules} locate the problem in the second- and third-family couplings of the down-type sector. Note that the quoted fit uses the matter metrics specified above. If the mass and kinetic bases are approximately aligned, rescaling the second and third $d^c$ fields by a relative factor of seven would change $m_s/m_b$ by roughly the same factor. This corresponds to a ratio of order fifty between their diagonal kinetic coefficients.

The quark mixing phase follows directly from the factorized form $M_f=v_f\det\im\Omega\,\alpha_f\,\mathcal{D}\,C_f\,\mathcal{D}$: since $\mathcal{D}=\diag(Y_3,Y_2,Y_5)$ is common to both quark sectors, its phases are removed by rephasing $q$, $u^c$ and $d^c$. With CP imposed (section~\ref{sec:CP}) $r_u$ and $r_d$ are real, both quark mass matrices become real at every point of moduli space, and the Jarlskog invariant vanishes. The \ac{CKM} phase of the fit in table~\ref{tab:quark-fit} thus comes entirely from the complex ratios $r_f$, i.e.\ from explicit CP violation. In the lepton sector $M_N^{-1}\propto\mathcal{D}/(Y_2Y_3Y_5)$, so the phases of $\mathcal{D}$ survive in the light-neutrino matrix and leptonic CP violation can still arise from the moduli.

\subsection{One coupling across all regions}\label{sec:global}
\begin{figure}[t]
\centering
\includegraphics[width=\textwidth]{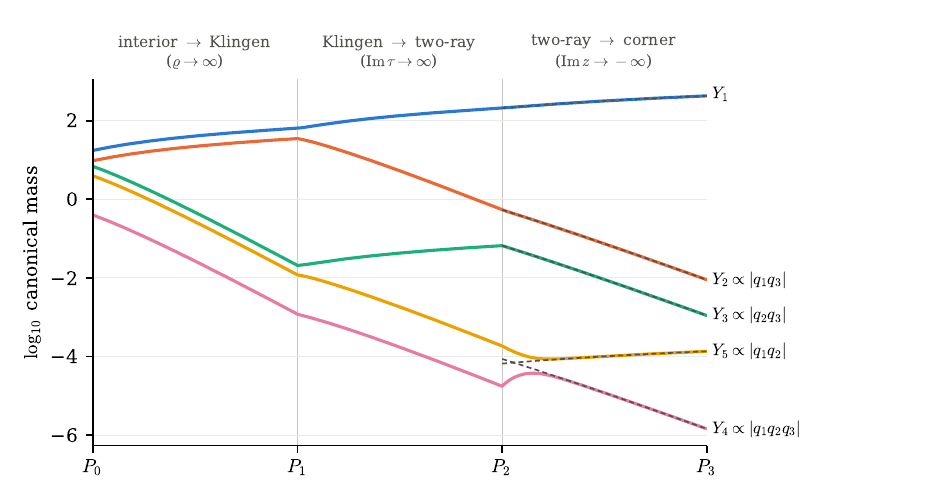}
\caption{Canonical masses of the Hessian coupling eq.~\eqref{eq:Hessian-assignment}, in units of $|\lambda_Hv_H|$, along a path through the boundary regions, with $\im\Omega=y_1Q_1+y_2Q_2+y_3Q_3$ and fixed real parts $(\re\tau,\re\sigma,\re z)=(0.10,-0.05,0.07)$. The nodes are $P_0=(0.9,1.1,0.2)$, $P_1=(0.9,4.5,0.2)$, $P_2=(3.5,4.5,0.2)$ and $P_3=(3.5,4.5,2.5)$ in terms of $(y_1,y_2,y_3)$. The three segments successively increase $y_2$, $y_1$ and $y_3$, with the other two cone coordinates fixed on each segment. Solid curves show the singular values of the full mass matrix. The dashed curves on the last segment give eq.~\eqref{eq:corner-Hessian} including the common normalization. Both spectra are ordered by mass at each point. The labels identify the dominant components at $P_3$.}\label{fig:hessian-path}
\end{figure}

The preceding models use the same Siegel modular forms with different assignments for the matter fields. In contrast, we now fix the matter fields and the superpotential in a toy model, and follow a single coupling of two matter-field quintets across moduli space. Recall that the Hessian eq.~\eqref{eq:Hessian-matrix} defines the unique coupling of weight 2 between two quintets of matter fields under the full Siegel modular group, so the model below has a single free parameter. With
\begin{equation}\label{eq:Hessian-assignment}
 L_5\sim(\mathbf{5},1)\,,\qquad R_5\sim(\mathbf{5},1)\,,\qquad \Phi_H\sim(\mathbf{1},0)\,,\qquad
 \mathscr{W}_H=\lambda_H \Phi_H L_5^\top\mathcal{H} R_5\,,
\end{equation}
where each pair denotes $(\mathbf{r},k_I)$, the minimal metric $G_5=\diag(1,2,2,2,2)$ and $\langle \Phi_H\rangle=v_H$, the physical mass matrix is
\begin{equation}\label{eq:Hessian-physical-mass}
 M_H=\lambda_Hv_H\det\im\Omega\begin{pmatrix}
12Y_1^2&12\sqrt{2}Y_2^2&12\sqrt{2}Y_3^2&12\sqrt{2}Y_4^2&12\sqrt{2}Y_5^2\\
12\sqrt{2}Y_2^2&24Y_1Y_2&24Y_4Y_5&24Y_3Y_5&24Y_3Y_4\\
12\sqrt{2}Y_3^2&24Y_4Y_5&24Y_1Y_3&24Y_2Y_5&24Y_2Y_4\\
12\sqrt{2}Y_4^2&24Y_3Y_5&24Y_2Y_5&24Y_1Y_4&24Y_2Y_3\\
12\sqrt{2}Y_5^2&24Y_3Y_4&24Y_2Y_4&24Y_2Y_3&24Y_1Y_5
\end{pmatrix}\,.
\end{equation}
This five-family toy example links the modular subsymmetries through a single interaction. Two of its blocks already occur in the three-family models: restricted to $(Y_1,Y_2,Y_3)$ it has the two-ray form eq.~\eqref{eq:curve-model-Y} with $(\alpha,\beta,\beta',\delta,\delta')=(12,24,24,48,48)$, and restricted to $(Y_3,Y_2,Y_5)$ the corner form eq.~\eqref{eq:corner-model-Y} with $\alpha=\beta=48$. By eq.~\eqref{eq:Hessian-chi10} it is non-degenerate everywhere except on the surface $z=0$ and its images. We follow its five masses through the regions (figure~\ref{fig:hessian-path}).

\paragraph{Interior.}
By eq.~\eqref{eq:H1-factorization} we have $Y_4=Y_5$ and $Y_1Y_4=Y_2Y_3$ on the split surface $z=0$, so the last two columns of $\mathcal{H}$ coincide and
\begin{equation}\label{eq:H1-kernel}
 \mathcal{H}\,(0,0,0,1,-1)^\top=0\,,\qquad
 \det\mathcal{H}_{\{1,2,3,4\}}=2^{14}3^8\,\eta(\tau)^{16}\eta(\sigma)^{16}\neq0\,.
\end{equation}
Exactly one state is massless on $z=0$, the odd combination of the fourth and fifth matter components, and by modular covariance the same holds on all images of this surface. On the equal-diagonal surface $\tau=\sigma$ the Hessian commutes with the exchange $Y_2\leftrightarrow Y_3$, and the odd state $(0,1,-1,0,0)$ has canonical mass $24|\lambda_Hv_H|\,\det\im\Omega\,|Y_1Y_2-Y_4Y_5|$, which is generically nonzero. Notice that these two surfaces have isomorphic level-3 finite modular groups, although only on $H_1$ does the Hessian have a massless state at every point.

\paragraph{Klingen region.}
As shown in section~\ref{sec:FN}, the two matter states in the doublet have unsuppressed masses after factoring out $|\lambda_H v_H|\det\im\Omega$. The three masses in the triplet start at order $e^{-2\pi\varrho/3}$ for fixed $\tau,z$, one Fourier order below the doublet (section~\ref{sec:hierarchies}), with the texture $\mathcal{Y}[Y^{(1,0)}_{\mathbf{2}^{(0)}},Y^{(1,3)}_{\mathbf{3}^{(0)}}]$, which is generically of full rank along the two-torsion sections (section~\ref{sec:weight2}).

\paragraph{Two-ray region.}
With eq.~\eqref{eq:Siegel-leading-theta} the coupling becomes nearly block diagonal, and the light block of $M_H/(\lambda_Hv_H\det\im\Omega)$ on $(Y_4,Y_5)$ is
\begin{equation}\label{eq:two-ray-block}
 24\begin{pmatrix}Y_1Y_4&Y_2Y_3\\Y_2Y_3&Y_1Y_5\end{pmatrix}\simeq24\,q_\tau q_\sigma\begin{pmatrix}P_4(\zeta)&9\\9&P_5(\zeta)\end{pmatrix}\,,\qquad
 P_4P_5-81=-72\sin^2(\pi z)\,.
\end{equation}
At leading order the canonical masses in the same units are
\begin{equation}\label{eq:two-ray-spectrum}
 (m_1,m_2,m_3,m_+,m_-)\simeq\bigl(12,\ 72|q_\tau|,\ 72|q_\sigma|,\ 432|q_\tau q_\sigma|,\ 96\pi^2|z|^2|q_\tau q_\sigma|\bigr)\,,\qquad |z|\ll1\,.
\end{equation}
The reflection $\DD$, i.e.\ $z\to-z$ or $\zeta\to\zeta^{-1}$, exchanges $Y_4$ and $Y_5$ and leaves the surface $z=0$ pointwise fixed. There the two light states are the combinations of the fourth and fifth matter components that are even ($m_+$) and odd ($m_-$) under this exchange, and for $|z|\ll1$ they stay close to these combinations. The exponential suppression from the boundary combines with a quadratic suppression in the distance from the surface $z=0$: the ratio
\begin{equation}\label{eq:two-ray-ratio}
 \frac{m_-}{m_+}\simeq\frac{2\pi^2}{9}\,|z|^2
\end{equation}
does not depend on $q_\tau$ and $q_\sigma$, so the boundary suppresses both light masses, while the proximity to $z=0$ further suppresses the odd one. The vanishing mass of the odd combination at $z=0$ also uses the factorization eq.~\eqref{eq:H1-factorization}.

\paragraph{Corner.}
With eq.~\eqref{eq:corner-leading-theta-main} all off-diagonal entries are suppressed, and the canonical masses, divided by $|\lambda_Hv_H|\det\im\Omega$, are
\begin{equation}\label{eq:corner-Hessian}
 (m_1,m_2,m_3,m_4,m_5)\simeq\bigl(12,\ 72|q_1q_3|,\ 72|q_2q_3|,\ 144|q_1q_2q_3|,\ 72|q_1q_2|\bigr)\,.
\end{equation}
The five states form a heavy singlet, a triplet permuted by the corner group, and a light singlet. Since there is a single coupling constant, the masses obey a parameter-free relation,
\begin{equation}\label{eq:corner-sum-rule}
 3\,m_1\,m_4^2=2\,m_2\,m_3\,m_5\,,
\end{equation}
valid at leading order for every approach to the corner, although the ordering of the triplet depends on the approach. For instance, $|q_i|=e^{-t\,(1,2,3)_i}$ gives $m_5>m_2>m_3$, and $|q_i|=e^{-t\,(3,2,1)_i}$ gives $m_3>m_2>m_5$. On the central ray of section~\ref{sec:meet} the triplet is degenerate.

The change in the spectrum follows the branching in table~\ref{tab:boundary-branching}. The Klingen triplet $(Y_3,Y_4,Y_5)$, the two-ray doublets $(Y_2,Y_3)$ and $(Y_4,Y_5)$, and the corner triplet $(Y_3,Y_2,Y_5)$ are different subsets of the same five states. Near $z=0$ the two light states approach the even and odd combinations of the fourth and fifth matter components. Toward the corner, on the other hand, they align with the individual components.

This toy model also shows how the constraints of interior loci and boundary regions combine. On an interior locus the pointwise stabilizer $H$ acts only on the fields and imposes exact relations along the whole locus, such as $Y_4=Y_5$ on $z=0$, and couplings that vanish there grow as powers of the distance from it. Near a boundary the Fourier orders of section~\ref{sec:hierarchies} set the exponential suppression. Isomorphic finite groups need not give the same spectrum, as the surfaces $z=0$ and $\tau=\sigma$ show, and the two types of suppression multiply, as in eq.~\eqref{eq:two-ray-ratio}.

\section{Remarks on string theory constructions}\label{sec:topdown}
So far our discussion has been purely bottom-up. Notice, however, that the same three moduli occur in heterotic compactifications with a Wilson line, where $\Sp(4,\Z)$ is part of the target-space duality group. This gives a string interpretation of the invariant subspaces and boundary regions studied above, together with their modular subsymmetries. In this section we collect several such connections, which also point to open questions.

\subsection{The heterotic moduli and their boundary limits}\label{sec:het-dictionary}
In heterotic compactifications on $T^2$, on $K3\times T^2$ or on orbifolds with a $T^2$ factor, a complex Wilson line $V$ joins the complex structure $U$ and the K\"ahler modulus $T$. The three moduli parametrize the connected component $\mathrm{SO}_0(2,3)/(\mathrm{SO}(2)\times \mathrm{SO}(3))\cong\Sp(4,\mathds{R})/\mathrm{U}(2)=\HH_2$. When the Wilson line lies along a lattice vector of norm two, such as a root of an $\mathrm{SU}(2)$ factor (Wilson-line index $t=1$), the integral Siegel group acts on these moduli, including the $T$ and $U$ modular transformations and their mirror exchange \cite{LopesCardoso:1996nc,Baur:2020yjl,Nilles:2021glx}. With
\begin{equation}\label{eq:het-Omega}
 \Omega=\begin{pmatrix}U&V\\V&T\end{pmatrix}\,,\qquad
 \mathscr{K}=-\log\det\im\Omega\,,\qquad
 \varrho=\im T-\frac{(\im V)^2}{\im U}=\frac{\det\im\Omega}{\im U}\,,
\end{equation}
we have $(\tau,z,\sigma)=(U,V,T)$,\footnote{Part of the literature places $T$ in the upper-left entry of $\Omega$. The two conventions differ by the mirror exchange $\PP$; ours makes the Klingen limit $\im\sigma\to\infty$ a large-volume limit.} the tree-level K\"ahler potential takes the familiar form, and $\varrho$ is the volume corrected by the Wilson line. Notice that here $\Omega$ is not the period matrix of a physical genus-two surface: the geometric picture of section~\ref{sec:moduli} only organizes the action of $\Sp(4,\Z)$ and of its boundary subgroups. The boundary regions of section~\ref{sec:cusps} then acquire a meaning in terms of the compactification torus.

\paragraph{Klingen region.}
Here $\im T\to\infty$ at fixed $U$ and $V$: a large volume at fixed shape and Wilson line. The surviving Jacobi group acts on $(U,V)$ through $\SL(2,\Z)_U$ and the integral shifts $V\to V+\lambda U+\mu$ of the Wilson line, together with the compensating transformation eq.~\eqref{eq:Delta} of $T$, and its central generator $\mathsf{Z}$ is the axionic shift $\re T\to\re T+1$. These Wilson-line shifts and the compensating transformation of $T$ already appear in the analysis of toroidal heterotic compactifications with a Wilson line in ref.~\cite{Baur:2020yjl}, where their products close only up to integer shifts of $T$; together with these shifts they form the integral Heisenberg group. The Jacobi index $m=3n_2$ is fixed by the Fourier order $n_2$ in $q_\sigma=e^{2\pi iT/3}$. A unit axion shift measures $n_2$ modulo three, and the \ac{FN} parameter $e^{-2\pi\varrho/3}$ of eq.~\eqref{eq:FN-identity} is an exponential suppression in the corrected volume, in the same way as worldsheet instantons suppress couplings by the exponential of an area. Jacobi forms with a Wilson-line argument already appear in heterotic threshold corrections \cite{LopesCardoso:1996nc,Nazaroglu:2013pea}; for couplings originating from a Siegel modular form, section~\ref{sec:derivation} identifies their Fourier--Jacobi expansion with a large-volume expansion.

\paragraph{Two-ray region.}
Here $\im U$ and $\im T$ are both large. The finite modular group of this region contains the exchange $\PP$, which is the mirror symmetry $T\leftrightarrow U$ of the $T^2$ compactification, and $\DD$, which is the reflection $V\to-V$. Both are standard elements of the heterotic duality group \cite{Baur:2020yjl,Nilles:2021glx}. Of the two doublets of this region in table~\ref{tab:boundary-branching}, $(Y_2,Y_3)$ is exchanged by mirror symmetry and $(Y_4,Y_5)$ by the Wilson-line reflection, up to signs, so the two are distinguished by these string symmetries.

\paragraph{Corner.}
Here $-\im V$ grows together with $\im U$ and $\im T$, with $\im\Omega$ kept positive definite. The three large parameters are $\im(U+V)$, $\im(T+V)$ and $-\im V$ in the basis of eq.~\eqref{eq:corner-coordinates}. In heterotic terms this is a large-volume limit in which the Wilson line grows as well, at a rate comparable to $\im U$ and $\im T$. The finite part $D_{12}$ of the corner symmetry contains the mirror exchange $\PP$ and the element $\CC$, which permutes the three large combinations cyclically. Unlike $\PP$ and $\DD$, $\CC$ mixes the Wilson line with both $U$ and $T$ and does not preserve the locus $V=0$, so it has no counterpart in a compactification without Wilson line. The corner is approached along its central ray by the hexagonal curve $U=T=-2V$ (section~\ref{sec:meet}).

\subsection{Two realizations of the same finite group}\label{sec:eclectic}
In $T^2/\Z_3$ orbifolds the eclectic flavor group combines the modular group of the K\"ahler modulus with a traditional flavor group $\Delta(54)$; without $R$-symmetries the result is isomorphic to $J_3=\Delta(27)\rtimes T'$ \cite{Baur:2019kwi,Nilles:2020nnc,Nilles:2020kgo,Baur:2024qzo}. This isomorphism has motivated the interpretation of the finite Jacobi group as an eclectic flavor group \cite{Feruglio:2026xxo}. The two realizations nevertheless differ physically. In the eclectic scheme the traditional flavor group acts on the matter fields but not on the moduli. It permutes the three orbifold fixed points and imposes additional restrictions on the matter K\"ahler metric, whereas modular covariance by itself still allows non-universal kinetic terms \cite{Chen:2019ewa,Nilles:2020kgo}. In the Klingen realization, $\Delta(27)$ consists of the Wilson-line translations $V\to V+\lambda U+\mu$ modulo three, together with the center. It acts on the moduli and is part of the modular symmetry, so the Wilson line behaves as a flavon for the translations that move it, while the center fixes $(U,V)$ and acts on the normal coordinate and on fields. At a generic value of $V$ the translations by nonzero three-torsion points move the vacuum, and the K\"ahler potential is then constrained only by modular covariance. Hence non-universal kinetic terms built from Jacobi forms, as used in the fits of Feruglio and Marrone (section~\ref{sec:FN}), remain allowed. Notice also that the derivation of the eclectic group from heterotic orbifolds involves no infinite Heisenberg group: its traditional part arises from outer automorphisms of the Narain space group \cite{Baur:2019kwi,Nilles:2020nnc}, and the orbifold has no continuous Wilson-line modulus to be translated. The common finite group $J_3$ thus appears to be a coincidence at level 3, and a common string-theoretic origin of the two realizations remains to be established.

The two realizations also take $T'$ from different factors. On the curve $\diag(\omega,u)$ of table~\ref{tab:interior}, that is $U=\omega$, $V=0$ and $T=u$, the moduli are those of the untwisted torus sector of a $T^2/\Z_3$ orbifold: the complex structure is frozen at the $\Z_3$-symmetric point, the Wilson line is off, and the K\"ahler modulus is free. Its finite group $\Z_6\times T'$ takes $T'$ from $\SL(2,\Z)_T$, as in the orbifold, while the factor $\Z_6$ fixes every point of the curve and acts there as a traditional symmetry, much as the orbifold twist acts on the twisted states. The Klingen $T'$ instead comes from $\SL(2,\Z)_U$, and the mirror exchange $\PP$ maps one onto the other. The orbifold moduli thus trace one of the elliptic invariant curves of section~\ref{sec:elliptic}, in line with the general observation that orbifolds arise as fixed loci of $\Sp(4,\Z)$ \cite{Nilles:2021glx}. The traditional $\Delta(54)$ also acts on the orbifold states, and it meets the modular group $T'$ in a $\Z_2$, so that the eclectic group has order $648$ rather than $54\times24$ \cite{Baur:2019kwi,Nilles:2020nnc}. Any identification beyond the level of finite groups would require comparing their generators on states and operators, including modular weights and scalar phases; appendix~\ref{app:reps} provides this information on the effective-theory side.

\subsection{Special loci and light states}\label{sec:light-states}
The surface $z=0$ and its images, the Humbert surface $H_1$, form the zero locus of $\chi_{10}$, eq.~\eqref{eq:Hessian-chi10}. In heterotic $K3\times T^2$ models with one Wilson line this is the locus $V=0$, where additional charged states can become massless. In the examples of ref.~\cite{LopesCardoso:1996nc}, with index $t=1$, $\mathrm{U}(1)_V$ is enhanced to $\mathrm{SU}(2)$ there; in $\mathcal{N}=4$ compactifications (heterotic on $T^6$, dual to type~II on $K3\times T^2$), the same cusp form $\chi_{10}$ arises as the Borcherds lift of the $K3$ elliptic genus~\cite{Dijkgraaf:1996it}. By contrast, the $\mathcal{N}=2$ model with 96 additional charged hypermultiplets has Wilson-line index two and a different integral duality group \cite{LopesCardoso:1996zj}. In both comparisons it is the local condition $V=0$ that coincides with the massless locus of the Hessian in section~\ref{sec:global}. Matching the states would require their string quantum numbers. The surface $\tau=\sigma$, which represents the Humbert surface $H_4$ and is the self-mirror locus $U=T$, has the same finite group at level 3 as $z=0$. A generic point of $H_4$ has a full-rank Hessian, while points of $H_4\cap H_1$ lie on its massless locus. A representative of $H_4$ reaches the Klingen region at the two-torsion points, where the Feruglio--Marrone textures have their zero mode (section~\ref{sec:weight2}). The Jacobi zero concerns a particular Fourier coefficient on the boundary of $H_4$, whereas the full Hessian becomes singular on $H_1$.

\subsection{Lattices and fluxes}\label{sec:lattices}
The level-3 structures have a lattice origin: as noted in footnote~\ref{fn:lattice}, the quartic invariant and its gradient are theta series of $E_8$ and $E_6$, tied together by the index-three sublattice $E_6\oplus A_2\subset E_8$. The breaking $E_8\to E_6\times \mathrm{SU}(3)$ is the gauge symmetry breaking of $\Z_3$ orbifolds with standard embedding~\cite{Dixon:1986jc}, and the order-three rotation of $A_2$ fixes no nonzero lattice vector, as the orbifold twist does on the compactification lattice. These facts suggest looking for the global coupling of section~\ref{sec:global} in a $\Z_3$ orbifold with a Wilson line, but we have no derivation. Moreover, an index-$t$ Wilson-line lattice leads to a paramodular group $K(t)$ \cite{LopesCardoso:1996zj,Nazaroglu:2013pea}. For a scalar Jacobi cusp form of weight $k$ and index $t$ with trivial character, the additive Gritsenko lift gives a weight-$k$ paramodular form with the original Jacobi form as its first Fourier--Jacobi coefficient \cite{Gritsenko:1995,Gritsenko:2015products}. Extending this construction to the multiplets and multipliers of the present models is a separate question.

A second route is through magnetized tori. On $T^2$ the zero modes in a flux background are Jacobi theta functions, and section~\ref{sec:derivation} showed that an index-$m$ Jacobi multiplet has the quasi-periodicity of the lowest Landau level with $2m$ flux quanta \cite{Cremades:2004wa}. On magnetized $T^4$ and $T^6$ the zero modes are Riemann theta functions, $\Sp(4,\Z)$ and $\Sp(6,\Z)$ act on them through the complex structure $\Omega$ of $T^4=\C^2/(\Z^2+\Omega\Z^2)$, which here plays the role of the period matrix of section~\ref{sec:moduli}, and at the origin they are Siegel modular forms \cite{Kikuchi:2022psj,Kikuchi:2023dow}; metaplectic multipliers of the kind carried by the Jacobi matter fields arise naturally there \cite{Almumin:2021fbk}. Our $A_2$ theta functions have weight 1, and we have not attempted to identify a flux configuration that produces them.

\subsection{Distances and dynamics}\label{sec:distances}
All boundary regions lie at infinite distance in moduli space. With the K\"ahler metric from $\mathscr{K}$ in eq.~\eqref{eq:het-Omega}, shifting $\varrho$ at fixed $U,V,\re T$ gives a canonical displacement $|\log(\varrho/\varrho_\star)|/\sqrt{2}$ relative to a reference height $\varrho_\star$. Phenomenologically useful hierarchies still occur at moderate excursions: for $n_2=1$, $e^{-2\pi\varrho/3}$ reaches $10^{-5}$ at $\varrho\simeq5.50$, about $1.2$ reduced Planck units from $\varrho_\star=1$. By the distance conjecture~\cite{Ooguri:2006in}, towers of states become light exponentially in the distance, i.e.\ as a power of $\varrho$, though the power law alone does not fix the mass of the first tower state without a concrete compactification.

A more basic question is how a vacuum can sit near a boundary region rather than in the interior. In heterotic orbifolds, a finite-order twist fixes moduli only at an interior elliptic invariant locus of $\Sp(4,\Z)$ (table~\ref{tab:interior})~\cite{Nilles:2021glx}. Because a boundary cusp is stabilized by an infinite parabolic subgroup rather than a finite elliptic element, a geometric orbifold twist cannot project the theory onto a boundary region, nor lock $\varrho$ at a large finite value while leaving $U$ and $V$ dynamical. Consequently, pure string orbifolds naturally lead to interior modular subsymmetries, whereas a vacuum near a boundary region requires dynamical moduli stabilization.

Existing studies of $\Sp(4,\Z)$ moduli stabilization, using Siegel modular superpotentials~\cite{Funakoshi:2024yxg} or gaugino condensation with moduli-dependent threshold corrections~\cite{Leedom:2026nby,Hernandez-Segura:2026oyy}, focus on the interior of moduli space. In the gaugino-condensation models of ref.~\cite{Hernandez-Segura:2026oyy}, negative powers of $\chi_{10}$, $\chi_{12}$ and $\chi_{35}$ drive the potential up exponentially toward every cusp, so vacua with order-one parameters naturally lie at interior fixed points (notably the isolated $S_4$ and $D_{12}$ points) or along interior supersymmetric Minkowski loci. Even so, two dynamical routes reach the boundary: the flat directions along $H_1$, $H_4$ and $H_1\cap H_4$ run directly into the cusps (table~\ref{tab:meet}), and a second mechanism uses the meromorphic modular function
\begin{equation}\label{eq:R-function}
 \mathcal{R}(\Omega):=\frac{\chi_{10}(\Omega)^6}{\chi_{12}(\Omega)^5}\,.
\end{equation}

When the leading Fourier coefficients of both cusp forms are nonzero, $\mathcal{R}$ decays exponentially: it scales as $e^{2\pi i T}$ at Klingen, $e^{2\pi i(T+U)}$ in the two-ray region and $e^{2\pi i(T+U+V)}$ at the corner. In a generic Klingen direction we can therefore write $\mathcal{R}=A(U,V)e^{2\pi i T}+\cdots$ with $0<|A(U,V)|<\infty$. Consider then the superpotential
\begin{equation}\label{eq:W-Dc}
 \mathscr{W}=W_0(S,\Omega)\,\bigl[\mathcal{R}(\Omega)-c\bigr]^2\,,\qquad |c|\ll 1\,,
\end{equation}
where $W_0$ carries the modular weight $-h$. Wherever $W_0$ is regular and nonzero and $d\mathcal{R}\neq 0$, the surface $\mathcal{D}_c=\{\mathcal{R}(\Omega)=c\}$ is a supersymmetric Minkowski locus of complex codimension one. Since it is defined by a single complex equation, this construction fixes only one complex combination of the moduli. At fixed $(U,V)$ it selects a large but finite imaginary part,
\begin{equation}\label{eq:ImT-Dc}
 \im T\simeq\frac{1}{2\pi}\log\frac{|A(U,V)|}{|c|}\,,
\end{equation}
in level-1 normalization, which moves arbitrarily close to the boundary as $c\to 0$, while $U$, $V$ and the dilaton remain flat. Notice that the poles of $\mathcal{R}$ cannot be avoided: by the Koecher principle every non-constant modular function of $\Sp(4,\Z)$ has poles in $\HH_2$, and here the poles along $\chi_{12}=0$ are inherited from the $\chi_{12}$ threshold factor. Stabilizing all three moduli near the boundary, for instance at the two-ray benchmark of section~\ref{sec:two-ray-model}, remains open; in a genus-two racetrack with competing condensates, it would involve hierarchical coefficients of order $e^{-2\pi y_i}$.

\section{Conclusions}\label{sec:conclusions}
In multi-modulus theories, residual modular symmetries can survive not only on interior invariant subspaces but also at the boundary of moduli space. In this work we have determined these residual symmetries for the genus-two Siegel modular group $\Sp(4,\Z)$ and worked out their finite images explicitly at level $N=3$. There $\Gamma_{2,3}\cong\Sp(4,\Z_3)$ is the double cover of Burkhardt's group, the five $A_2$ theta functions furnish the weight-1 quintet $Y^{(1)}_{\mathbf{5}}$, and the Hessian of its quartic invariant has determinant proportional to $\chi_{10}$. Alongside the interior invariant surfaces and curves (table~\ref{tab:interior} and appendix~\ref{app:subsym}), we have analyzed all boundary limits, which geometrically pinch one, two or three cycles of the genus-two surface: the Klingen cusp, the Siegel cusp, and its two-ray and corner toroidal strata. For each region we have determined the modular subsymmetry, its finite modular group at level 3, the branching of $Y^{(1)}_{\mathbf{5}}$ and the generalized CP extension by $\Z_2$ (tables~\ref{tab:cusp-groups} and \ref{tab:boundary-branching}).

The Klingen region reproduces the level-3 Jacobi flavor symmetry of Feruglio and Marrone, who already linked their Jacobi group to this cusp. On the one hand, the integral Jacobi group fixes a primitive generator of the Klingen cusp line and maps to $J_3$ at level 3. On the other hand, the shift of $\sigma$ under the embedded Jacobi transformations supplies the Jacobi automorphy factor and ties the Fourier order $n_2$ to the index $m=3n_2$. As a consequence, all four weight-1 Jacobi form multiplets of Feruglio and Marrone arise from a single Fourier--Jacobi expansion of the weight-1 Siegel modular form $Y^{(1)}_{\mathbf{5}}$, and canonical normalization reproduces the Gaussian factor of their K\"ahler potential and attaches the \ac{FN} factor $e^{-2\pi n_2\varrho/3}$ to the order $n_2$. In this sense each Jacobi coupling of their construction keeps a single order of the Fourier--Jacobi expansion of a Siegel modular form. At weight 2, however, level-3 Siegel covariance is more restrictive: it fixes the relative Dirac coefficient to $r_f=-6$, omits the index-9 singlet, and excludes their Majorana coupling because its Dedekind eta multiplier is incompatible with level 3; higher-level theta series supply these structures independently (appendix~\ref{app:coefficient}). Because $\Delta(27)$ acts here by modulus translations, the K\"ahler potential is constrained only by modular covariance, leaving room for non-universal kinetic terms.

Near the Siegel cusp, selecting a direction in the cone of $\im\Omega$ yields distinct three-family textures. In the two-ray region, a singlet--doublet model with generalized CP accommodates the quark mass ratios and mixing parameters with real, though hierarchical, coefficients and a \ac{CKM} phase coming entirely from $\re\Omega$. At the corner, the distinct-triplet model has factorized Yukawa matrices, so CP-conserving couplings yield no quark CP violation and a minimal-metric fit leaves $m_s/m_b$ too small by roughly a factor of seven. The Hessian of Burkhardt's quartic links all regions through a single global coupling: it has one massless state on $z=0$, its Klingen block contains the index-12 combination with $r_f=-6$ at order $q_\sigma^4$, near $z=0$ in the two-ray region its two light masses have the ratio $(2\pi^2/9)|z|^2$, independent of $q_\tau$ and $q_\sigma$ at leading order, and at the corner its leading masses obey $3m_1m_4^2=2m_2m_3m_5$.

More generally, the boundary regions realize an \ac{FN}-like hierarchy mechanism in which the moduli replace the flavons (section~\ref{sec:hierarchies}): the expansion parameters are the exponentials $|q_a|$, with charges fixed modulo three by translations and bounded below by positivity. Moderate imaginary parts of order $2$ to $4$ already give suppressions of $10^{-2}$ to $10^{-4}$. At the Klingen cusp all entries of a three-family Yukawa matrix carry the same charge modulo three, whereas in the two-ray and corner regions different families can carry different charges. At level 3 the smallest allowed power of each parameter is therefore $0$, $1$ or $2$, except where positivity raises it at the corner as shown in appendix~\ref{sec:corner-rule}, and higher levels allow a wider range.

Several open questions remain. Unlike a vacuum on an interior elliptic fixed locus, which can be selected geometrically by a finite-order orbifold twist, a vacuum near the boundary requires dynamical moduli stabilization. Existing $\Sp(4,\Z)$ stabilization studies focus on the interior \cite{Funakoshi:2024yxg,Leedom:2026nby,Hernandez-Segura:2026oyy}, while the superpotential eq.~\eqref{eq:W-Dc} fixes only one complex combination of the moduli near the boundary. Moreover, on the string side, the Klingen limit is a large-volume limit of heterotic compactifications with a Wilson line, and the decomposition $E_8\supset E_6\oplus A_2$ points to $\Z_3$ orbifolds with Wilson lines as a possible microscopic origin. Last but not least, the same methods extend to higher levels and more general finite quotients of $\Sp(4,\Z)$ \cite{Liu:2021gwa}, as well as to $\Sp(2g,\Z)$, where the parabolic stabilizers eq.~\eqref{eq:parabolic} label the boundary components and the Fourier--Jacobi coefficients are Jacobi forms of higher degree \cite{Ziegler:1989}. We leave these directions for future work.

\acknowledgments
The author thanks Hans Peter Nilles and Sa\'ul Ramos-S\'anchez for helpful discussions. This work was partly supported by the Universidad Nacional Aut\'onoma de M\'exico Postdoctoral Program (POSDOC). The author used the AI models Claude (Anthropic), ChatGPT (OpenAI) and Gemini (Google) to assist with editing the text, with part of the calculations, and with the preparation of the figures. All results were independently verified by the author, who takes full responsibility for the content of this article.

\appendix
\addtocontents{toc}{\protect\setcounter{tocdepth}{1}}
\section{The finite Siegel modular group \texorpdfstring{$\Gamma_{2,3}$}{Gamma(2,3)}}\label{app:group}

\subsection{Generators and defining relations}\label{app:words}
All generators used in this paper are words in the four generators $\mathsf{S},\mathsf{T}_1,\mathsf{T}_2,\mathsf{T}_3$ of eq.~\eqref{eq:generators}. Besides the partial inversions $\si,\sii$ of eq.~\eqref{eq:partial-words}, we use the $\GL(2,\Z)$ transformations $\gamma_A=\diag(A,A^{-\top})$
\begin{equation}\label{eq:words}
 \begin{array}{@{}ll@{\qquad}ll@{}}
 \EE=\sii\mathsf{T}_3\sii^{-1}, & A_{\EE}=\begin{pmatrix}1&1\\0&1\end{pmatrix}, &
 \FF=\si\mathsf{T}_3\si^{-1}, & A_{\FF}=\begin{pmatrix}1&0\\1&1\end{pmatrix}\,,\\[8pt]
 \RR=\EE\FF^{-1}\EE, & A_{\RR}=\begin{pmatrix}0&1\\-1&0\end{pmatrix}, &
 \DD=\sii^2, & A_{\DD}=\begin{pmatrix}1&0\\0&-1\end{pmatrix}\,,\\[8pt]
 \PP=\DD\RR, & A_{\PP}=\begin{pmatrix}0&1\\1&0\end{pmatrix}, &
 \CC=\RR^{-1}\EE, & A_{\CC}=\begin{pmatrix}0&-1\\1&1\end{pmatrix}\,.
 \end{array}
\end{equation}
Thus $\PP$ exchanges $\tau\leftrightarrow\sigma$, $\DD$ reflects $z\to-z$, and $\CC$ has order six with $\CC^3=-I_4$. The partial inversions are
\begin{equation}
 \si=\begin{pmatrix}0&0&1&0\\0&1&0&0\\-1&0&0&0\\0&0&0&1\end{pmatrix}\,,\qquad
 \sii=\begin{pmatrix}1&0&0&0\\0&0&0&1\\0&0&1&0\\0&-1&0&0\end{pmatrix}\,.
\end{equation}

\paragraph{Presentations of $\Sp(4,\Z)$ and $\Gamma_{2,3}$.}
A compact presentation of $\Sp(4,\Z)$ is due to Bender \cite{Bender:1980}. With
\begin{equation}\label{eq:Bender-words}
 \mathsf{K}=\FF\DD\,,\qquad \mathsf{L}=\mathsf{S}^{-1}\mathsf{T}_1\,,\qquad \mathsf{H}=\mathsf{K}\mathsf{L}^5\mathsf{K}\mathsf{L}^7\mathsf{K}\,,
\end{equation}
the two generators $\mathsf{K},\mathsf{L}$ obey the eight defining relations, with the commutator convention $[g,h]=ghg^{-1}h^{-1}$,
\begin{gather}
 \mathsf{K}^2=\mathsf{L}^{12}=1\,,\qquad (\mathsf{K}\mathsf{L}^7\mathsf{K}\mathsf{L}^5\mathsf{K})\mathsf{L}=\mathsf{L}\mathsf{H}\,,\qquad [\mathsf{L}^2\mathsf{K}\mathsf{L}^4,\mathsf{H}]=[\mathsf{L}^3\mathsf{K}\mathsf{L}^3,\mathsf{H}]=1\,,\notag\\
 (\mathsf{L}^2\mathsf{H})^2=(\mathsf{H}\mathsf{L}^2)^2\,,\qquad [\mathsf{L},(\mathsf{L}^6\mathsf{H})^2]=1\,,\qquad (\mathsf{K}\mathsf{L}^5)^5=(\mathsf{L}^6\mathsf{H})^2\,.\label{eq:Bender}
\end{gather}
In terms of the generators in eq.~\eqref{eq:words}, $\mathsf{H}=\PP$. The partial inversions are recovered from $\sii^{-1}=\mathsf{L}^9\mathsf{H}\mathsf{L}^6\mathsf{H}$ and $\si^{-1}=\mathsf{H}\sii^{-1}\mathsf{H}$. Hence
\begin{equation}
 \mathsf{S}=\si\sii\,,\qquad \mathsf{T}_1=\mathsf{S}\mathsf{L}\,,\qquad \mathsf{T}_2=\mathsf{H}\mathsf{T}_1\mathsf{H}\,,\qquad \mathsf{T}_3=\sii^{-1}(\mathsf{L}^5\mathsf{K}\mathsf{L})\sii\,.
\end{equation}

Adding $\mathsf{T}_1^3=\mathsf{T}_2^3=\mathsf{T}_3^3=1$ to eq.~\eqref{eq:Bender} gives a presentation of $\Gamma_{2,3}$; the shorter presentation eq.~\eqref{eq:G23-presentation} in the original generators is equivalent. Adding $\mathsf{S}^2=1$ gives the Burkhardt group.

\paragraph{Embedded $\SL(2,\Z)$ and translation relations.}
For every embedded copy of $\SL(2,\Z)$ below, with generators $\mathsf{s},\mathsf{t}$ corresponding to $\left(\begin{smallmatrix}0&1\\-1&0\end{smallmatrix}\right)$ and $\left(\begin{smallmatrix}1&1\\0&1\end{smallmatrix}\right)$, we use the defining relations
\begin{equation}\label{eq:SL2-relations}
 \mathsf{s}^4=1\,,\qquad (\mathsf{s}\mathsf{t})^3=1\,,\qquad [\mathsf{s}^2,\mathsf{t}]=1\,.
\end{equation}
Adding $\mathsf{t}^3=1$ gives $T'=\SL(2,\Z_3)$.

The groups at the cusps contain the commuting translations $\mathsf{T}_i$. For a $\GL(2,\Z)$ transformation $\gamma_A$ their conjugation relations have the common form
\begin{equation}\label{eq:translation-conj}
 [\mathsf{T}_i,\mathsf{T}_j]=1\,,\qquad \gamma_A\mathsf{T}_i\gamma_A^{-1}=\mathsf{T}_1^{b'_{11}}\mathsf{T}_2^{b'_{22}}\mathsf{T}_3^{b'_{12}}\,,\qquad B'=AB_iA^\top\,.
\end{equation}

\subsection{Representations and tensor products}\label{app:group-reps}
In the notation $[\mathbf{d},n]$ of section~\ref{sec:finite-group}, the $34$ irreducible representations of $\Gamma_{2,3}$ are
\begin{align}\label{eq:G23-irreps}
 &[\mathbf{1},1],\ [\mathbf{4},2],\ [\mathbf{5},2],\ [\mathbf{6},1],\ [\mathbf{10},2],\ [\mathbf{15},2],\ [\mathbf{20},6],\ [\mathbf{24},1]\,,\notag\\
 &[\mathbf{30},3],\ [\mathbf{36},2],\ [\mathbf{40},2],\ [\mathbf{45},2],\ [\mathbf{60},4],\ [\mathbf{64},2],\ [\mathbf{80},1],\ [\mathbf{81},1]\,.
\end{align}
The $20$ irreducible representations of the Burkhardt group are listed in eq.~\eqref{eq:Burkhardt-irreps}. The quintet matrices are given in eqs.~\eqref{eq:rho-T}--\eqref{eq:rho-S}; the two quintets $\mathbf{5}$ and $\bar{\mathbf{5}}$ are complex conjugate. The products used in the text are
\begin{gather}
 \mathbf{5}\otimes\mathbf{5}=\mathbf{10}_A\oplus\mathbf{15}_S\,,\qquad \wedge^2\mathbf{5}=\mathbf{10}\,,\qquad \mathrm{Sym}^2\mathbf{5}=\mathbf{15}\,,\qquad \mathbf{5}\otimes\bar{\mathbf{5}}=\mathbf{1}\oplus\mathbf{24}\,,\notag\\
 \mathrm{Sym}^3\mathbf{5}=\bar{\mathbf{5}}\oplus\mathbf{30}\,,\qquad \mathrm{Sym}^4\mathbf{5}=\mathbf{1}\oplus\mathbf{15}\oplus\mathbf{24}\oplus\mathbf{30}^{\prime}\,.\label{eq:quintet-products}
\end{gather}
The singlet in $\mathrm{Sym}^4\mathbf{5}$ is $I_4$ and the $\bar{\mathbf{5}}$ in $\mathrm{Sym}^3\mathbf{5}$ is the gradient of $I_4$. The singlet in $\mathbf{5}\otimes\bar{\mathbf{5}}$ is the $G_5$-invariant Hermitian product, which gives the matter metric.

\section{Modular subsymmetries of the invariant subspaces}\label{app:subsym}
We use the generator words of appendix~\ref{app:words}. Each integral presentation is followed by the additional relations at level $N=3$. As in eq.~\eqref{eq:G23-irreps}, $[\mathbf{d},n]$ denotes $n$ inequivalent irreducible representations of dimension $d$. For each interior locus we also give the relations satisfied by the weight-1 quintet $Y^{(1)}_{\mathbf{5}}$ and the multiplets of $N_3(H)$ formed by its remaining components, as table~\ref{tab:boundary-branching} does for the boundary regions.

\subsection{Interior surfaces and curves}\label{app:interior}

\paragraph{The split surface.}
Here $\Omega=\left(\begin{smallmatrix}u&0\\0&v\end{smallmatrix}\right)$.
The pointwise stabilizer is $H=\{\pm I_4,\pm\DD\}\cong\Z_2^2$.
The two pairs $(\si,\mathsf{T}_1)$ and $(\sii,\mathsf{T}_2)$ act on $u$ and $v$, respectively, and $\PP$ exchanges them.
The defining relations are eq.~\eqref{eq:SL2-relations} for $(\si,\mathsf{T}_1)$ and for $(\sii,\mathsf{T}_2)$, and
\begin{equation}
 [\si,\sii]=[\si,\mathsf{T}_2]=[\mathsf{T}_1,\sii]=[\mathsf{T}_1,\mathsf{T}_2]=1\,,\qquad \PP^2=1\,,\qquad \PP\si\PP=\sii\,,\qquad \PP\mathsf{T}_1\PP=\mathsf{T}_2\,.
\end{equation}

At level 3 one imposes $\mathsf{T}_1^3=\mathsf{T}_2^3=1$, giving $N_3(H)\cong(T'\times T')\rtimes\Z_2\cong[1152,155468]$.

The irreducible representations are $[\mathbf{1},6]$, $[\mathbf{2},3]$, $[\mathbf{4},15]$, $[\mathbf{6},3]$, $[\mathbf{8},3]$, $[\mathbf{9},2]$, $[\mathbf{12},3]$. On this surface the quintet obeys $Y_4=Y_5$ and $Y_1Y_4=Y_2Y_3$. In the unitary basis $Y^{(1)}_{\mathbf{4}}(u,v)=(2Y_4,\,-\sqrt{2}\,Y_2,\,-\sqrt{2}\,Y_3,\,Y_1)^\top = Y^{(1,0)}_{\mathbf{2}^{(0)}}(u)\otimes Y^{(1,0)}_{\mathbf{2}^{(0)}}(v)$ of eq.~\eqref{eq:H1-factorization}, the generators act as
\begin{gather}
 \rho(\si)=-\rho_{\mathbf{2}^{(0)}}(\mathsf{S})\otimes I_2\,,\qquad \rho(\mathsf{T}_1)=\rho_{\mathbf{2}^{(0)}}(\mathsf{T})\otimes I_2\,,\notag\\
 \rho(\sii)=-I_2\otimes\rho_{\mathbf{2}^{(0)}}(\mathsf{S})\,,\qquad \rho(\mathsf{T}_2)=I_2\otimes\rho_{\mathbf{2}^{(0)}}(\mathsf{T})\,,\qquad
 \rho(\PP)=-\begin{pmatrix}1&0&0&0\\0&0&1&0\\0&1&0&0\\0&0&0&1\end{pmatrix}\,,
\end{gather}
with $\rho_{\mathbf{2}^{(0)}}(\mathsf{S})$ and $\rho_{\mathbf{2}^{(0)}}(\mathsf{T})$ as in eq.~\eqref{eq:rho2} (note $\rho_{\mathbf{2}^{(0)}}(\mathsf{S})^{-1}=-\rho_{\mathbf{2}^{(0)}}(\mathsf{S})$).

\paragraph{The equal-diagonal surface.}
Here $\Omega=\left(\begin{smallmatrix}u&v\\v&u\end{smallmatrix}\right)$.
The pointwise stabilizer is $H=\{\pm I_4,\pm\PP\}\cong\Z_2^2$.
The generators are $\mathsf{S}$, $\mathsf{t}=\mathsf{T}_1\mathsf{T}_2$, $\mathsf{T}_3$ and $\DD$. With $\mathsf{a}=\mathsf{t}\mathsf{T}_3$, $\mathsf{b}=\mathsf{S}\mathsf{a}\mathsf{S}^{-1}$ and $\mathsf{z}_0=\mathsf{t}\mathsf{b}\mathsf{t}^{-1}\mathsf{a}\mathsf{b}=-\PP$, the word $\mathsf{a}$ is a translation of $u+v$ alone, $\mathsf{b}$ is its inverted counterpart and $\mathsf{z}_0$ is the central sign on that coordinate.
The defining relations are eq.~\eqref{eq:SL2-relations} for $(\mathsf{S},\mathsf{t})$, and
\begin{gather}
 \mathsf{z}_0^2=1,\quad [\mathsf{z}_0,\mathsf{a}]=[\mathsf{z}_0,\mathsf{b}]=[\mathsf{z}_0,\mathsf{S}]=[\mathsf{z}_0,\mathsf{t}]=1,\quad \mathsf{S}\mathsf{a}\mathsf{S}^{-1}=\mathsf{b},\quad \mathsf{S}\mathsf{b}\mathsf{S}^{-1}=\mathsf{a}\,,\notag\\
 \mathsf{t}\mathsf{a}\mathsf{t}^{-1}=\mathsf{a},\quad \mathsf{t}\mathsf{b}\mathsf{t}^{-1}=\mathsf{z}_0\mathsf{b}^{-1}\mathsf{a}^{-1},\quad \DD^2=1,\quad [\DD,\mathsf{S}]=[\DD,\mathsf{t}]=1\,,\notag\\
 \DD\mathsf{a}\DD=\mathsf{t}^2\mathsf{a}^{-1},\quad \DD\mathsf{b}\DD=\mathsf{S}\mathsf{t}^2\mathsf{S}^{-1}\mathsf{b}^{-1},\quad \DD\mathsf{z}_0\DD=\mathsf{S}^2\mathsf{z}_0\,.
\end{gather}
They describe pairs of modular matrices that agree modulo two, followed by the interchange of the two coordinates.\footnote{The kernel of the reduction $\SL(2,\Z)\to\SL(2,\Z_2)$ is generated by $\mathsf{T}^2$, $\mathsf{S}\mathsf{T}^2\mathsf{S}^{-1}$ and the central $-I_2$; the first two are free generators after dividing by the central sign.}

At level 3 one imposes $\mathsf{t}^3=\mathsf{a}^3=1$, giving $N_3(H)\cong(T'\times T')\rtimes\Z_2\cong[1152,155468]$.
The representation dimensions are the same as on the split surface. Here the quintet obeys $Y_2=Y_3$. In the $G_5$-normalized basis $Y^{(1)}_{\mathbf{4}}(u,v)=(Y_1,\,2Y_2,\,\sqrt{2}\,Y_4,\,\sqrt{2}\,Y_5)^\top$, the surviving irreducible quartet transforms unitarily as
\begin{gather}
 \rho(\mathsf{S})=\frac{1}{3}\begin{pmatrix}-1&-2&-\sqrt{2}&-\sqrt{2}\\-2&-1&\sqrt{2}&\sqrt{2}\\-\sqrt{2}&\sqrt{2}&1&-2\\-\sqrt{2}&\sqrt{2}&-2&1\end{pmatrix}\,,\qquad
 \rho(\mathsf{t})=\diag(1,\omega,\omega^2,\omega^2)\,,\notag\\
 \rho(\mathsf{T}_3)=\diag(1,1,\omega,\omega^2)\,,\qquad \rho(\DD)=-\bigl(I_2\oplus\sigma_1\bigr)\,.
\end{gather}

\paragraph{The curves with one elliptic modulus fixed.}
Here $\Omega=\left(\begin{smallmatrix}i&0\\0&u\end{smallmatrix}\right)$ and $\Omega=\left(\begin{smallmatrix}\omega&0\\0&u\end{smallmatrix}\right)$.
The pointwise stabilizer is $H=\langle \mathsf{h}\rangle\times\{\pm I_4\}$ with $\mathsf{h}=\si$ (order four) or $\mathsf{h}=\si^{-1}\mathsf{T}_1$ (order six),
\begin{equation}
 \si^{-1}\mathsf{T}_1=\begin{pmatrix}0&0&-1&0\\0&1&0&0\\1&0&1&0\\0&0&0&1\end{pmatrix}\,.
\end{equation}
The generators are $\sii$, $\mathsf{T}_2$ and $\mathsf{h}$, with $n=4$ or $n=6$.
The defining relations are eq.~\eqref{eq:SL2-relations} for $(\sii,\mathsf{T}_2)$, and $\mathsf{h}^n=1$, $[\mathsf{h},\sii]=[\mathsf{h},\mathsf{T}_2]=1$.

At level 3 one imposes $\mathsf{T}_2^3=1$, giving $N_3(H)\cong\Z_4\times T'\cong[96,69]$, respectively $\Z_6\times T'\cong[144,156]$.

The irreducible representations are $[\mathbf{1},12]$, $[\mathbf{2},12]$, $[\mathbf{3},4]$, respectively $[\mathbf{1},18]$, $[\mathbf{2},18]$, $[\mathbf{3},6]$. On both curves $Y_4=Y_5$, $Y_1=c\,Y_2$ and $Y_3=c\,Y_4$, with $c=\theta_L(\tau_0)/\theta_{L+\mu_0}(\tau_0)$, namely $c=1+\sqrt{3}$ at $\tau_0=i$ and $c=1+i\sqrt{3}$ at $\tau_0=\omega$. The surviving doublet $Y^{(1)}_{\mathbf{2}}(u)=(-\sqrt{2}\,Y_4,\,Y_2)^\top=\theta_{L+\mu_0}(\tau_0)\,Y^{(1,0)}_{\mathbf{2}^{(0)}}(u)$ transforms as
\begin{gather}
 \rho(\sii)=-\rho_{\mathbf{2}^{(0)}}(\mathsf{S})\,,\qquad \rho(\mathsf{T}_2)=\rho_{\mathbf{2}^{(0)}}(\mathsf{T})=\diag(\omega,1)\,,\notag\\
 \rho(\mathsf{h})=\begin{cases}i\,I_2,&\tau_0=i\ (\mathsf{h}=\si)\,,\\-\omega\,I_2,&\tau_0=\omega\ (\mathsf{h}=\si^{-1}\mathsf{T}_1)\,.\end{cases}
\end{gather}

\paragraph{The scalar diagonal.}
Here $\Omega=\left(\begin{smallmatrix}u&0\\0&u\end{smallmatrix}\right)$.
The pointwise stabilizer consists of the eight signed permutation matrices $\gamma_A$, generated by the exchange $\PP$ and the reflection $\DD$, so $H\cong D_8$.
The generators are $\mathsf{S}$, $\mathsf{t}=\mathsf{T}_1\mathsf{T}_2$, $\PP$ and $\DD$.
The defining relations are eq.~\eqref{eq:SL2-relations} for $(\mathsf{S},\mathsf{t})$, and
\begin{equation}
 \PP^2=\DD^2=(\PP\DD)^4=1\,,\qquad \mathsf{S}^2=(\PP\DD)^2\,,\qquad [\mathsf{S},\PP]=[\mathsf{S},\DD]=[\mathsf{t},\PP]=[\mathsf{t},\DD]=1\,.
\end{equation}

At level 3 one imposes $\mathsf{t}^3=1$, giving $N_3(H)\cong(T'\times D_8)/\Z_2\cong[96,202]$.

The irreducible representations are $[\mathbf{1},12]$, $[\mathbf{3},4]$, $[\mathbf{4},3]$. On this curve $Y_2=Y_3$, $Y_4=Y_5$ and $Y_1Y_4=Y_2^2$, since $(Y_1,Y_2,Y_4)=(\theta_L^2,\theta_L\theta_{L+\mu_0},\theta_{L+\mu_0}^2)$ at argument $u$. In the $G_5$-normalized basis $Y^{(1)}_{\mathbf{3}}(u)=(2Y_4,\,-2Y_2,\,Y_1)^\top$, which is the unitary symmetric square of $Y^{(1,0)}_{\mathbf{2}^{(0)}}(u)$, the generators act by
\begin{equation}
 \rho(\mathsf{S})=\frac{1}{3}\begin{pmatrix}-1&-2&-2\\-2&-1&2\\-2&2&-1\end{pmatrix}\,,\qquad
 \rho(\mathsf{t})=\diag(\omega^2,\omega,1)\,,\qquad
 \rho(\PP)=\rho(\DD)=-I_3\,.
\end{equation}

\paragraph{The curve with $z=1/2$.}
Here $\Omega=\left(\begin{smallmatrix}u&1/2\\1/2&u\end{smallmatrix}\right)$.
The pointwise stabilizer is $H\cong D_8$, generated by $\mathsf{h}_1=\PP$ and $\mathsf{h}_2=\mathsf{T}_3\DD$.
The generators are $\mathsf{h}_1$, $\mathsf{h}_2$, $\mathsf{t}=\mathsf{T}_1\mathsf{T}_2$ and the Fricke lift
\begin{equation}
 \mathsf{w}=\mathsf{T}_1\si^{-1}\mathsf{T}_2\sii^{-1}\EE\RR^{-1}\mathsf{T}_3\mathsf{T}_2^2\sii=\begin{pmatrix}1&0&-1&-1\\1&0&-1&0\\1&1&-1&-1\\1&-1&0&0\end{pmatrix}\,.
\end{equation}
On $t'=u-1/2$, $\mathsf{t}$ acts as $t'\to t'+1$ and $\mathsf{w}$ as the Fricke map $t'\to-1/(2t')$.
The defining relations are
\begin{gather}
 \mathsf{h}_1^2=\mathsf{h}_2^2=(\mathsf{h}_1\mathsf{h}_2)^4=1\,,\qquad [\mathsf{t},\mathsf{h}_1]=[\mathsf{t},\mathsf{h}_2]=1\,,\notag\\
 \mathsf{w}\mathsf{h}_1\mathsf{w}^{-1}=\mathsf{h}_2\,,\qquad \mathsf{w}\mathsf{h}_2\mathsf{w}^{-1}=\mathsf{h}_1\,,\qquad \mathsf{w}^2=(\mathsf{h}_1\mathsf{h}_2)^2\,,\qquad (\mathsf{w}\mathsf{t})^4=(\mathsf{h}_1\mathsf{h}_2)^2\,.
\end{gather}
After dividing by $\langle \mathsf{h}_1,\mathsf{h}_2\rangle$ these are the relations $\mathsf{w}^2=(\mathsf{w}\mathsf{t})^4=1$ of $\Gamma_0(2)^+$.

At level 3 one imposes $\mathsf{t}^3=1$, giving $N_3(H)\cong(T'\rtimes\Z_4)\rtimes\Z_2\cong[192,989]$.

The irreducible representations are $[\mathbf{1},4]$, $[\mathbf{2},5]$, $[\mathbf{3},4]$, $[\mathbf{4},2]$, $[\mathbf{6},1]$, $[\mathbf{8},1]$. On this curve $Y_2=Y_3$ and $Y_4=\omega Y_5$. In the $G_5$-normalized basis $Y^{(1)}_{\mathbf{3}}(t')=(Y_1,\,2Y_2,\,2Y_4)^\top$, the surviving irreducible triplet transforms unitarily as
\begin{equation}
 \rho(\mathsf{h}_1)=\rho(\mathsf{h}_2)=-I_3\,,\qquad \rho(\mathsf{t})=\diag(1,\omega,\omega^2)\,,\qquad
 \rho(\mathsf{w})=\frac{1}{3}\begin{pmatrix}-1&-2\omega&-2\\-2\omega^2&2&-\omega^2\\-2&-\omega&2\end{pmatrix}\,.
\end{equation}

\paragraph{The hexagonal curve.}
Here $\Omega=\frac{u}{2}\left(\begin{smallmatrix}2&1\\1&2\end{smallmatrix}\right)$.
The pointwise stabilizer is $H\cong D_{12}$, the automorphism group of the hexagonal form, generated by $\mathsf{r}=\RR\EE^{-1}=\gamma_{\left(\begin{smallmatrix}0&1\\-1&1\end{smallmatrix}\right)}$ (order six) and $\PP$.
The generators are $\mathsf{r}$, $\PP$, $\mathsf{w}=\DD\mathsf{S}$ and $\mathsf{t}=\mathsf{T}_1^2\mathsf{T}_2^2\mathsf{T}_3$. On $t'=u/2$, $\mathsf{t}$ acts as $t'\to t'+1$ and $\mathsf{w}$ as the Fricke map $t'\to-1/(3t')$.
The defining relations are
\begin{gather}
 \mathsf{r}^6=\PP^2=1\,,\qquad \PP\mathsf{r}\PP=\mathsf{r}^{-1}\,,\qquad [\mathsf{t},\mathsf{r}]=[\mathsf{t},\PP]=1\,,\notag\\
 \mathsf{w}\mathsf{r}\mathsf{w}^{-1}=\mathsf{r}^{-1}\,,\qquad \mathsf{w}\PP\mathsf{w}^{-1}=\mathsf{r}^3\PP\,,\qquad \mathsf{w}^2=(\mathsf{w}\mathsf{t})^6=\mathsf{r}^3\,.
\end{gather}
After dividing by the pointwise group these are the relations $\mathsf{w}^2=(\mathsf{w}\mathsf{t})^6=1$ of $\Gamma_0(3)^+$.

At level 3 one imposes $\mathsf{t}^3=1$ and $(\mathsf{w}\mathsf{t})\mathsf{x}(\mathsf{w}\mathsf{t})^{-1}=\mathsf{x}^{-1}$, with $\mathsf{x}=\mathsf{w}\mathsf{t}\mathsf{w}^{-1}\mathsf{t}^{-1}$, giving $N_3(H)\cong\Z_3\times((\Z_6\times S_3)\rtimes\Z_2)\cong[216,121]$.

The irreducible representations are $[\mathbf{1},12]$, $[\mathbf{2},27]$, $[\mathbf{4},6]$. On this curve $Y_2=Y_3=Y_4$. The three surviving components split into the $G_5$-orthogonal singlet $Y^{(1)}_{\mathbf{1}}(t')=Y_1+2Y_5$ and unitary doublet $Y^{(1)}_{\mathbf{2}}(t')=(Y_1-Y_5,\,3Y_2)^\top$. In the basis $(Y^{(1)}_{\mathbf{1}};\,Y^{(1)}_{\mathbf{2}})$, the generators act by
\begin{equation}
 \rho(\mathsf{r})=1\oplus I_2\,,\qquad \rho(\PP)=(-1)\oplus(-I_2)\,,\qquad
 \rho(\mathsf{t})=1\oplus\diag(1,\omega^2)\,,\qquad \rho(\mathsf{w})=1\oplus\sigma_1\,.
\end{equation}

Notice the difference between these decompositions and those of table~\ref{tab:boundary-branching}. On an interior locus $\mathcal{F}_H$ the pointwise stabilizer imposes $Y^{(1)}_{\mathbf{5}}=j(h,\Omega)\,\rho_{\mathbf{5}}(h)\,Y^{(1)}_{\mathbf{5}}$ for all $h\in H$, so the components outside the joint $+1$ eigenspace vanish identically on $\mathcal{F}_H$ and reappear only away from it. Near a cusp no such pointwise stabilizer exists: in an open neighborhood all five functions remain linearly independent and form complete multiplets, separated by their Fourier orders in $q_a$, and only in the strict limit $q_a\to0$ does the number of nonvanishing components drop, to two at Klingen and to one at the Siegel cusp.

\subsection{Cusp regions}\label{app:cusp-groups}

\paragraph{Klingen region.}
The generators are $\mathsf{s}_J=\si^{-1}$, $\mathsf{t}_J=\mathsf{T}_1$, $\mathsf{X}=\FF$, $\mathsf{Y}=\mathsf{T}_3$, $\mathsf{Z}=\mathsf{T}_2$, together with the central $\mathsf{S}^2=-I_4$.
The defining relations are
\begin{gather}
 \mathsf{s}_J^4=1,\quad (\mathsf{s}_J\mathsf{t}_J)^3=\mathsf{s}_J^2,\quad [\mathsf{s}_J^2,\mathsf{t}_J]=1,\quad [\mathsf{X},\mathsf{Z}]=[\mathsf{Y},\mathsf{Z}]=1,\quad \mathsf{X}\mathsf{Y}=\mathsf{Z}^2\mathsf{Y}\mathsf{X}\,,\notag\\
 \mathsf{s}_J\mathsf{X}\mathsf{s}_J^{-1}=\mathsf{Y},\quad \mathsf{s}_J\mathsf{Y}\mathsf{s}_J^{-1}=\mathsf{X}^{-1},\quad [\mathsf{s}_J,\mathsf{Z}]=1\,,\notag\\
 \mathsf{t}_J\mathsf{X}\mathsf{t}_J^{-1}=\mathsf{Z}\mathsf{X}\mathsf{Y}^{-1},\quad [\mathsf{t}_J,\mathsf{Y}]=[\mathsf{t}_J,\mathsf{Z}]=1\,.
\end{gather}
The extra generator $\mathsf{S}^2$ has order two and commutes with all five Jacobi generators. The factor $\mathsf{s}_J^2$ in the second relation reflects the sign of the partial inversion relative to eq.~\eqref{eq:SL2-relations}. For a general embedded modular matrix $\gamma_1\in\SL(2,\Z)$ the action on the Heisenberg group is
\begin{equation}
 G_{\gamma_1} H(\lambda,\mu,\kappa)G_{\gamma_1}^{-1}=H(\lambda',\mu',\kappa')\,,\qquad
 \binom{\lambda'}{\mu'}=\gamma_1^{-\top}\binom\lambda\mu\,,\qquad \kappa'=\kappa-\lambda\mu+\lambda'\mu'\,.
\end{equation}

At level 3 one imposes $\mathsf{t}_J^3=\mathsf{X}^3=\mathsf{Y}^3=\mathsf{Z}^3=1$. The group is $J_3=\Delta(27)\rtimes T'\cong[648,533]$, and with $\mathsf{S}^2$ it is $J_3\times\Z_2\cong[1296,2895]$.

The irreducible representations of $J_3$ are $[\mathbf{1},3]$, $[\mathbf{2},3]$, $[\mathbf{3},7]$, $[\mathbf{6},6]$, $[\mathbf{8},3]$, $[\mathbf{9},2]$.

For the metaplectic lifts $\tilde{\mathsf{s}}_J,\tilde{\mathsf{t}}_J$ \cite{Stromberg:2013} the relation $\mathsf{s}_J^4=1$ is replaced by $\tilde{\mathsf{s}}_J^8=1$, the other relations being unchanged. A faithful triplet multiplied by $\eps$ has a finite image with the additional relations $\tilde{\mathsf{t}}_J^3=\mathsf{u}$, $\mathsf{u}^8=1$, $\tilde{\mathsf{s}}_J^4=\mathsf{u}^4$ and $\mathsf{u}$ central; since $\eps(\mathsf{s}_J)=e^{-\pi i/4}$ and $\eps(\mathsf{t}_J)=e^{\pi i/12}$, $\mathsf{u}=e^{\pi i/4}I_3$ and the image is $J_3\times\Z_8$.

\paragraph{Siegel cusp.}
The generators are $\mathsf{T}_1,\mathsf{T}_2,\mathsf{T}_3$ and $\RR, \EE, \DD$.
The relations are eq.~\eqref{eq:translation-conj} together with
\begin{equation}
 \RR^4=(\RR\EE)^3=[\RR^2,\EE]=1\,,\qquad \DD^2=1\,,\qquad \DD\RR\DD=\RR^{-1}\,,\qquad \DD\EE\DD=\EE^{-1}\,,
\end{equation}
which define $\GL(2,\Z)$.

At level 3 one imposes $\mathsf{T}_i^3=\EE^3=1$, giving $\Z_3^3\rtimes\GL(2,\Z_3)\cong[1296,3085]$. Adding $\RR^2=1$ gives the projective group $\Z_3^3\rtimes S_4\cong[648,704]$.

The irreducible representations of the projective group are $[\mathbf{1},2]$, $[\mathbf{2},1]$, $[\mathbf{3},2]$, $[\mathbf{4},4]$, $[\mathbf{6},4]$, $[\mathbf{8},2]$, $[\mathbf{12},2]$.

\paragraph{Two-ray region.}
The generators are $\mathsf{T}_1,\mathsf{T}_2,\mathsf{T}_3$ and $\PP, \DD$.
The relations are eq.~\eqref{eq:translation-conj} together with $\PP^2=\DD^2=(\PP\DD)^4=1$.

At level 3 one imposes $\mathsf{T}_i^3=1$, giving $\Z_3^3\rtimes D_8\cong\Z_3\times((\Z_6\times S_3)\rtimes\Z_2)\cong[216,121]$ (isomorphic to the hexagonal-curve group of table~\ref{tab:interior}); dividing by $(\PP\DD)^2=-I_4$ gives $\Z_3\times S_3\times S_3\cong[108,38]$.

The irreducible representations of the projective group are $[\mathbf{1},12]$, $[\mathbf{2},12]$, $[\mathbf{4},3]$.

\paragraph{Corner.}
The generators are $\mathsf{T}_1,\mathsf{T}_2,\mathsf{T}_3$ and $\CC, \PP$.
The relations are eq.~\eqref{eq:translation-conj} together with $\CC^6=\PP^2=1$, $\PP\CC\PP=\CC^{-1}$.

At level 3 one imposes $\mathsf{T}_i^3=1$, giving $\Z_3^3\rtimes D_{12}\cong[324,68]$; adding $\CC^3=1$ gives $\Z_3^3\rtimes S_3\cong[162,10]$.

The irreducible representations of the projective group are $[\mathbf{1},6]$, $[\mathbf{2},3]$, $[\mathbf{3},12]$, $[\mathbf{6},1]$.

\section{Representations used in the flavor models}\label{app:reps}
Table~\ref{tab:matter-images} gives the image of the finite modular group on the matter fields in each model. The matter triplets in the corner and Jacobi models are faithful representations of $\Z_3^3\rtimes S_3$ and $J_3$, so their matrix images coincide with the groups in table~\ref{tab:cusp-groups}. The $\mathbf{1}\oplus\mathbf{2}$ assignments in the interior and two-ray models have nontrivial kernels, and their matrix images are the smaller groups $[18,3]$ and $[36,12]$ in the first two rows.

\begin{table}[t]
\centering\small\tablesetup[1.3]
\begin{tabular}{lll}
\toprule
flavor model & matter representation & finite image group\\
\midrule
interior, section~\ref{sec:interior-model} & $\mathbf{1}_0\oplus\mathbf{2}_1$ & $\Z_3\times S_3\cong[18,3]$\\
two-ray, section~\ref{sec:two-ray-model} & $\mathbf{1}_3\oplus\mathbf{2}_5$ & $\Z_6\times S_3\cong[36,12]$\\
corner, sections~\ref{sec:corner-common} and \ref{sec:factorised} & $\mathbf{3}^-_{a,b}$ & $\Z_3^3\rtimes S_3\cong[162,10]$\\
Jacobi, section~\ref{sec:FN} & $\mathbf{3}^{(p)}$ & $\Delta(27)\rtimes T'\cong[648,533]$\\
Jacobi with odd $\eps$ multiplier & $\mathbf{3}^{(p)}_{\eps^{\pm1}}$ & $J_3\times\Z_8$\\
\bottomrule
\end{tabular}
\caption{Finite images of the representation matrices used for matter fields in each flavor model. The last row includes the scalar eta multiplier (appendix~\ref{app:cusp-groups}); there is no SmallGroups identifier at that order. They should not be confused with the finite modular groups of tables~\ref{tab:interior} and \ref{tab:cusp-groups}. The table lists the image of $\rho$; in the two-ray model the complete weight-1 field transformation $j^{-1}\rho$ has image $[18,3]$.}\label{tab:matter-images}
\end{table}
\subsection{The finite Jacobi group \texorpdfstring{$J_3$}{J3}}\label{app:J3}
We use the generators $\mathsf{s}_J,\mathsf{t}_J,\mathsf{X},\mathsf{Y},\mathsf{Z}$ of appendix~\ref{app:cusp-groups}.
\paragraph{Singlets and doublets.}
For $\mathbf{1}^{(p)}$, $p=0,1,2$, $\rho(\mathsf{t}_J)=\omega^p$, $\rho(\mathsf{s}_J)=\rho(\mathsf{X})=\rho(\mathsf{Y})=\rho(\mathsf{Z})=1$.
The doublet $\mathbf{2}^{(0)}$ has $\rho(\mathsf{s}_J)$ and $\rho(\mathsf{t}_J)$ as in eq.~\eqref{eq:rho2} and $\rho(\mathsf{X})=\rho(\mathsf{Y})=\rho(\mathsf{Z})=I_2$; the twisted doublets $\mathbf{2}^{(p)}$ ($p=1,2$) have $\rho(\mathsf{t}_J)$ multiplied by $\omega^p$ with all other matrices unchanged.
\paragraph{Triplets and sextets.}
The triplet $\mathbf{3}^{(0)}$ has the matrices eq.~\eqref{eq:rho3}. The twisted triplets $\mathbf{3}^{(p)}$ have $\rho(\mathsf{t}_J)$ multiplied by $\omega^p$, all other matrices unchanged, and $[\mathbf{3}^{(p)}]^*=\bar{\mathbf{3}}^{(-p)}$, where barred triplets have $\rho(\mathsf{Z})=\omega^2I_3$.
For $\mathbf{6}^{(p)}$ the matrix is $\rho_{\mathbf{2}^{(0)}}(g)\otimes\rho_{\mathbf{3}^{(p)}}(g)$. We order the components as $(x_1v_1,x_1v_2,x_1v_3,x_2v_1,x_2v_2,x_2v_3)$.
\paragraph{Tensor products.}
In the conventions of Feruglio and Marrone the superscripts are added modulo three. The products used here are
\begin{gather}
 \mathbf{2}^{(p)}\otimes\mathbf{2}^{(q)}=\mathbf{1}^{(p+q+1)}\oplus\mathbf{3}_{\rm mod}\,,\qquad
 \mathbf{2}^{(p)}\otimes\mathbf{3}^{(q)}=\mathbf{6}^{(p+q)}\,,\notag\\
 \mathbf{3}^{(p)}\otimes\mathbf{3}^{(q)}=\bar{\mathbf{3}}_A^{(p+q+2)}\oplus\bar{\mathbf{6}}_S^{(p+q)}\,,\qquad
 \mathbf{3}^{(p)}\otimes\bar{\mathbf{3}}^{(q)}=\mathbf{1}^{(p+q)}\oplus\mathbf{8}^{(p+q)}\,.
\end{gather}
Here $\mathbf{3}_{\rm mod}=\mathrm{Sym}^2\mathbf{2}^{(0)}$, which needs no superscript since $\mathbf{3}_{\rm mod}\otimes\mathbf{1}^{(r)}\cong\mathbf{3}_{\rm mod}$, is neutral under $\mathsf{X},\mathsf{Y},\mathsf{Z}$ and is the triplet of $A_4$ up to a change of basis. The singlet contraction of two doublets is $x_1y_2-x_2y_1$, and the symmetric triplet is spanned by $x_1y_1$, $(x_1y_2+x_2y_1)/\sqrt{2}$, $x_2y_2$.

The symmetric sextet in two triplets that enters the mass terms is
\begin{equation}
 B(\chi,\psi)=\Bigl(-\tfrac{\chi_2\psi_3+\chi_3\psi_2}{\sqrt{2}},\ -\tfrac{\chi_1\psi_3+\chi_3\psi_1}{\sqrt{2}},\ -\tfrac{\chi_1\psi_2+\chi_2\psi_1}{\sqrt{2}},\ \chi_1\psi_1,\ \chi_2\psi_2,\ \chi_3\psi_3\Bigr)^\top\,.
\end{equation}
For a sextet $R$, contraction with this vector gives $R^\top B(\chi,\psi)=\chi^\top\mathcal{Y}[R]\psi$, with the matrix in eq.~\eqref{eq:MXY}. Scalar eta multipliers multiply these representations in the complete field law; in particular two $\mathbf{3}^{(1)}_{\eps^{-1}}$ fields require a $\mathbf{6}^{(1)}_{\eps^2}$ in the Majorana term.

\subsection{Matter singlet and doublet on the equal-diagonal surface}\label{app:reps-interior}
The matrices eq.~\eqref{eq:interior-matter-rep} generate $\Z_3\times S_3\cong[18,3]$ with presentation
\begin{equation}
 \Z_3\times S_3=\langle \mathsf{c},\mathsf{r},\mathsf{p}:\ \mathsf{c}^3=\mathsf{r}^3=\mathsf{p}^2=1,\ [\mathsf{c},\mathsf{r}]=[\mathsf{c},\mathsf{p}]=1,\ \mathsf{p}\mathsf{r}\mathsf{p}=\mathsf{r}^{-1}\rangle\,,
\end{equation}
where $\mathsf{c}=\rho(\mathsf{G}_1)$, $\mathsf{r}=\rho(\mathsf{G}_2)$ and $\mathsf{p}=\rho(\mathsf{G}_4)$. The singlets $\mathbf{1}_a,\mathbf{1}'_a$ carry $\Z_3$ charge $a$ and $\mathsf{p}$-parity $\pm1$, while the doublets $\mathbf{2}_a$ have
\begin{equation}\label{eq:Z3S3-doublets}
 \rho_{\mathbf{2}_a}(\mathsf{c})=\omega^a I_2\,,\qquad \rho_{\mathbf{2}_a}(\mathsf{r})=\diag(\omega,\omega^2)\,,\qquad \rho_{\mathbf{2}_a}(\mathsf{p})=\sigma_1\,.
\end{equation}
With charges added modulo three, the tensor products and \ac{CG} contractions for $x=(x_1,x_2)^\top\sim\mathbf{2}_a$ and $y=(y_1,y_2)^\top\sim\mathbf{2}_b$ are
\begin{align}
 \mathbf{1}_a\otimes\mathbf{2}_b&=\mathbf{2}_{a+b}\,,\qquad \mathbf{2}_a\otimes\mathbf{2}_b=\mathbf{1}_{a+b}\oplus\mathbf{1}'_{a+b}\oplus\mathbf{2}_{a+b}\,,\notag\\
 \mathbf{1}_{a+b}&:\ x_1y_2+x_2y_1\,,\qquad \mathbf{1}'_{a+b}:\ x_1y_2-x_2y_1\,,\qquad \mathbf{2}_{a+b}:\ (x_2y_2,\,x_1y_1)^\top\,.
\end{align}
For matter $\mathbf{1}_0\oplus\mathbf{2}_1$, the weight-4 forms $\mathbf{1}_0\oplus\mathbf{1}_1\oplus\mathbf{2}_2$ of eq.~\eqref{eq:interior-quartics} yield the four coefficients of eq.~\eqref{eq:interior-texture}; the two orientations of the singlet--doublet term are independent for Dirac couplings and coincide for Majorana couplings.

\subsection{Matter singlet and doublet in the two-ray model}\label{app:reps-tworay}
The matrices eq.~\eqref{eq:curve-model-rep} generate $\Z_6\times S_3\cong[36,12]$ with commuting generator $\mathsf{c}=\rho(\DD\mathsf{T}_1\mathsf{T}_2)=\diag(-1,-\omega,-\omega)$ and $S_3$ generators $\mathsf{r}=\rho(\mathsf{T}_1\mathsf{T}_2^{-1})$, $\mathsf{p}=\rho(\PP\DD)$. The labels $\mathbf{1}_a,\mathbf{2}_a$ specify the $\mathsf{c}$-eigenvalue $e^{\pi ia/3}$ with the same $\mathsf{r},\mathsf{p}$ matrices as above, so the matter fields transform as $\mathbf{1}_3\oplus\mathbf{2}_5$ and the tensor products of appendix~\ref{app:reps-interior} hold with charges modulo six.

The weight-2 forms on which the kernel of the matter representation acts trivially decompose as
\begin{equation}
 \mathbf{1}_0:\ Y_1^2\,,\qquad \mathbf{1}_2:\ Y_4Y_5\,,\qquad \mathbf{1}_4:\ Y_2Y_3\,,\qquad \mathbf{2}_2:\ (Y_1Y_2,Y_1Y_3)\,,\qquad \mathbf{2}_4:\ (Y_3^2,Y_2^2)\,.
\end{equation}
The matter bilinear decomposes as
\begin{equation}\label{eq:tworay-bilinear}
 (\mathbf{1}_3\oplus\mathbf{2}_5)^{\otimes2}=\mathbf{1}_0\oplus 2\cdot\mathbf{2}_2\oplus\mathbf{1}_4\oplus\mathbf{1}'_4\oplus\mathbf{2}_4\,.
\end{equation}
Pairing with the conjugate forms leaves five Dirac invariants and four symmetric Majorana invariants (the form $Y_2Y_3$ in $\mathbf{1}_4$ does not pair), reproducing the parameter count of eq.~\eqref{eq:curve-model-Y}. Including the weight-1 factor $j^{-1}$ in the field transformation cancels the common sign and reduces the image to $[18,3]$.

\subsection{Matter triplets in the corner models}\label{app:reps-corner}
We use the translations $\mathsf{T}_B$ of eq.~\eqref{eq:corner-action}, with phases specified by $t_1,t_2,t_3$. For $a,b\in\Z_3$ with $a\neq b$ and a sign $\epsilon=\pm1$, the twelve triplets $\mathbf{3}^\epsilon_{a,b}$ of $\Z_3^3\rtimes S_3$ are given by
\begin{gather}
 \rho_{\mathbf{3}^\epsilon_{a,b}}(\mathsf{T}_B)=\diag\Bigl(\omega^{at_i+b\sum_{j\neq i}t_j}\Bigr)_{i=1,2,3}\,,\notag\\
 \rho_{\mathbf{3}^\epsilon_{a,b}}(\PP)=\epsilon\begin{pmatrix}0&1&0\\1&0&0\\0&0&1\end{pmatrix}\,,\qquad
 \rho_{\mathbf{3}^\epsilon_{a,b}}(\CC)=\begin{pmatrix}0&0&1\\1&0&0\\0&1&0\end{pmatrix}\,.\label{eq:corner-triplets}
\end{gather}
The remaining irreducible representations are six singlets $\mathbf{1}^\pm_c$ (character $\omega^{c\sum t_i}$ and $\PP=\pm1$), three doublets $\mathbf{2}_c$ (character $\omega^{c\sum t_i}$ times the $S_3$ doublet) and one sextet $\mathbf{6}$ carrying the six permutations of $(0,1,2)$. Complex conjugation sends $(a,b)\to(-a,-b)$ and preserves $\epsilon$.

The theta functions of weight 1 decompose as $Y_1\sim\mathbf{1}^-_0$, $Y_4\sim\mathbf{1}^-_1$ and $(Y_3,Y_2,Y_5)^\top\sim\mathbf{3}^-_{0,1}$. The space of weight 2 decomposes into three singlets and four triplets:
\begin{align}
 \mathbf{1}^+_0&:\ Y_1^2\,, & \mathbf{1}^+_1&:\ Y_1Y_4\,, & \mathbf{1}^+_2&:\ Y_4^2\,,\notag\\
 \mathbf{3}^+_{0,1}&:\ Y_1(Y_3,Y_2,Y_5)\,, & \mathbf{3}^+_{1,2}&:\ Y_4(Y_3,Y_2,Y_5)\,,\notag\\
 \mathbf{3}^+_{0,2}&:\ (Y_3^2,Y_2^2,Y_5^2)\,, & \mathbf{3}^+_{2,1}&:\ (Y_2Y_5,Y_3Y_5,Y_2Y_3)\,.
\end{align}
The triplet products $\mathbf{3}^\epsilon_{a,b}\otimes\mathbf{3}^\epsilon_{a,b}=\mathbf{3}^+_{2a,2b}\oplus\mathbf{3}^+_{2b,a+b}\oplus\mathbf{3}^-_{2b,a+b}$ and $\mathbf{3}^-_{2,1}\otimes\mathbf{3}^-_{1,0}=\mathbf{3}^+_{0,1}\oplus\mathbf{3}^+_{1,2}\oplus\mathbf{3}^-_{1,2}$ use the same \ac{CG} coefficients,
\begin{equation}
 (x_1y_1,\,x_2y_2,\,x_3y_3)^\top\,,\qquad \frac{1}{\sqrt{2}}\bigl(x_2y_3\pm x_3y_2,\ x_3y_1\pm x_1y_3,\ x_1y_2\pm x_2y_1\bigr)^\top\,.
\end{equation}

Common $\mathbf{3}^-_{0,1}$ matter pairs with $Y_1(Y_3,Y_2,Y_5)$ and $Y_4(Y_3,Y_2,Y_5)$, giving eq.~\eqref{eq:corner-model-Y}. The mixed product $\mathbf{3}^-_{2,1}\otimes\mathbf{3}^-_{1,0}$ pairs with $(Y_3^2,Y_2^2,Y_5^2)$ and $(Y_2Y_5,Y_3Y_5,Y_2Y_3)$, giving eq.~\eqref{eq:factorised-Y}. For two right-handed-neutrino triplets $\mathbf{3}^-_{2,1}$ only $(Y_2Y_5,Y_3Y_5,Y_2Y_3)$ is available in the symmetric channel, which gives the single diagonal Majorana coupling.

\section{Explicit Jacobi forms and conventions}\label{app:forms}
\paragraph{Explicit multiplets.}
We write $q=e^{2\pi i\tau}$ and $\xi=e^{2\pi iw}$. The index-0 doublet $Y^{(1,0)}_{\mathbf{2}^{(0)}}$ is given in eq.~\eqref{eq:D-doublet}. Define the odd Jacobi theta function and the theta components of index $m$,
\begin{align}
 \vartheta(\tau,w)&=\sum_{n\in\Z}(-1)^nq^{(n+1/2)^2/2}\xi^{n+1/2}\,,\notag\\
 \vartheta_{m,r}(\tau,w)&=\sum_{n\in\Z}q^{(2mn+r)^2/(4m)}\xi^{2mn+r}\,,\qquad r\bmod2m\,.
\end{align}
Feruglio and Marrone build the triplets from two theta blocks \cite{Gritsenko:2026tb},
\begin{equation}\label{eq:theta-blocks}
 Q_{1,1}(\tau,w)=\frac{\vartheta(\tau,w)^2\vartheta(\tau,2w)}{\eta(\tau)}\,,\qquad
 Q_{2,2}(\tau,w)=Q_{1,1}(\tau,2w)\,.
\end{equation}
To project onto definite translation characters, introduce
\begin{equation}\label{eq:Jacobi-projectors}
 (\mathcal{U}f)(\tau,w)=f(\tau,w+1/3)\,,\qquad
 \mathcal{P}_a=\tfrac{1}{3}\sum_{j=0}^2\omega^{-aj}\mathcal{U}^j\,,
\end{equation}
and
\begin{equation}
 \mathcal{V}_mf(\tau,w)=e^{2\pi im(\tau/9+2w/3)}f(\tau,w+\tau/3)\,.
\end{equation}
For $a=1,2$ set $m=3a^2$. The triplets $Y^{(1,3)}_{\mathbf{3}^{(0)}}$ and $Y^{(1,12)}_{\mathbf{3}^{(0)}}$ are respectively
\begin{equation}\label{eq:triplet-seed}
 Y^{(1,m)}_{\mathbf{3}^{(0)}}=\begin{pmatrix}\mathcal{P}_0\mathcal{V}_mQ_{a,a}\\\mathcal{P}_1Q_{a,a}\\-\mathcal{P}_2Q_{a,a}\end{pmatrix}\,.
\end{equation}
For the doublet of index $m=9$, set $C_r(\tau)=\vartheta_{3,r}(\tau,0)$, $A_9=\vartheta_{9,0}+\vartheta_{9,6}+\vartheta_{9,12}$ and $B_9=\vartheta_{9,3}+\vartheta_{9,9}+\vartheta_{9,15}$, with arguments $(\tau,w)$. Then
\begin{equation}\label{eq:Eexplicit}
 Y^{(1,9)}_{\mathbf{2}^{(0)}}=\begin{pmatrix}-\sqrt{2}\,(C_1B_9+C_2A_9)\\C_0A_9+C_3B_9\end{pmatrix}\,,\qquad Y^{(1,9)}_{\mathbf{2}^{(0)}}(\tau,0)=Y^{(1,0)}_{\mathbf{2}^{(0)}}(\tau)\,.
\end{equation}
These expressions follow from the lattice sums eq.~\eqref{eq:Theta} by writing $\ell$ in terms of two orthogonal vectors of $A_2$, of length squared two and six, which turns the two-dimensional sums into products of one-dimensional theta series \cite{Skoruppa:2008cw}. For the triplets one uses the product formula
\begin{align}\label{eq:A2-product}
 f_1(\tau,3w)-f_2(\tau,3w)
 &=q^{1/3}(\xi+\xi^{-1}-2)(\xi-\xi^{-1})\notag\\
 &\quad\times\prod_{n\ge1}\left[(1-q^n)^2
 \prod_{\epsilon=\pm1}(1-q^n\xi^\epsilon)^2(1-q^n\xi^{2\epsilon})\right]\,,
\end{align}
whose right-hand side is $Q_{1,1}$ by the Jacobi triple product. The projections in eq.~\eqref{eq:triplet-seed} then give eq.~\eqref{eq:FG-f}, including signs.

\section{Fourier coefficients at weight 2}\label{app:coefficient}
\paragraph{The missing singlet and the fixed Yukawa ratio.}
The restrictions of section~\ref{sec:weight2} can be seen from the first Fourier coefficients alone. Writing a scalar Jacobi function as $f=\sum c_f(n,\ell)\,q^n\xi^\ell$, define
\begin{equation}\label{eq:low-coefficient-test}
 L_9(f)=c_f(\tfrac{1}{3},0)-4\,c_f(\tfrac{1}{3},3)\,,\qquad
 L_{12}(f)=2\,c_f(\tfrac{1}{3},0)-3\,c_f(\tfrac{1}{3},3)\,.
\end{equation}
For all fifteen products of two theta functions, $L_9$ vanishes on the coefficient of $q_\sigma^3$ and $L_{12}$ on the coefficient of $q_\sigma^4$. On the other hand $L_9(Y^{(2,9)}_{\mathbf{1}^{(1)}})=-6\sqrt{2}$, $L_{12}(Y^{(1,0)}_{\mathbf{2}^{(0)},1}Y^{(1,12)}_{\mathbf{3}^{(0)},1})=-6\sqrt{2}$ and $L_{12}(Y^{(1,9)}_{\mathbf{2}^{(0)},1}Y^{(1,3)}_{\mathbf{3}^{(0)},1})=-\sqrt{2}$. Hence $Y^{(2,9)}_{\mathbf{1}^{(1)}}$ is not a coefficient, and $L_{12}(\alpha Y^{(1,0)}_{\mathbf{2}^{(0)},1}Y^{(1,12)}_{\mathbf{3}^{(0)},1}+\beta Y^{(1,9)}_{\mathbf{2}^{(0)},1}Y^{(1,3)}_{\mathbf{3}^{(0)},1})=-\sqrt{2}(6\alpha+\beta)$ vanishes only for $\beta=-6\alpha$, which is the ratio $-6$ of eq.~\eqref{eq:fixed-sextet}.

\paragraph{The neutrino sextet.}
The Majorana coupling of Feruglio and Marrone has weight $k=2$, index $m=12$ and the multiplier $\eps(\gamma_1)^2$ of $\eta(\tau)^2$, with $\eps(\mathsf{T})=e^{\pi i/12}$ and $\eps(\mathsf{S})=e^{-\pi i/4}$ \cite{Almumin:2021fbk}. In appendix~C of ref.~\cite{Feruglio:2026xxo}, each component is a sum of products of theta functions of indices $2,4,6,12$ with characteristics $(a,b,c,r)$ satisfying $6a^2+3b^2+2c^2+r^2\equiv 4\pmod{16}$, so all Fourier exponents lie in $\tfrac{1}{12}+\tfrac{1}{3}\Z$ (for instance, the second component begins with $\sqrt{2}\,q^{1/12}(\xi-\xi^{-2})$). Hence
\begin{equation}\label{eq:T3twist}
 Y^{(2,12)}_{\mathbf{6}^{(1)}}[\eps^2](\tau+3,w)=i\,Y^{(2,12)}_{\mathbf{6}^{(1)}}[\eps^2](\tau,w)\,.
\end{equation}
Since $\tau\to\tau+3$ lies in $\Gamma_2(3)$, every Fourier--Jacobi coefficient of a level-3 Siegel form with trivial multiplier is invariant under it, excluding the sextet from this image; a term $q^{1/12}$ requires a principal level divisible by 12.

\paragraph{Theta series at levels 12 and 48.}
The level-3 restrictions disappear for larger lattices. For an even positive-definite lattice $M$ the theta series
\begin{equation}\label{eq:general-theta}
 \Theta^M_{\alpha,\beta}(\Omega)=\sum_{\ell\in M+\alpha}\ \sum_{\ell'\in M+\beta}e^{\pi i\tau\ell^2+2\pi iz\,\ell\cdot\ell'+\pi i\sigma\ell'^2}
\end{equation}
form a vector-valued Siegel modular form of weight $\operatorname{rank}(M)/2$ \cite{Mueller:2024thesis}. Take $M_{12}=2(L\oplus L)$, of rank four and level $12$. Let $\ell$ run over $(L+j\mu_0)\oplus(L+k\mu_0)$ and $\ell'$ over a single coset $v+M_{12}$. For the cosets $v=(0,-2\mu_0)$ and $v=(\alpha_1,\mu_0)$, with $\alpha_1$ a root of $A_2$ and $z=3w$, the lowest vectors have $\ell'^2=8/3$, and the leading Fourier--Jacobi coefficients are $3\,\widetilde{Y}^{(1,0)}_{\mathbf{2}^{(0)}}\otimes Y^{(1,12)}_{\mathbf{3}^{(0)}}$ and $-2\,\widetilde{Y}^{(1,9)}_{\mathbf{2}^{(0)}}\otimes Y^{(1,3)}_{\mathbf{3}^{(0)}}$, where the doublets are normalized as $\widetilde{X}=(-X_1/\sqrt{2},X_2)$. These twelve functions are independent, so $r_f$ becomes free. For $v=(0,\alpha_1)$ the lowest vectors have $\ell'^2=2$, and the leading coefficient $2\,\widetilde{Y}^{(1,0)}_{\mathbf{2}^{(0)}}\otimes\widetilde{Y}^{(1,9)}_{\mathbf{2}^{(0)}}$ contains $Y^{(2,9)}_{\mathbf{1}^{(1)}}$. For the neutrino sextet the corresponding Siegel modular form is a weight-2 theta series of the rank-four lattice $\langle4\rangle\oplus\langle8\rangle\oplus\langle12\rangle\oplus\langle24\rangle$ of level $48$, in which the multiplier $\eps^2$ is part of the transformation of the theta series. These examples give explicit Siegel realizations. Their minimal possible levels remain open.

\section{Allowed Fourier powers at the corner}\label{sec:corner-rule}
At the corner every Fourier term is a monomial $q_1^{n_1}q_2^{n_2}q_3^{n_3}$ in the three small parameters, see eq.~\eqref{eq:corner-monomial}. The exponents are restricted by two conditions.

\paragraph{Positivity.}
Since $\ell=n_1+n_2-n_3$, the condition $\ell^2\le4n_1n_2$ of eq.~\eqref{eq:full-Fourier} becomes
\begin{equation}\label{eq:Heron}
 4n_1n_2-\ell^2=(s_1+s_2+s_3)(-s_1+s_2+s_3)(s_1-s_2+s_3)(s_1+s_2-s_3)\ge0\,,\qquad s_i=\sqrt{n_i}\,.
\end{equation}
The right-hand side is Heron's formula, sixteen times the squared area of a triangle with sides $s_i$. Hence
\begin{equation}\label{eq:triangle}
 q_1^{n_1}q_2^{n_2}q_3^{n_3}\ \text{can appear only if}\ \sqrt{n_1},\ \sqrt{n_2},\ \sqrt{n_3}\ \text{satisfy the triangle inequalities,}
\end{equation}
degenerate triangles included. For example, no power of a single $q_i$ can appear, and a monomial in two of the $q_i$ must have equal exponents. The rule is symmetric under permutations of the $q_i$, as it must be, since the corner group permutes them. Geometrically, a positive semi-definite matrix $T$ is the Gram matrix of two vectors $u,v$, so the exponents $n_1,n_2,n_3$ are proportional to the squared side lengths $|u|^2,|v|^2,|u-v|^2$ of the triangle they span. In the theta-series exponent $\im[\tau\ell^2+2z\,\ell\cdot\ell'+\sigma\ell'^2]=y_1|\ell|^2+y_2|\ell'|^2+y_3|\ell-\ell'|^2$, the cone coordinates $y_i$ act as Schwinger proper times on the three edges of a two-loop theta graph (the tropical limit of a genus-two worldsheet~\cite{Tourkine:2013rda,DHoker:2014oxd}, and the dual graph of the corner degeneration in figure~\ref{fig:degenerations}); the vectors $\ell,\ell^{\prime}$ and $\ell-\ell^{\prime}$ obey momentum conservation at a vertex, and their lengths satisfy eq.~\eqref{eq:triangle}. The group $D_{12}\cong S_3\times\Z_2$ is the automorphism group of the graph.

\paragraph{Translation charges.}
By eq.~\eqref{eq:corner-action}, a translation multiplies $q_1^{n_1}q_2^{n_2}q_3^{n_3}$ by $\omega^{n_1t_1+n_2t_2+n_3t_3}$. A component of a multiplet with a definite translation character therefore contains only monomials whose exponents have fixed values modulo three, $(n_1,n_2,n_3)\equiv(\hat{n}_1,\hat{n}_2,\hat{n}_3)$ with $\hat{n}_i\in\{0,1,2\}$. We call $\hat{\mathbf{n}}$ the \emph{class} of the component. The theta functions have the classes
\begin{equation}\label{eq:theta-classes}
 Y_1:(0,0,0)\,,\qquad Y_2:(1,0,1)\,,\qquad Y_3:(0,1,1)\,,\qquad Y_4:(1,1,1)\,,\qquad Y_5:(1,1,0)\,,
\end{equation}
and the class of a product is the sum of the classes modulo three.

Combining the two conditions gives a simple rule. In every class there is a smallest allowed monomial, which divides all the others:
\begin{equation}\label{eq:least-monomial}
 \mathbf{n}_*=\begin{cases}\hat{\mathbf{n}}&\text{if }\hat{\mathbf{n}}\text{ satisfies eq.~\eqref{eq:triangle}}\,,\\
 \hat{\mathbf{n}}\text{ with every zero entry replaced by }3&\text{otherwise.}\end{cases}
\end{equation}
To see this, note that a vector in $\{0,1,2\}^3$ fails eq.~\eqref{eq:triangle} exactly when it has a zero entry and its other two entries differ. A zero exponent would then require two exponents with different residues to be equal, which is impossible, so that exponent must be at least three. The resulting vectors $(3,1,2)$, $(3,3,1)$, $(3,3,2)$ and their permutations satisfy the triangle inequalities. All classes are listed in table~\ref{tab:corner-rule}. Every component of a coupling therefore has the form
\begin{equation}\label{eq:least-power-factor}
 Y^{(k)}_{\mathbf{r},\hat{\mathbf{n}}}=q_1^{n_{*1}}q_2^{n_{*2}}q_3^{n_{*3}}\bigl[c_{\hat{\mathbf{n}}}+\mathcal{O}(q_1^3,q_2^3,q_3^3)\bigr]\,,
\end{equation}
where the corrections vanish at the corner. If $c_{\hat{\mathbf{n}}}$ is nonzero, this monomial dominates along every approach with all $q_i\to0$. Its coefficient is determined by the chosen modular form and its permutation law.

\begin{table}[t]
\centering\small\setlength{\tabcolsep}{4pt}\tablesetup[1.3]
\begin{tabular}{cccccc}\toprule
class $\hat{\mathbf{n}}$ & number of classes & least allowed monomial & degree & theta example & representations\\\midrule
$(0,0,0)$ & 1 & $1$ & 0 & $Y_1$ & $\mathbf{1}_0^\pm,\mathbf{2}_0$\\
$(0,1,1)$ & 3 & $q_2q_3$ & 2 & $Y_3$ & $\mathbf{3}_{0,1}^\pm$\\
$(1,1,1)$ & 1 & $q_1q_2q_3$ & 3 & $Y_4$ & $\mathbf{1}_1^\pm,\mathbf{2}_1$\\
$(0,2,2)$ & 3 & $q_2^2q_3^2$ & 4 & $Y_3^2$ & $\mathbf{3}_{0,2}^\pm$\\
$(1,1,2)$ & 3 & $q_1q_2q_3^2$ & 4 & $Y_2Y_3$ & $\mathbf{3}_{2,1}^\pm$\\
$(1,2,2)$ & 3 & $q_1q_2^2q_3^2$ & 5 & $Y_3Y_4$ & $\mathbf{3}_{1,2}^\pm$\\
$(0,1,2)$ & 6 & $q_1^3q_2q_3^2$ & 6 & $Y_2^2Y_5$ & $\mathbf{6}$\\
$(2,2,2)$ & 1 & $q_1^2q_2^2q_3^2$ & 6 & $Y_4^2$ & $\mathbf{1}_2^\pm,\mathbf{2}_2$\\
$(0,0,1)$ & 3 & $q_1^3q_2^3q_3$ & 7 & $Y_4Y_5^2$ & $\mathbf{3}_{1,0}^\pm$\\
$(0,0,2)$ & 3 & $q_1^3q_2^3q_3^2$ & 8 & $Y_4^2Y_5$ & $\mathbf{3}_{2,0}^\pm$\\\bottomrule
\end{tabular}
\caption{The Fourier-power restriction. The $27$ classes, i.e.\ the characters of the translations, are grouped into ten sets related by permutations; one representative is shown. For the classes $(0,1,2)$, $(0,0,1)$ and $(0,0,2)$ the triangle inequality raises the degree. The theta example is a theta monomial in the class with a non-zero leading coefficient. The last column lists the representations of the corner group $\Z_3^3\rtimes S_3$ in the notation of appendix~\ref{app:reps-corner} whose translation characters lie in the set. Altogether the group has six singlets, three doublets, twelve triplets and one sextet.}\label{tab:corner-rule}
\end{table}

The rule fixes the least allowed power; whether its coefficient $c_{\hat{\mathbf{n}}}$ is non-zero depends on the coupling. A permutation that maps a class to itself can force the coefficient to vanish. For example, the weight-3 combination $Y_2^3-Y_3^3$ has class $(0,0,0)$ but is odd under $q_1\leftrightarrow q_2$, and
\begin{equation}\label{eq:corner-cancellation-example}
 Y_2^3-Y_3^3=27\,q_3^3\bigl(q_1^3-q_2^3\bigr)+\dots\,,
\end{equation}
so that its leading term depends on whether $|q_1|$ or $|q_2|$ is larger. The theta functions themselves have non-zero leading coefficients,
\begin{equation}\label{eq:corner-leading-theta}
 Y_1\simeq1\,,\qquad Y_2\simeq3q_1q_3\,,\qquad Y_3\simeq3q_2q_3\,,\qquad Y_4\simeq6q_1q_2q_3\,,\qquad Y_5\simeq3q_1q_2\,,
\end{equation}
where the numbers count the shortest lattice vectors.

\apptocmd{\thebibliography}{\setlength{\itemsep}{3pt plus 0.2ex}}{}{}
\bibliographystyle{JHEP}
\bibliography{references}
\end{document}